\documentclass[a4paper,11pt]{article}
\pdfoutput=1 

\usepackage{jheppub} 
\usepackage{amsmath}
\usepackage{graphicx,tabularx}
\usepackage{url}
\usepackage{footmisc}
\usepackage{amsfonts}
\usepackage{cancel}
\usepackage{color}
\usepackage{multirow} 
\usepackage{amssymb}
\usepackage{pifont}
\usepackage{subfigure}
\usepackage{epstopdf}
\usepackage{slashed}
\usepackage{comment}
\usepackage{booktabs} 
\usepackage{natbib}
\usepackage{hyperref}
\usepackage{array}
\usepackage{feynmp}
\newcommand{\met}{\ensuremath{{\not\mathrel{E}}_T}}

\def\met{\ensuremath{\slashed{E}_T}}
\def\n{\ensuremath{\chi^0_1}}
\def\nn{\ensuremath{\chi^0_2}}

\def\cha{\ensuremath{\chi^\pm_1}}

\title{Unblinding the Dark Matter Blind Spots}
\author[a,b,c]{Tao Han,}
\author[d,e,f]{Felix Kling,}
\author[d]{Shufang Su,}
\author[b,a]{Yongcheng Wu}
\affiliation[a]{PITT PACC, Department of Physics and Astronomy, University of Pittsburgh, Pittsburgh, PA 15260, USA }
\affiliation[b]{Department of Physics, Tsinghua University, Beijing, 100086, China}
\affiliation[c]{Collaborative Innovation Center of Quantum Matter, Beijing, 100086, China}
\affiliation[d]{Department of Physics, University of Arizona, Tucson, Arizona 85721, USA}
\affiliation[e]{Fermilab, P.O. Box 500, Batavia, IL 60510, USA}
\affiliation[f]{Department of Physics and Astronomy, University of California, Irvine, CA 92697, USA}
\emailAdd{than@pitt.edu}
\emailAdd{fkling@uci.edu}
\emailAdd{shufang@email.arizona.edu}
\emailAdd{wuyongcheng12@mails.tsinghua.edu.cn}

\preprint{
\begin{flushright}
FERMILAB-PUB-16-565-T\\
PITT-PACC-1618\\
UCI-HEP-TR-2016-23
\end{flushright}
}

\abstract{
The dark matter (DM) blind spots in the Minimal Supersymmetric Standard Model (MSSM) refer to the parameter regions where the couplings of the DM particles to the $Z$-boson or the Higgs boson are almost zero, leading to vanishingly small signals for the DM direct detections. 
In this paper, we carry out comprehensive analyses for the DM searches under the blind-spot scenarios in MSSM. Guided by the requirement of acceptable DM relic abundance,  we explore the complementary coverage for the theory parameters at the LHC, the projection for the future underground DM direct searches, and the indirect searches from the relic DM annihilation into photons and neutrinos. 
We find that  
(i) the spin-independent (SI) blind spots may be rescued by the spin-dependent (SD) direct detection in the future underground experiments, and possibly by the indirect DM detections from IceCube and SuperK neutrino experiments;
(ii) the detection of gamma rays from Fermi-LAT may not reach the desirable sensitivity for searching for the DM blind-spot regions; 
(iii) the SUSY searches at the LHC will substantially extend the discovery region for the blind-spot  parameters. The dark matter blind spots thus may be unblinded with the collective efforts in future DM searches. } 

\begin{document}

\titlepage

\maketitle

\newpage


\flushbottom

\section{Introduction}

The weakly interacting massive particles (WIMPs), which appear in many theories beyond the Standard Model (SM), remain to be one of the most attractive candidates for cold dark matter (DM) to explain the observed energy budget in the universe. The relic abundance of dark matter particles is set by their annihilation cross section $\sigma \propto g_{\rm eff}^4/M_{\rm DM}^2$ in the early universe \cite{Lee:1977ua,Goldberg:1983nd} 
\begin{equation}
\Omega_{DM} h^2 = 0.11 \times \left(\frac{2.2 \times 10^{-26}~{\rm cm}^3/{\rm s}}{\langle \sigma v \rangle_{{\rm freeze}}}\right) .
\label{eq:relic}
\end{equation}
To avoid over-closure of the universe, today's relic abundance $\Omega_{DM} h^2 \sim 0.1$ translates to a limit on the dark matter mass as
\begin{equation}
M_{\rm DM} < 1.8~{\rm TeV} \left(\frac{g^2_{\rm eff}}{0.3}\right) .
\end{equation}
The electroweak coupling strength and the TeV mass scale naturally appear, leading to the notion of the ``WIMP miracle''. This strongly motives the search for the WIMP dark matter in the underground laboratories \cite{CDEX,XENON1T,XENON1T1512,PICO-2L,PICO-60,LUX-SD,LUX-SI,PandaX-II-SI,PandaX-II-SD,LZ-CDR,1509-08767},
in collider experiments \cite{CMSmonojetPublished,ATLASmonojet}, as well as indirect detections via gamma rays, positrons and neutrinos \cite{AMS02-Ele1,AMS02-Ele2,AMS02-Ele3,IC79-2016,SuperK,ANTARES-SUN-DM,FermiLAT-6yr-Gamma,Conrad:2014tla}.  

With the impressive improvement of sensitivities in the underground experiments \cite{PandaX-II-SI,PandaX-II-SD,LUX-SI,LUX-SD} for the dark matter direct detection, the null results have put stringent limits on the dark matter-nucleon scattering cross sections, excluding much of the parameter region for many WIMP dark matter models and thus challenging the WIMP miracle paradigm. On the other hand, the WIMP DM interactions with the SM particles and the mass spectra may be rather subtle. The annihilation cross section that governs the relic abundance and the dark matter-nucleon scattering cross section that controls the direct detection may not be from the same set of diagrams. It is therefore prudent to explore scenarios with suppressed dark matter-nucleon scattering cross sections.

Supersymmetry (SUSY) remains to be the strongest contender for theories beyond the SM. One of the desirable features for SUSY is the existence of a WIMP dark matter candidate, the lightest supersymmetric particle (LSP), typically the neutralino ($\chi_1^0$). It has been realized recently that there are regions in the SUSY parameter space where the direct detection cross section is highly suppressed due to subtle cancelations of the couplings. These regions are dubbed as the  ``blind spots'' \cite{blindspots} for the DM direct detection. 
%
It has been shown \cite{blindspots} that the DM coupling to the $Z$-boson $Z\chi_1^0\chi_1^0$
can be almost zero and thus the spin-dependent (SD) scattering amplitude will be vanishingly small. Analogously, the DM coupling to the Higgs boson $h\chi_1^0\chi_1^0$
can be almost zero, leading to the spin-independent (SI) scattering amplitude to be vanishingly small. These would be very unfortunate scenarios as far as the DM direct detections are concerned. One would wish to seek for other possible means to search for the WIMP dark matter in those parameter regions. 

In this paper, we carry out comprehensive analyses for the DM searches under the blind-spot scenarios in MSSM. In particular, we explore the complementary coverage for the theory parameters among the different searching schemes for the direct detections, the indirect detections with astro-physical means, as well as the collider searches for SUSY signals \cite{1405-6716,1409-6322,1412-5952,1501-06357,1508-01173,1509-05076,1509-05771}.

We find that the SI scattering blind spots may be rescued by SD scattering searches in the future direct detection experiments. The neutrino detections from IceCube and SuperK could approach the sensitivity on the SD scattering cross section for certain blind spots, while the detection of gamma rays from Fermi-LAT may not reach the desirable sensitivity for searching for the DM blind-spot regions.
Furthermore, the SUSY searches at the LHC, in particular the future upgrade to higher luminosities (HL-LHC), will substantially extend the coverage for the blind-spot scenarios to large parameter regions. 
The dark matter blind spots thus may be unblinded with the collective efforts in future DM searches.

This paper is organized as follows. In Sec.~\ref{sec:BS}, we define the blind spots for the spin-independent and spin-dependent scatterings in DM searches. We further study the constraints from the relic abundance for those scenarios in Sec.~\ref{sec:RD}. We discuss the DM direct detection via the spin-dependent scattering, and quantify the DM indirect searches at the Fermi-LAT from gamma rays and at the neutrino experiments in Sec.~\ref{sec:search}. Before presenting our collider studies at the LHC, we first examine the existing bounds on the SUSY parameters in Sec.~\ref{sec:LEP}. Our main results for the collider coverage are presented in Sec.~\ref{sec:collider}. We summarize our results in Sec.~\ref{sec:conclusion}.  


\section{Blind Spots}
\label{sec:BS}

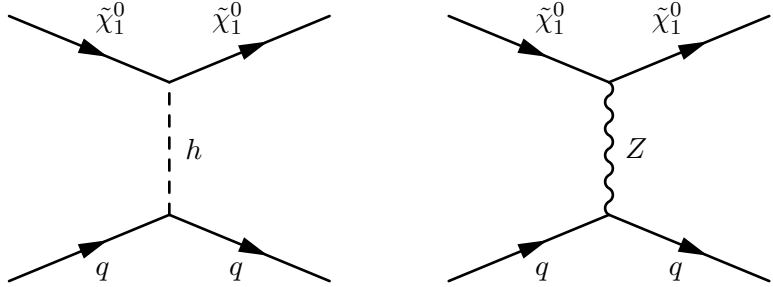
\begin{figure}[tb]
\centering
\begin{fmffile}{SI}
\begin{fmfgraph*}(150,100)
\fmfleft{i1,i2}
\fmfright{o1,o2}
\fmf{fermion,label=$\tilde{\chi}_1^0$,label.side=left}{i2,v2}
\fmf{fermion,label=$\tilde{\chi}_1^0$,label.side=left}{v2,o2}
\fmf{dashes,label=$h$}{v1,v2}
\fmf{fermion,label=$q$,label.side=right}{i1,v1}
\fmf{fermion,label=$q$,label.side=right}{v1,o1}
\end{fmfgraph*}
\end{fmffile}
\begin{fmffile}{SD}
\begin{fmfgraph*}(150,100)
\fmfleft{i1,i2}
\fmfright{o1,o2}
\fmf{fermion,label=$\tilde{\chi}_1^0$,label.side=left}{i2,v2}
\fmf{fermion,label=$\tilde{\chi}_1^0$,label.side=left}{v2,o2}
\fmf{boson,label=$Z$}{v1,v2}
\fmf{fermion,label=$q$,label.side=right}{i1,v1}
\fmf{fermion,label=$q$,label.side=right}{v1,o1}
\end{fmfgraph*}
\end{fmffile}
\caption{Feynman diagrams for direct detection of SUSY neutralino LSP dark matter.}
\label{fig:feynDD}
\end{figure}

Direct detections of the SUSY WIMP dark matter ($\chi_1^0$) in the underground laboratories usually go through two classes of scattering diagrams via the Higgs and $Z$-boson exchanges, 
as shown in Fig.~\ref{fig:feynDD}. The WIMP scattering cross section sensitively depends on the couplings of $h\chi_1^0\chi_1^0$ and $Z\chi_1^0\chi_1^0$, which are governed by the components of the $\chi_1^0$ admixture. 
It is sometimes informative to think about the limiting cases, that for large $|M_1|$, $|M_2|$, $|\mu|\gg m_Z$, the lightest neutralino $\chi_1^0$ is bino-like, wino-like or Higgsino-like, with mass being approximately $M_1$, $M_2$, $\pm \mu$, whichever one is the smallest, respectively.
Furthermore, the neutralino LSP-nucleus scattering via the axial-vector interaction $Z\chi_1^0\chi_1^0$ couples to the spin of the nucleus (spin-dependent, SD) 
and that via the scalar interaction $h\chi_1^0\chi_1^0$ is independent of the spin (spin-independent, SI). 
The scattering cross section off a heavy nuclear target of an atomic number $A$ for the SI interactions will be proportional to $A^2$ due to the coherent effect of the nucleons.
DM direct detections are thus more sensitive to the SI interactions due to this enhancement. On the other hand, the SD interactions may still be significant because of the stronger gauge interactions via the $Z$ exchange.

\subsection{SI blind spots}

\begin{table}[tb]
\centering
\begin{tabular}{|c|c|c|}
\hline 
$m_{\chi_1^0}$ & condition & signs \\ 
\hline 
$M_1 (<M_2, |\mu|)$ & $M_1+\mu\sin2\beta=0$ & $\text{sign}(M_1/\mu)=-1$ \\ 
\hline 
$M_2 (<M_1, |\mu|)$ & $M_2+\mu\sin2\beta=0$ & $\text{sign}(M_2/\mu)=-1$ \\ 
\hline 
$|\mu| ( < M_1, M_2)$ & $\tan\beta=1$ & $\text{sign}(M_{1,2}/\mu)=-1$ \\ 
\hline 
$M_{1,2} (< |\mu|)$ & $M_1=M_2,\quad |\mu|>|{M_{1,2}}/{\sin2\beta}|$ & $\text{sign}(M_{1,2}/\mu)=-1$ \\ 
\hline 
\end{tabular}
\caption{\label{tab:SIBS}The SI blind-spot mass relations.  }
\end{table} 

The spin-independent (SI) blind spots correspond to a vanishing $h\chi_1^0\chi_1^0$ coupling.
Using the low-energy Higgs theorem, we obtain the following condition on the theory parameters \cite{blindspots}:
\begin{equation}
(m_{\chi_1^0}+\mu\sin2\beta)\left(m_{\chi_1^0}-\frac{1}{2}(M_1+M_2+ (M_1-M_2) \cos2\theta_{\text{W}})\right)=0.
\label{eq:SI_condition}
\end{equation}
From this relation, we can extract the tree level blind-spot conditions as given in Table~\ref{tab:SIBS}. The first three conditions are obtained by requiring the first bracket in Eq.~(\ref{eq:SI_condition}) to be zero. In this case, the LSP mass is exactly equal to the mass
of a pure gaugino or Higgsino state ($M_1$, $M_2$ or $|\mu|$) as listed in the first column of Table~\ref{tab:SIBS}. The last condition is obtained by requiring the second bracket to be zero. In this case the neutralino mass is equal to the gaugino mass parameters $m_{\chi_1^0}=M_1=M_2$. The additional condition on $|\mu|$ guarantees that the neutralino state with mass $M_1=M_2$ is the lightest neutralino and therefore the LSP. 

Two remarks are now in order. First, we note that loop corrections to the LSP mass will slightly shift the exact location of the blind spots, but will not affect their existence.
Second, if there is BSM new physics that has significant couplings to the LSP and the SM quarks, then the additional contributions could change the above blind-spot conditions. Examples include  the heavy Higgs boson \cite{blindspotsHeavyH}, light squark \cite{blindspotsStops}, and the singlet scalar in NMSSM \cite{blindspotsNMSSM}. We will not discuss those cases further. 


\subsection{SD blind spots}
\label{sec:bs}

\begin{table}[tb]
\centering
\begin{tabular}{|c|c|c|}
\hline 
$m_\chi$ & condition & signs \\ 
\hline 
$M_{1,2}$ & $M_1=M_2,\quad |\mu|>|{M_{1,2}}/{\sin2\beta}| $ & $\text{Sign}({M_{1,2}/\mu})=-1$ \\ 
\hline 
-- & $\tan\beta=1$ & -- \\ 
\hline 
\end{tabular}
\caption{\label{tab:SDBS}
The SD blind-spot mass relations.}
\end{table} 

The spin-dependent (SD) blind spots correspond to vanishing $Z\chi_1^0\chi_1^0$ coupling.  For a pure Higgsino LSP, $\chi_{1,2}^0=\frac{1}{\sqrt{2}}(\tilde{H}_u^0\pm \tilde{H}_d^0)$, the only non-zero coupling is $Z\chi_1^0\chi_2^0$, with vanishing $Z\chi_1^0\chi_1^0$ and $Z\chi_2^0\chi_2^0$ couplings.  Pure bino or a pure wino states have no interaction with $Z$-boson. For a mixed LSP state, the $Z\chi_1^0\chi_1^0$ coupling must come from the Higgsino component in $\chi_1^0$. For $\tan\beta=1$ with restored symmetry of $u\leftrightarrow d$, the LSP $\chi_1^0$ also has vanishing coupling with $Z$-boson as in the case of a pure states \cite{blindspots}.   

We have identified the following 
mass relation, corresponding to a vanishing Higgsino component in the LSP $-$ a photino-like LSP, 
which also leads to a SD blind spot:
\begin{equation}
M_1=M_2,\quad |\mu|>\left|\frac{M_{1,2}}{\sin2\beta}\right| ,\quad \text{Sign}(\frac{M_{1,2}}{\mu})=-1.
\label{eq:mostb}
\end{equation}
Such a case has not been pointed out in the existing blind-spot studies~\cite{blindspots}. In fact,
this is the same condition of a vanishing $h \chi_1^0 \chi_1^0$ coupling as in the last line of Table \ref{tab:SIBS}. Thus this region is both SD blind spots and SI blind spots $-$ it is the ``most blind'' of all! 
We note that, Eq.~(\ref{eq:mostb}) with Sign$(M_{1,2}/{\mu})=+1$ would also lead to a SD blind spot. However in this case, $M_{1,2}$ will be smaller than $M_Z/2$, and we will not study it further because of the conflict with the collider bounds on the chargino/neutralino masses.
The SD blind-spot conditions under our consideration are listed in Table~\ref{tab:SDBS}. 

In our analyses below, we choose the following benchmark cases which correspond to SI and/or SD blind-spot regions:
\begin{description}
\item[Case A (SI Blind Spots):] $M_1+\mu\sin2\beta=0,\ m_{\chi_1^0}=M_1$ with $M_2$ decoupled, sign$(\frac{M_1}{\mu})=-1$.
\item[Case B (SI Blind Spots):] $M_2+\mu\sin2\beta=0,\ m_{\chi_1^0}=M_2$ with $M_1$ decoupled, sign$(\frac{M_2}{\mu})=-1$.
\item[Case C (SI and SD Blind Spots):] $M_1=M_2,\ |\mu|>|\frac{M_{1,2}}{\sin2\beta}|$, sign$(\frac{M_{1,2}}{\mu})=-1$.
\end{description}
For each case, we use SuSpect~\cite{suspect} to generate the corresponding parameter points.

For our collider analyses, we focus on Cases A and B in details.  The collider phenomenology of Case C  would be similar to those of Case A and B given the nearly degenerate LSP and NLSPs made of bino and winos, with heavier Higgsino states. For the DM relic density and indirect constraints, we also study Case C.   

In fact, there is a condition that would lead to another blind spot both for SI and SD, namely 
$\tan\beta=1$, as already listed in Tables~\ref{tab:SIBS} and \ref{tab:SDBS}. 
We will not discuss this scenario any further since this value of $\tan\beta$ is disfavored given the observed Higgs mass and because the phenomenological features are similar to the usual electroweakino LSP studies, with no characteristic mass scale for the NLSP states. 
  
\section{Dark Matter Relic Density}
\label{sec:RD}
 
Today's relic density of DM from global fits to a variety of observations is~\cite{PlanckXIII}
\begin{equation}
\Omega_{DM} h^2 \approx 0.1184\pm 0.0012.
\label{eq:relicb}
\end{equation}
The relic density for light wino and Higgsino are always under-abundant due to the relatively large ${\rm SU(2)}_L$ coupling, while the relic density in the bino case is mostly over-abundant due to the suppressed annihilation cross section. 

\begin{figure}[tb]
\centering
\includegraphics[width=0.49\textwidth]{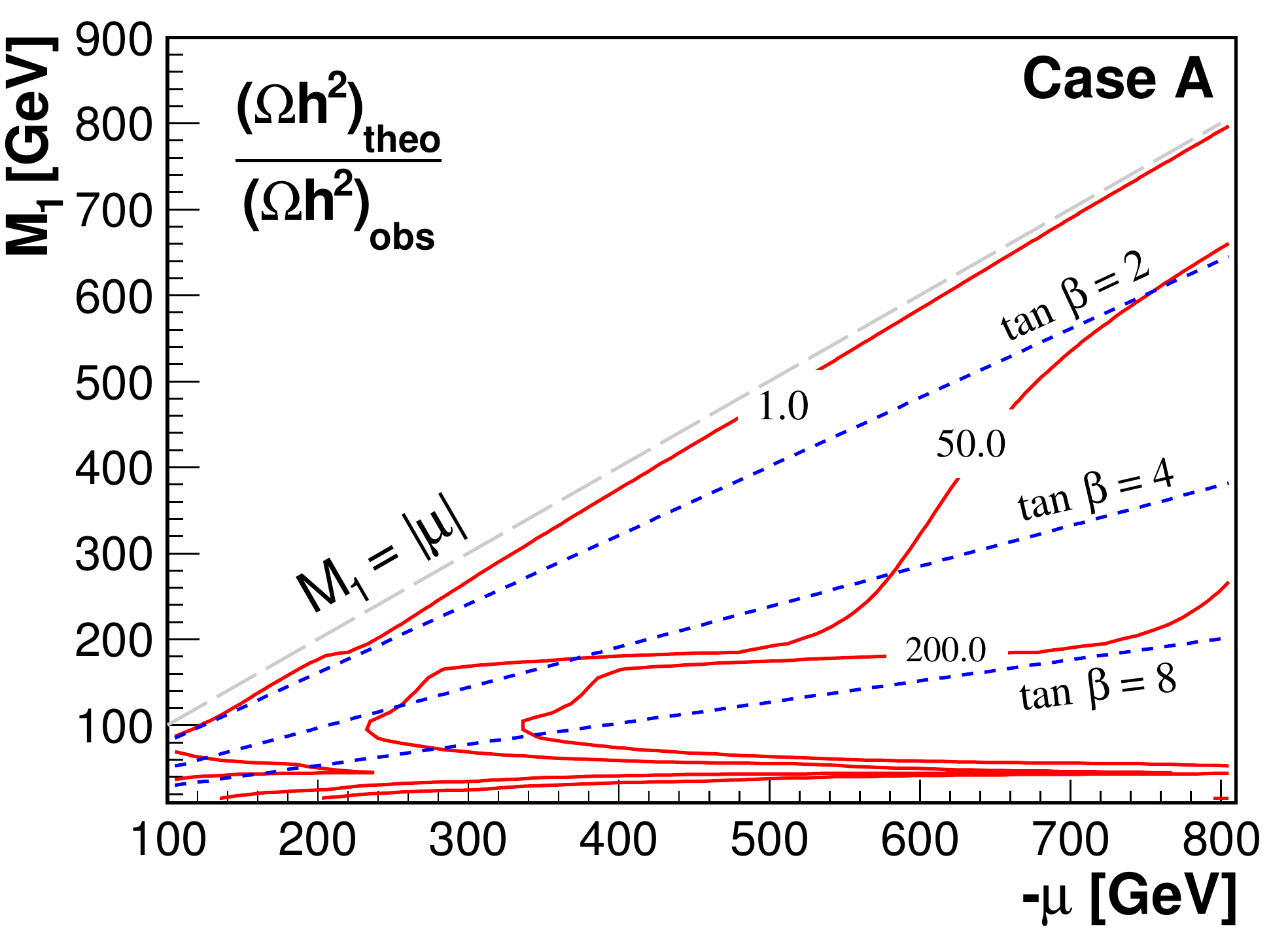}
\includegraphics[width=0.49\textwidth]{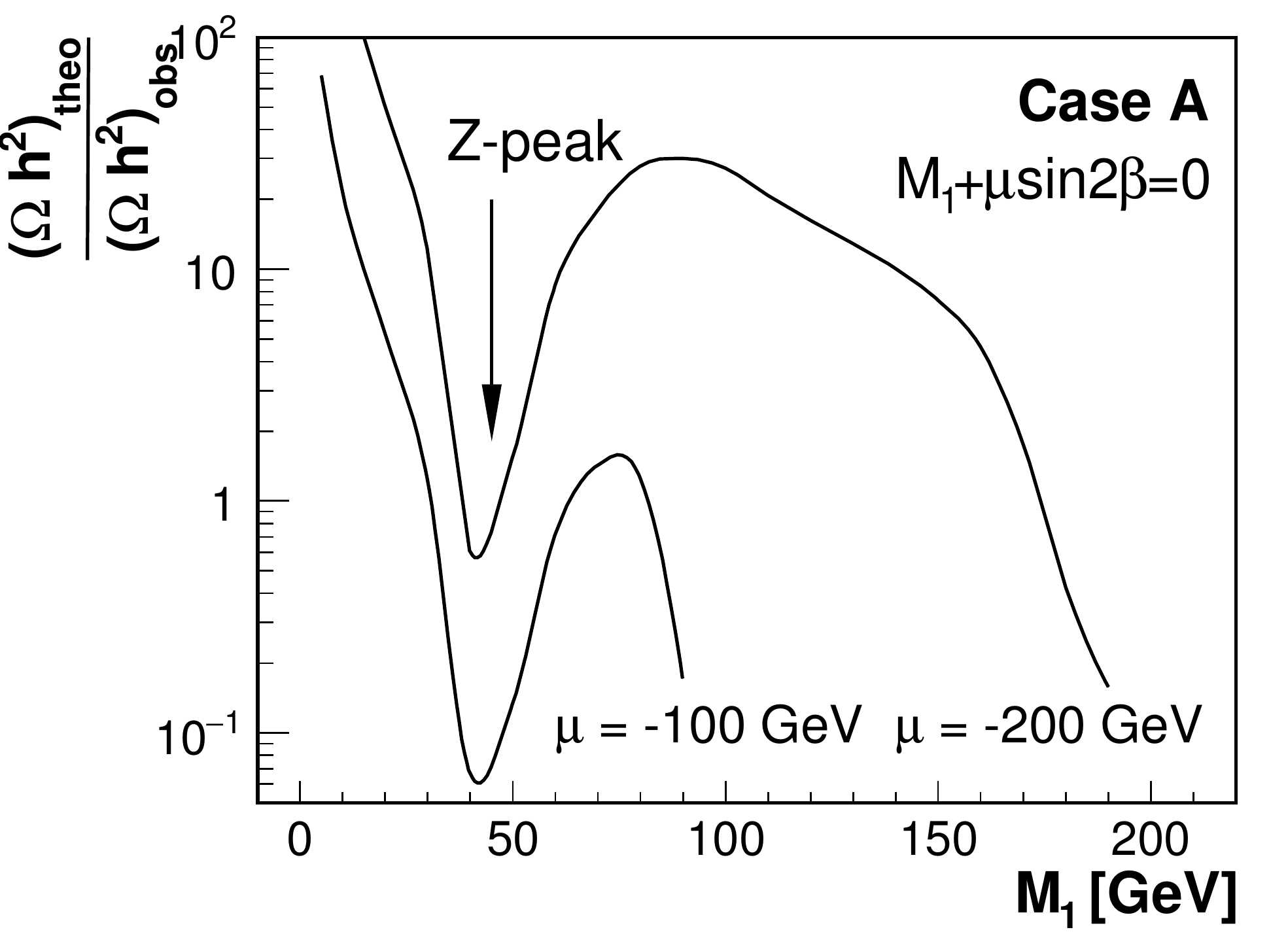}
\caption{Case A: dark matter relic density normalized to the observed value. Left panel: contour plot in the $M_1-\mu$ plane for various values of the relics. Right panel: as a function of $M_1$ for $\mu = -100$ and $-200$ GeV.  
The values of  $\tan\beta$ are fixed by the blind-spot relation 
$M_1+\mu \sin2 \beta=0$ for both plots.  They are also indicated by the dashed lines on the left-panel.
}
\label{fig:Relic_CaseA}
\end{figure}

Case A of the blind-spot region with the bino-like LSP is generically disfavored by relic density analysis, except for the $Z$-pole region of $M_1\sim m_Z/2$ with small $|\mu|$, or $M_1\sim |\mu|$ 
with small $\tan\beta$ and considerable amount of bino-Higgsino mixing, as shown by the various contours labelled by the relic density values normalized to the observed value in the left panel of Fig.~\ref{fig:Relic_CaseA}. We calculate the relic density and the direct/indirect detection cross section by the package micrOMEGAs~\cite{micromegas1305,micromegas1005,micromegas0803,micromegas0607}. 
In the right panel, we show the normalized relic density as a function of the dark matter mass parameter $M_1$, which clearly indicates the viable region near the $Z$-pole. Lower values of $|\mu |$ are more favorable. 

\begin{figure}[tb]
\centering
\includegraphics[width=0.49\textwidth]{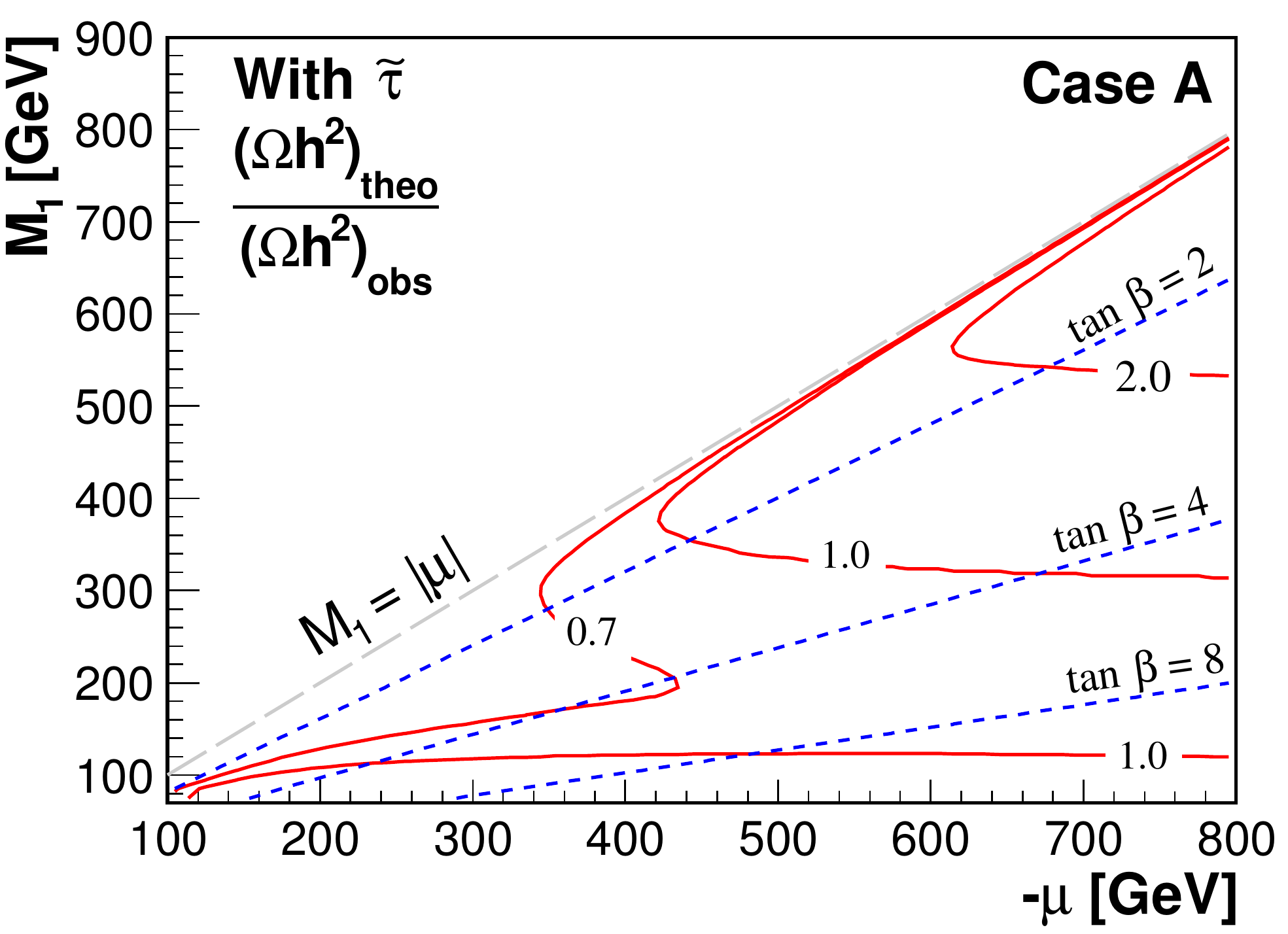}
\includegraphics[width=0.49\textwidth]{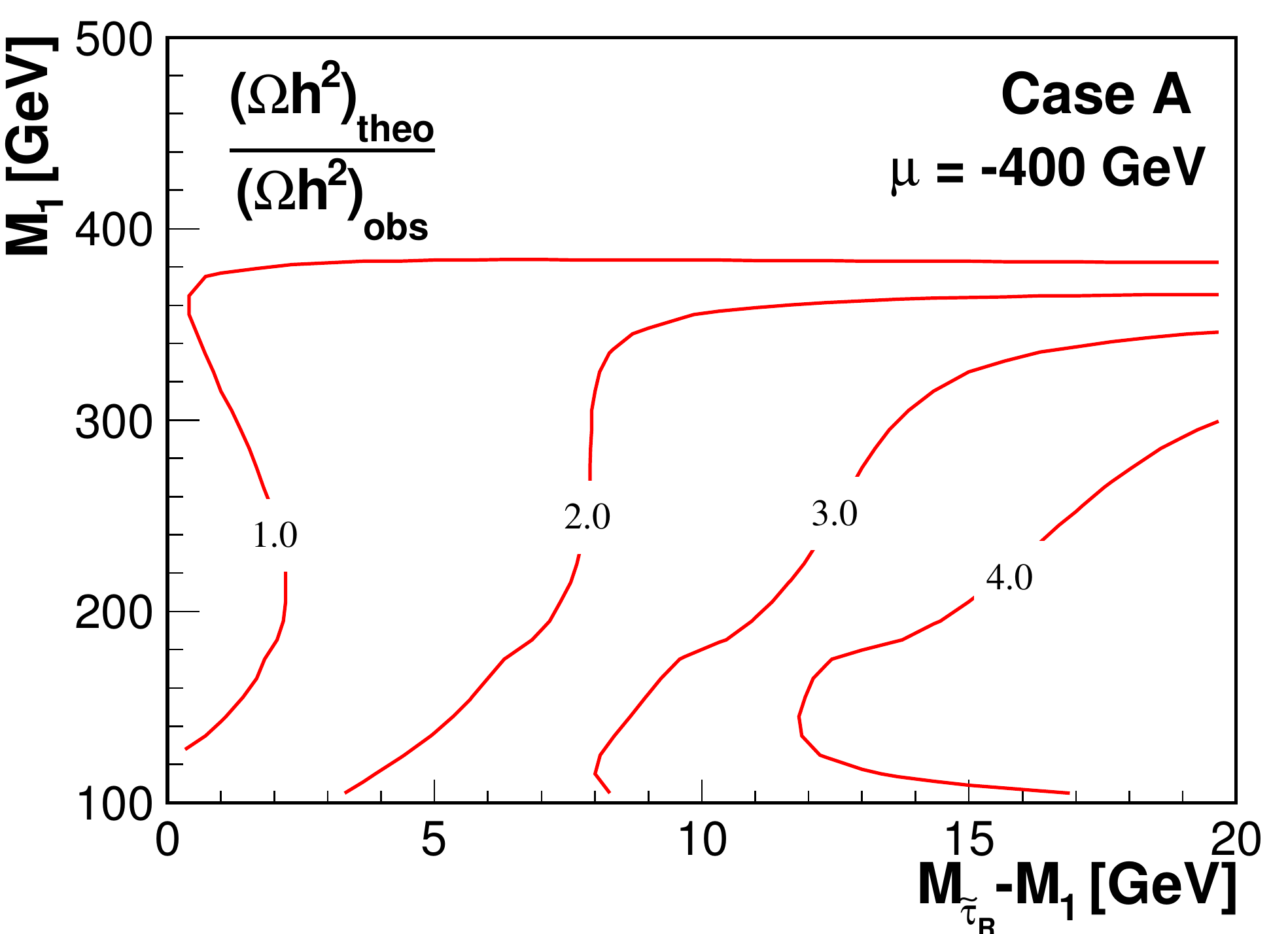}
\caption{Case A: dark matter relic density normalized to the observed value. Left panel: contour plot in the $M_1-\mu$ plane for various values of the relics and $\tan\beta$, including the co-annihilation effect with stau:  $M_{\tilde\tau_R}=M_1$. Right panel: contour plot in the $M_1$ versus $\delta M = M_{\tilde{\tau}_R}-M_{1}$ plane for $\mu = -400$ GeV. }
\label{fig:Relic_CaseA_stau}
\end{figure}

A possible situation to allow a large viable parameter space for Case A is to introduce co-annihilation for the LSP \cite{Ellis:1999mm,Han:2013gba}. In the left panel of Fig.~\ref{fig:Relic_CaseA_stau}, we present the normalized relic density for Case A with $M_{\tilde\tau_R}=M_1$ allowing the co-annihilation of the right-handed stau with the bino-like LSP. The observed relic density can be achieved in most of the $M_1$ versus $-\mu$ parameter space. The right panel of Fig.~\ref{fig:Relic_CaseA_stau} shows the effect of the 
$\delta M=M_{\tilde\tau_R}-M_1$ on the relic density. Only when the stau is nearly degenerate with LSP with $\delta M < 3$ GeV, the effect of stau co-annihilation is strong enough to suppress the relic density to achieve the observed value. Since such a nearly degenerate right-hand stau would hardly affect our further results on electroweakino sector,  we will always include such stau to accommodate acceptable relic density for Case A for any further study. 

\begin{figure}[tb]
\centering
\includegraphics[width=0.49\textwidth]{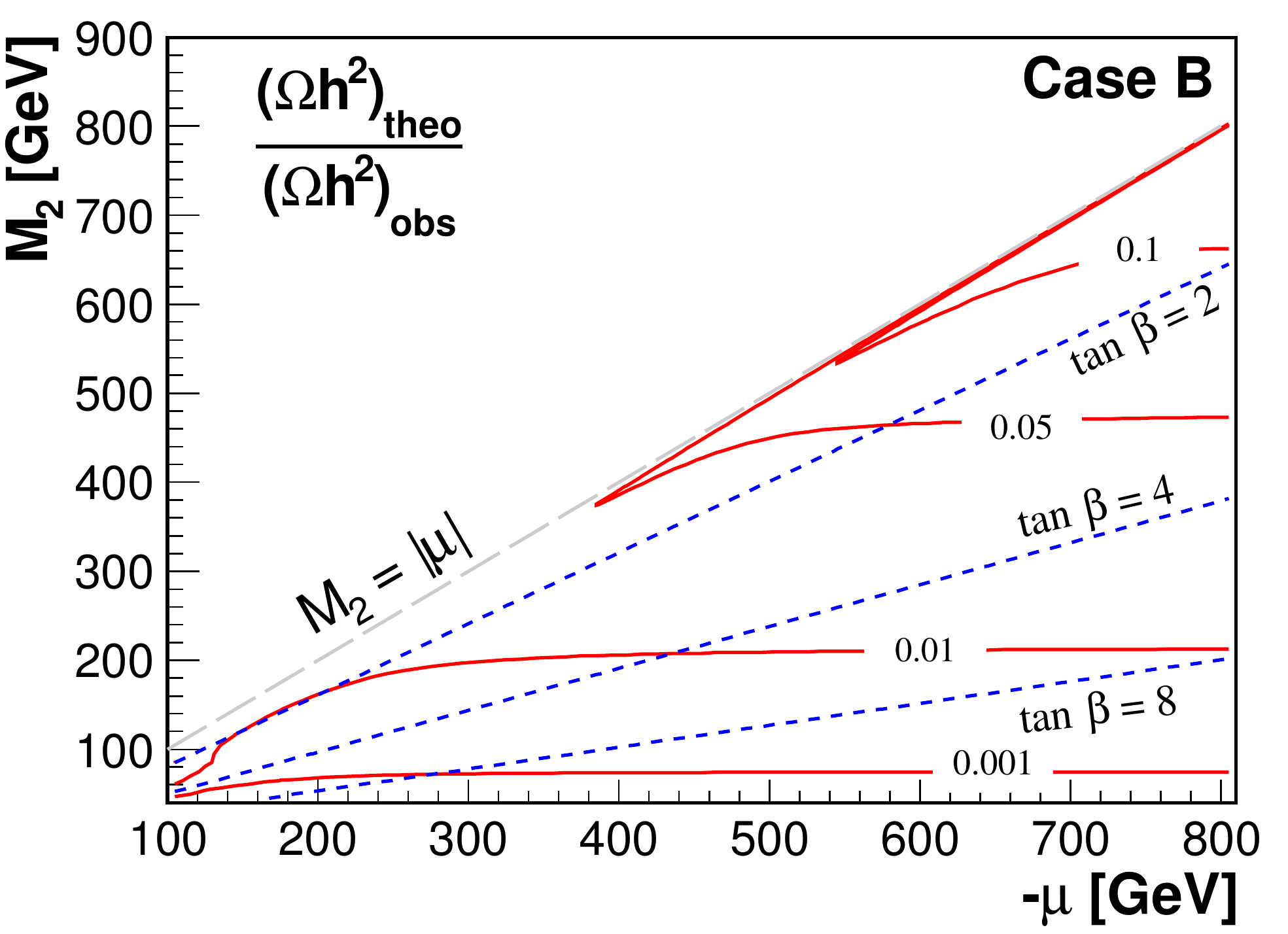}
\includegraphics[width=0.49\textwidth]{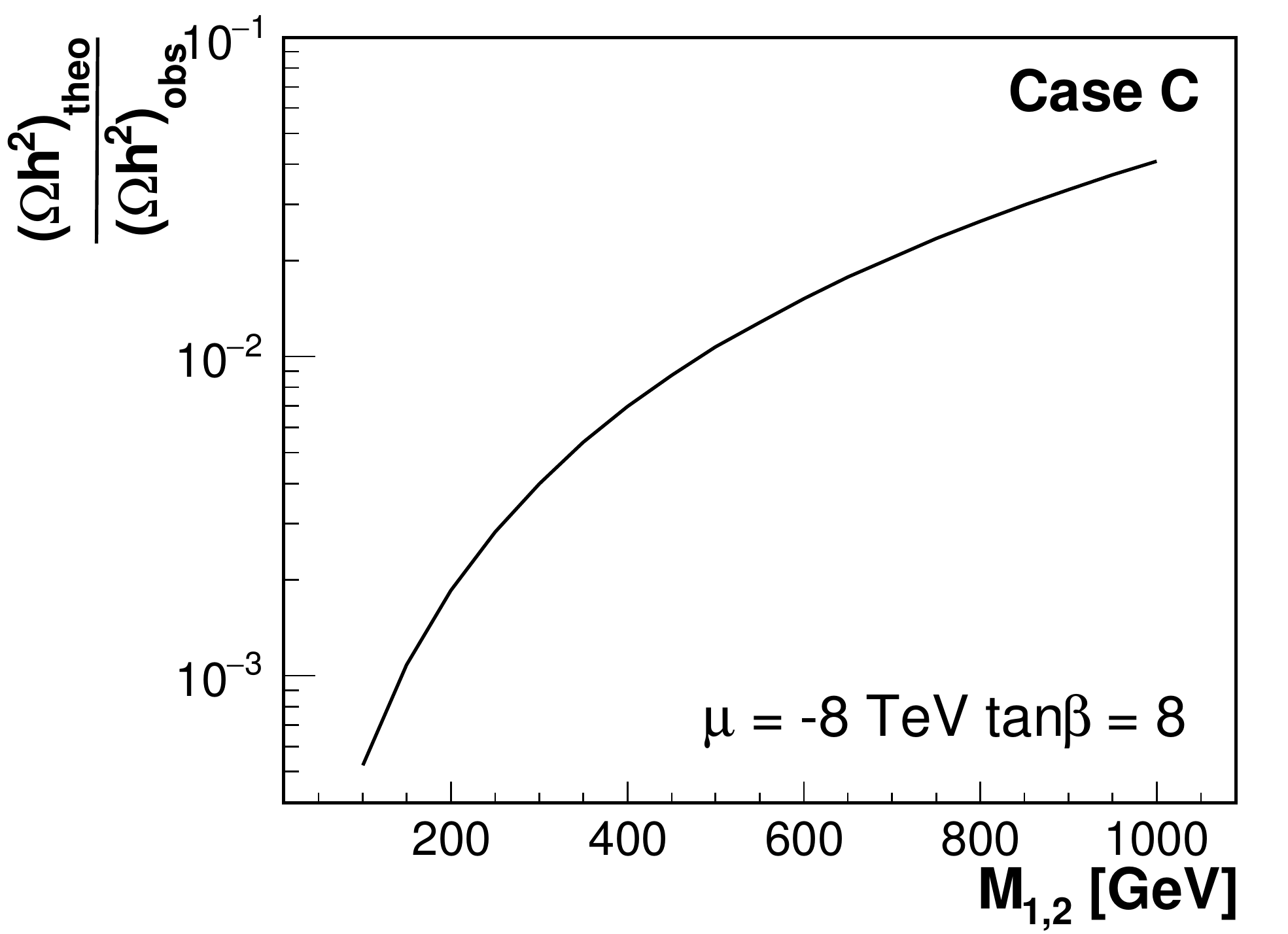}
\caption{Dark matter relic density normalized to the observed value. Left panel for Case B, in the $M_2 - \mu$ plane  for various values of the relics.
Right panel for Case C, versus $M_{1,2}$ for $\mu = -8$ TeV and $\tan\beta = 8$.
\label{fig:Relic_CaseBC} }
\end{figure}

However, in the wino-like LSP or Higgsino-like LSP case, the relic density is acceptably under-abundant. In Fig.~\ref{fig:Relic_CaseBC}, we show the relic density compared to observed value for Case B in $M_2$ versus $(-\mu)$ plane (left panel) and for Case C with $\mu=-8$ TeV and $\tan\beta = 8$ (right panel). For Case B,  the relic density in almost all the parameter region is acceptably under-abundant, only when the mass of the wino-like LSP is heavier than about 1.5 TeV, the relic density will achieve the observed value. For Case C as shown in the right panel of Fig.~\ref{fig:Relic_CaseBC}, the situation is similar to Case B. Since the wino annihilation dominates in this case, the relic density does not sensitively depend on the value of $\mu$ and $\tan\beta$.  

In the rest of the paper, whenever the relic density is concerned, we will adopt the value in Eq.~(\ref{eq:relicb}) for Case A, and will use the predicted results of the under-abundant relic densities for Cases B and C.


\section{Dark Matter Direct and Indirect Detections}
\label{sec:search}

There have been considerable efforts in searching for DM particles in the underground 
experiments \cite{PICO-2L,PICO-60,LUX-SD,LUX-SI,PandaX-II-SI,PandaX-II-SD,LZ-CDR}.
Since our Case A and Case B parameter regions are only SI-blind but have unsuppressed SD  cross sections, direct detection experiments aiming at constraining spin-dependent dark matter-nucleon scattering cross section~\cite{PICO-2L,PICO-60,LUX-SD,PandaX-II-SD,LZ-CDR} could provide important information for the theory parameter space.

The indirect detection of dark matter aims mainly at three kinds of detectable objects: gamma rays, charged-particle cosmic ray and neutrinos, from the relic DM annihilations. Constraints from indirect detection via gamma rays and charged-particle cosmic ray can be translated into upper bounds on the thermal averaged annihilation cross section for different dark matter annihilation channels.\footnote{This only works for an $s$-wave. For velocity dependent thermal averaged annihilation cross section, the indirect detection only constraints $\langle\sigma{v} \rangle_{\rm today}$, not $\langle\sigma{v} \rangle_{\rm freeze\ out}$.} Since limits from the charged-particle cosmic ray depend heavily on the propagation model,  we only  consider the constraints from gamma-ray detection~\cite{FermiLAT-6yr-Gamma}.  Furthermore, indirect dark matter detection via neutrino coming from the Sun can be used to put upper limits on the spin-dependent dark matter-proton scattering cross section~\cite{IC79-2016,SuperK,ANTARES-SUN-DM} under the equilibrium assumption. 
 

\subsection{Dark matter direct detection via SD scattering}

\begin{figure}
\centering
\includegraphics[width=0.49\textwidth]{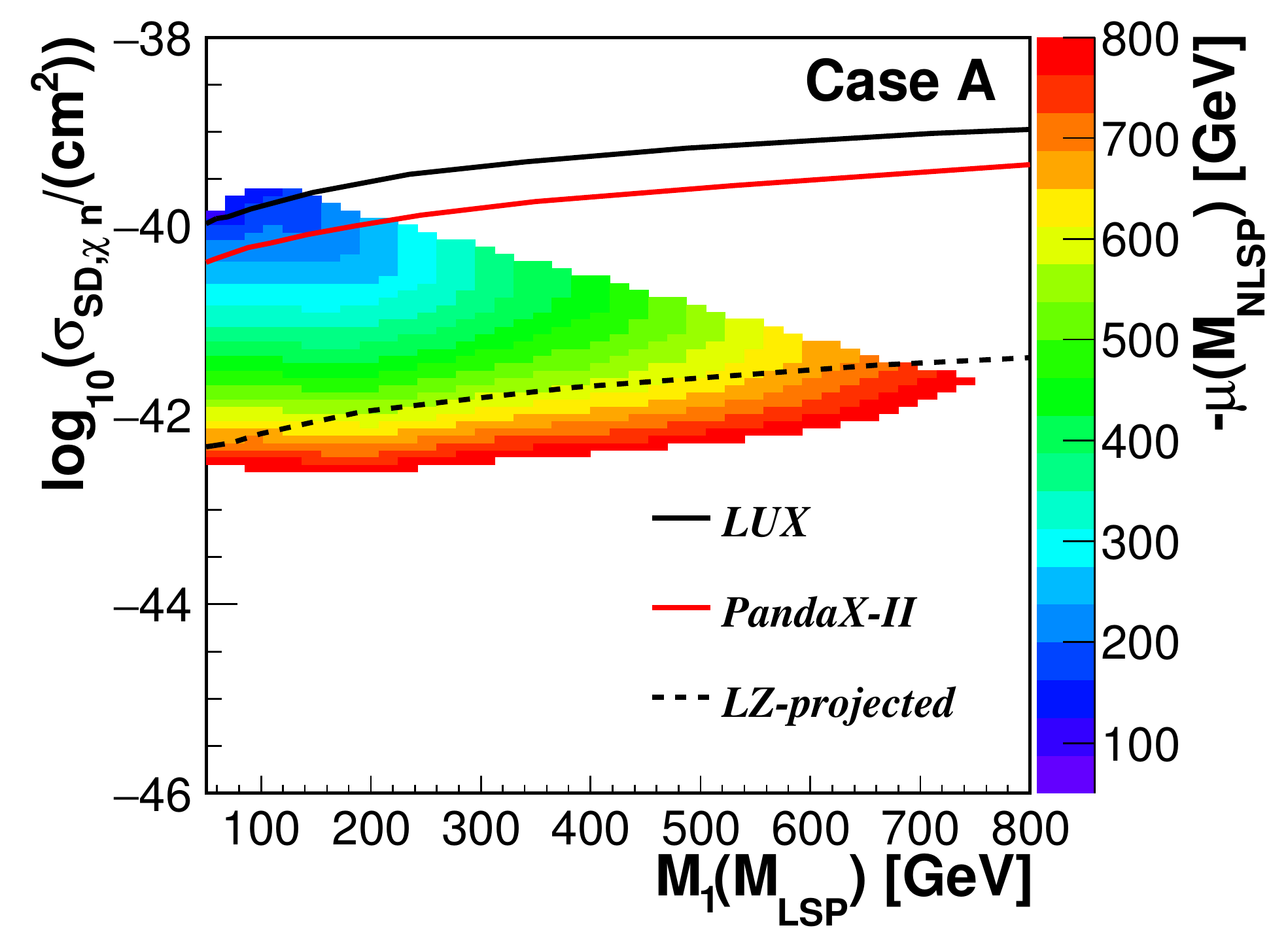}
\includegraphics[width=0.49\textwidth]{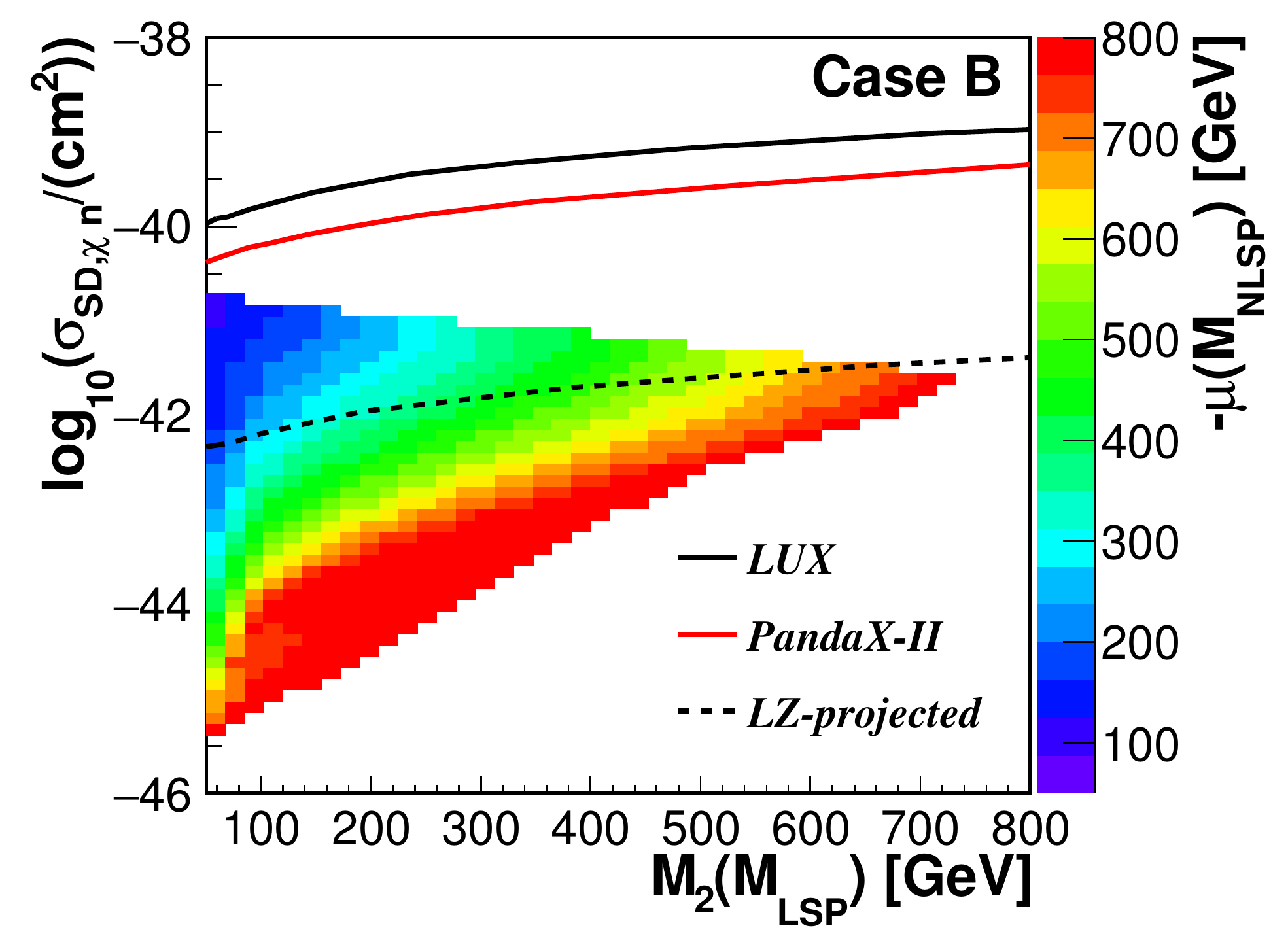}
\caption{
Spin-dependent DM-neutron scattering cross section: left panel for Case A and right panel for Case B.
The $90\%$ CL. limits from LUX, PandaX II, and the LZ projection are also shown. 
The color code indicates the $\mu$ values as labeled on the right-handed vertical axis.}
\label{fig:SigmaSDn}
\end{figure}

The DM search sensitivity has been improved substantially over the years \cite{1509-08767}. By design, the blind-spot scenarios are the most difficult situations for the direct detection. However, our Cases A and B are blind spots only for SI scattering, with unsuppressed cross sections for SD scattering. 
The spin-dependent DM-neutron scattering cross section $\sigma_{SD, \chi n}$ versus the LSP mass are presented in Fig.~\ref{fig:SigmaSDn} for Case A (left panel) and Case B (right panel).  
Note that, when calculating the cross section involving dark matter, we have properly treated the relic density for each case as stated at the end of Sec.~\ref{sec:RD}.
The constraints obtained from the direct detection are also shown there for LUX (upper solid curves), PandaX II (middle solid curves) and the LZ perspective (bottom dotted curves). 
From these figures, we find that LUX just reached the sensitivity near $M_1\sim 100$ GeV in Case A, and is still nearly one order of magnitude away in Case B. 
The bounds report by the PandaX II experiment \cite{PandaX-II-SD} can exclude a region in the parameter space up to $-\mu\approx 250$ GeV in Case A, but remains insensitivity in Case B.
%
The projected reach of the future DM direct detection experiment LZ \cite{LZ-CDR} for spin-dependent scattering is expected to cover parts of the blind-spot parameter regions with up to 700 GeV on $M_1$ for Case A and on $M_2$ for Case B as seen by the dotted curves.

\begin{figure}
\centering
\includegraphics[width=0.49\textwidth]{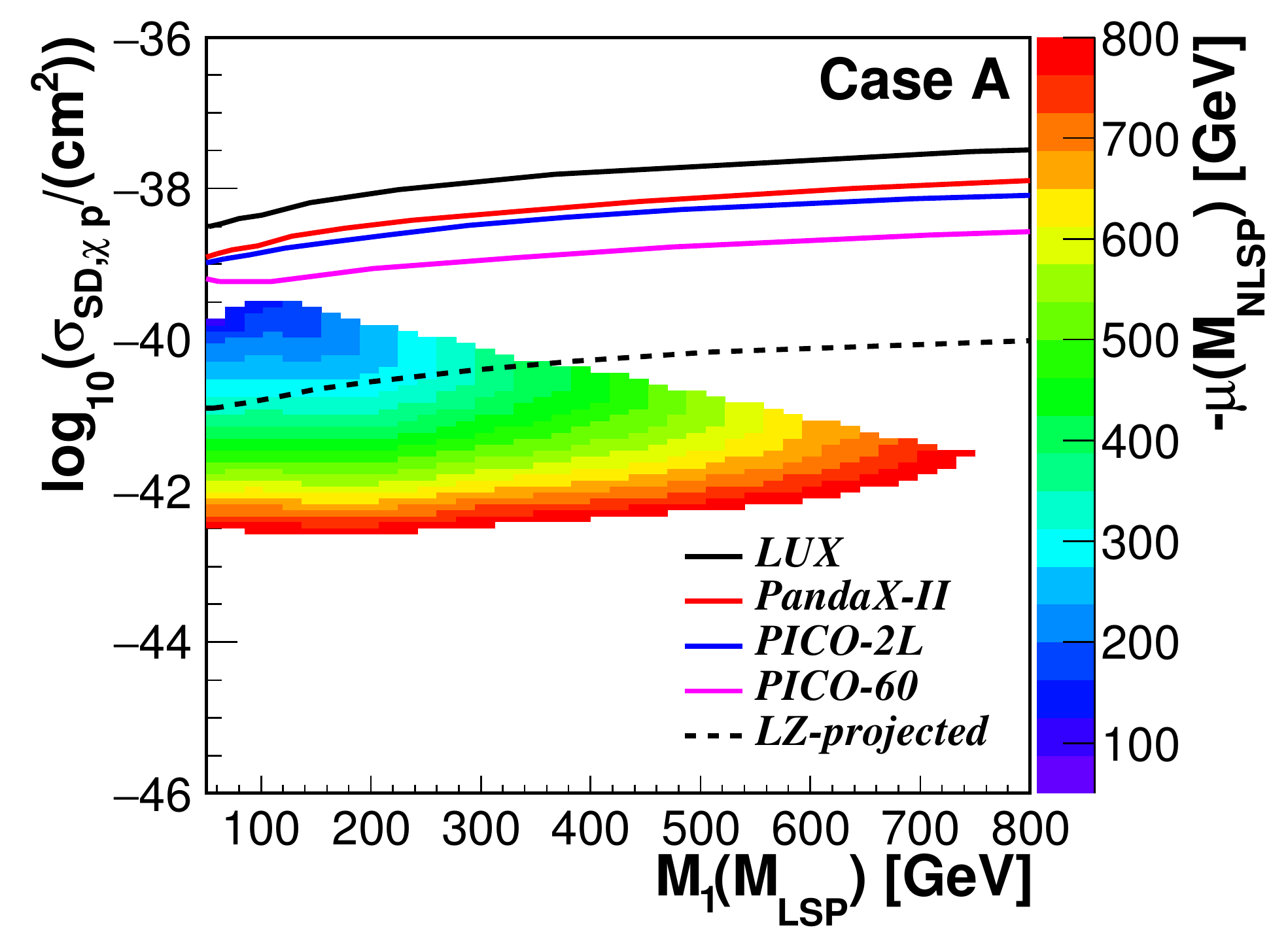}
\includegraphics[width=0.49\textwidth]{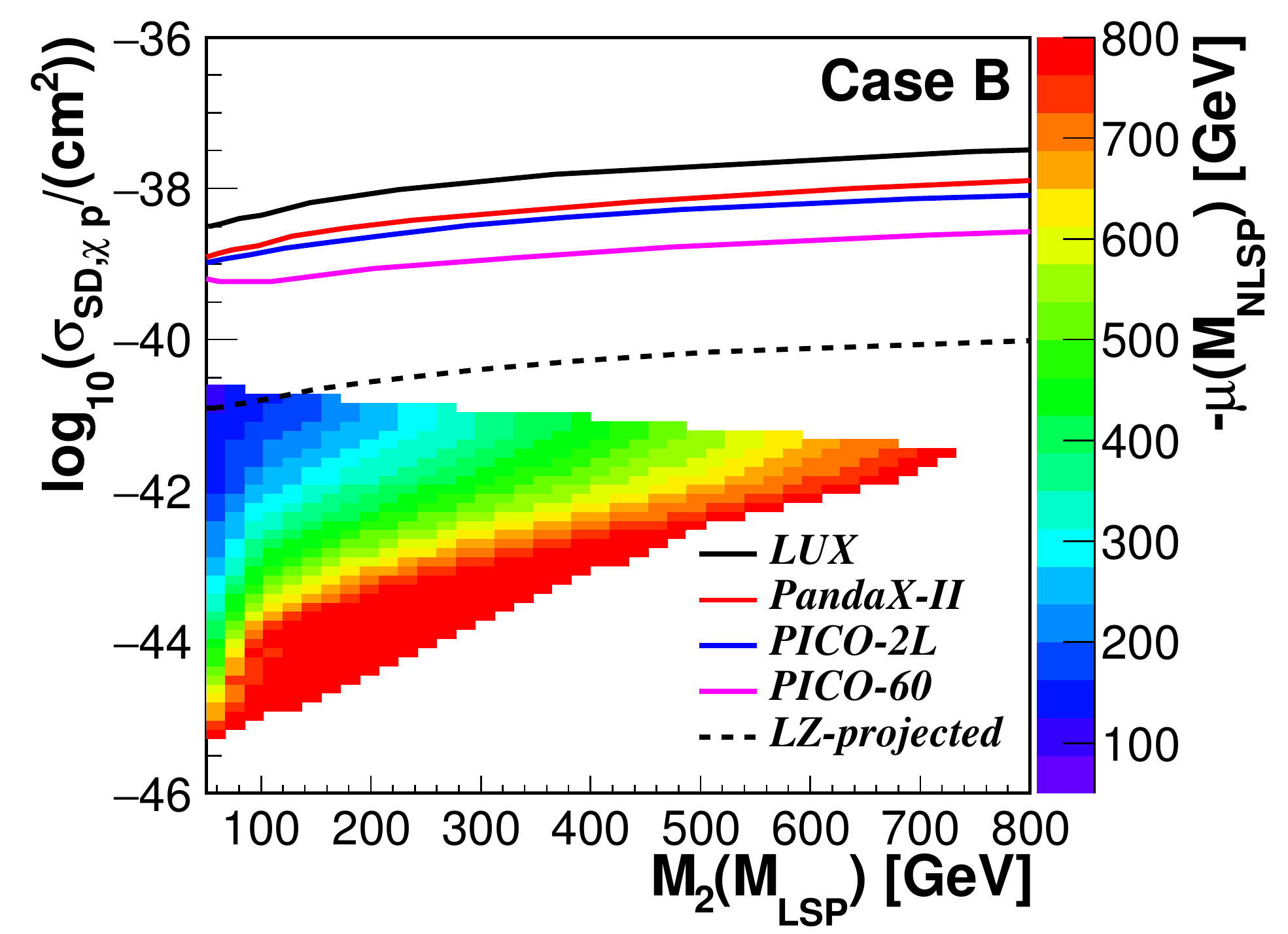}
\caption{
Spin-dependent DM-proton scattering cross section: left panel for Case A and right panel for Case B.
The $90\%$ CL. limits from LUX, PandaX II, PICO, and the LZ projection are also shown. 
The color code indicates the $\mu$ values as labeled on the right-handed vertical axis. 
}
\label{fig:SigmaSDp}
\end{figure}

On the other hand, the spin-dependent DM-proton scattering cross section $\sigma_{SD, \chi p}$ is known to be somewhat smaller than that for DM-neutron scattering, and is shown in Fig.~\ref{fig:SigmaSDp} for Case A (left panel) and Case B (right panel). The constraints obtained from the current direct detection experiments are not quite sensitive enough yet as shown by the solid curves in the figure. The future projection for the LZ experiment may be able to cover up to $M_1\sim 400$ GeV in Case A, and 100 GeV in Case B, as seen by the dotted curves. 

\subsection{Neutrino detection}

\begin{figure}
\centering
\includegraphics[width=0.49\textwidth]{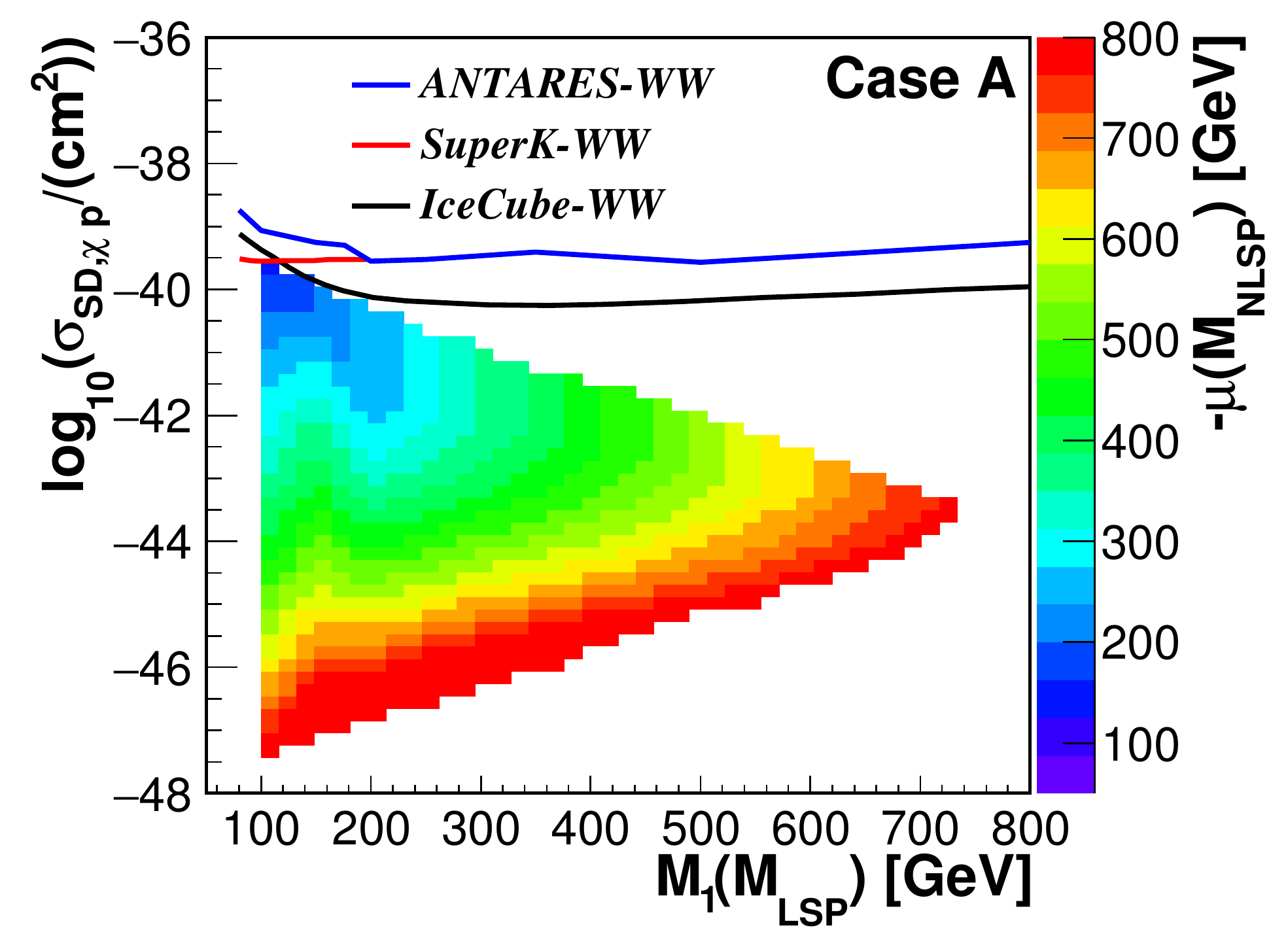}
\includegraphics[width=0.49\textwidth]{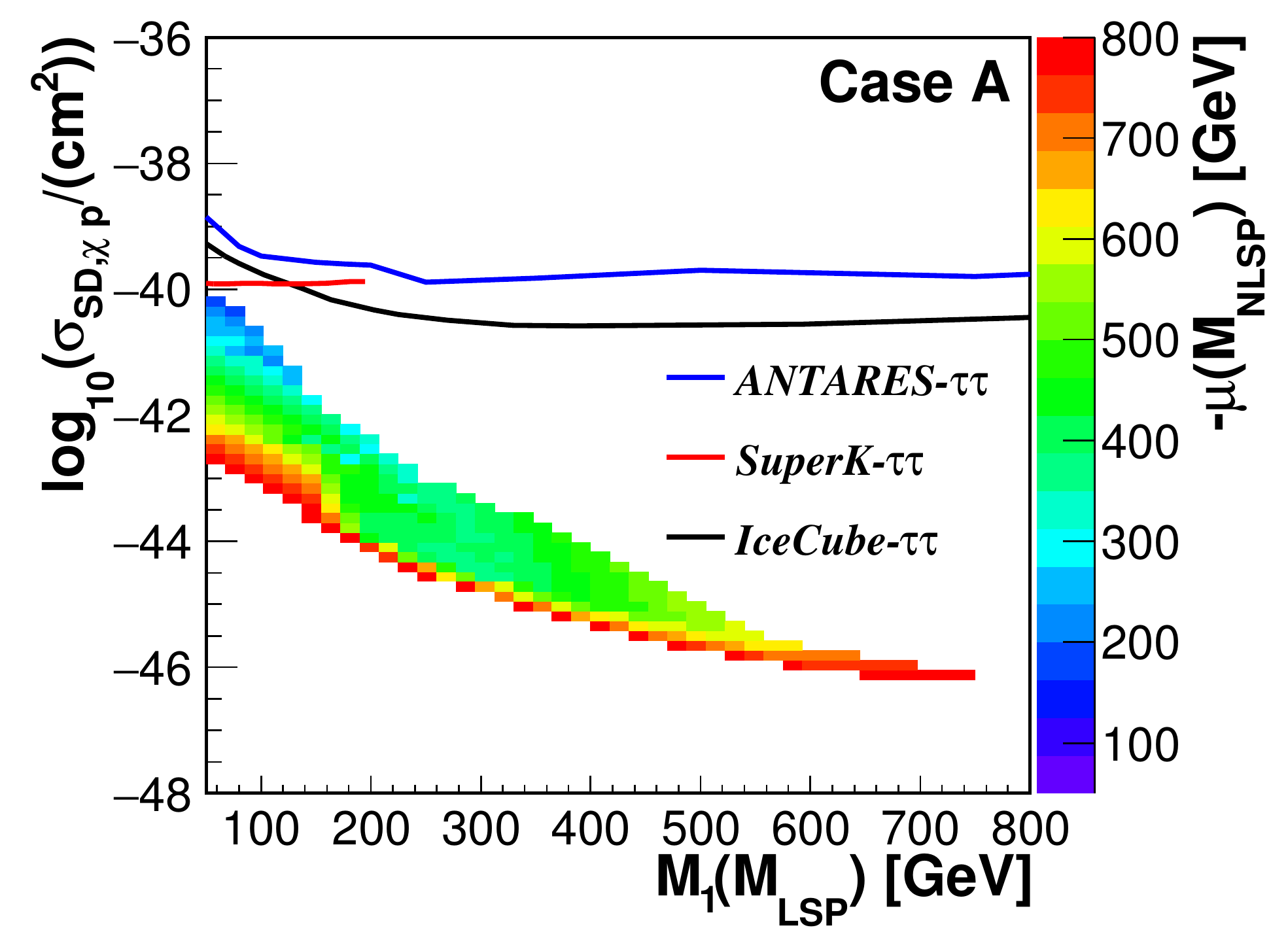}
\caption{Case A: Scaled spin-dependent DM-proton scattering cross section for DM annihilation channels: left panel $WW$ mode and right panel $\tau\tau$ mode. Experimental sensitivities from IceCube, SuperK and ANTARES are shown. The color code indicates the $\mu$ values as labeled on the right-handed vertical axis.}
\label{fig:SigmaSDpCaseA}
\end{figure}

\begin{figure}
\centering
\includegraphics[width=0.49\textwidth]{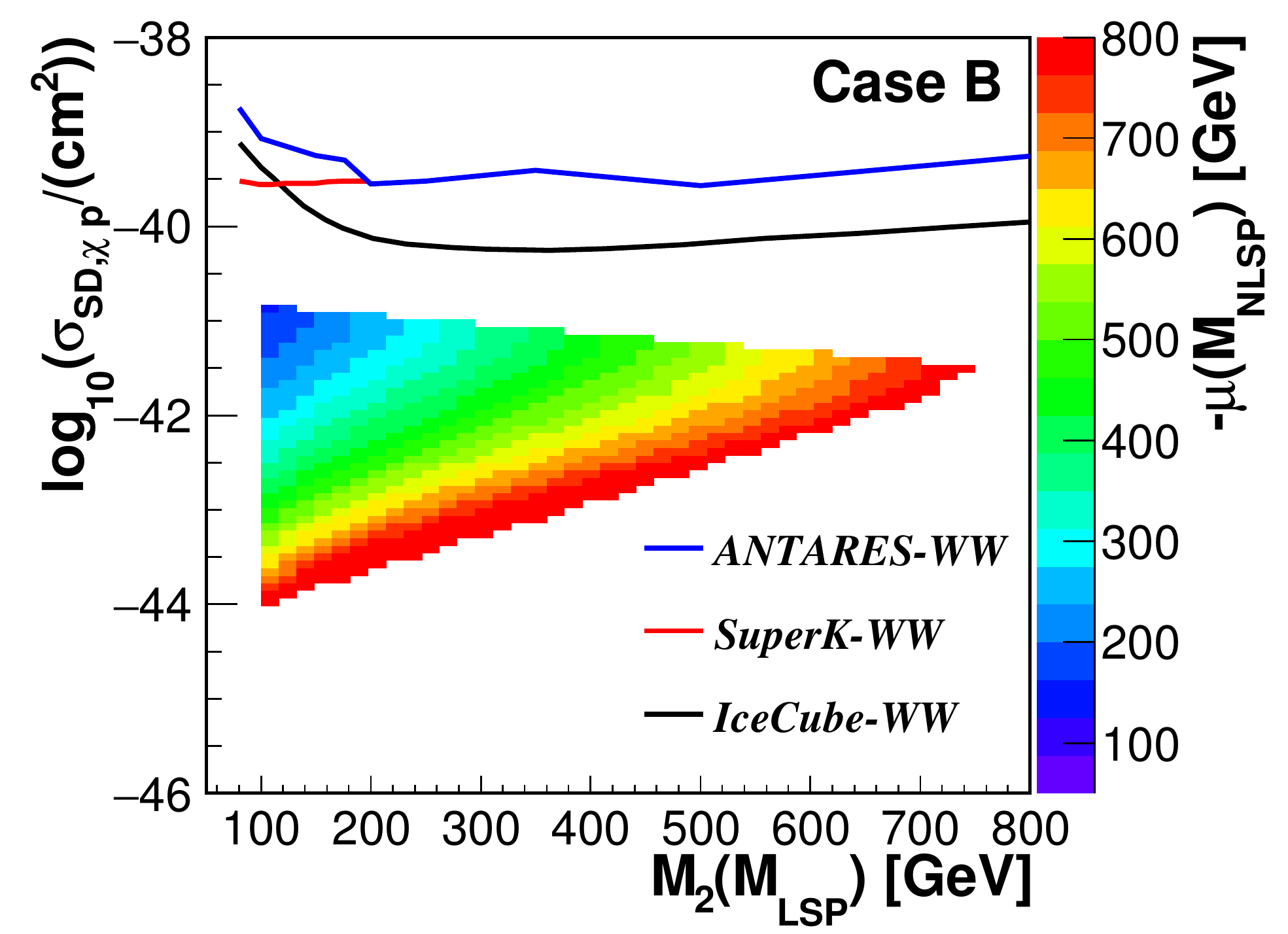}
\includegraphics[width=0.49\textwidth]{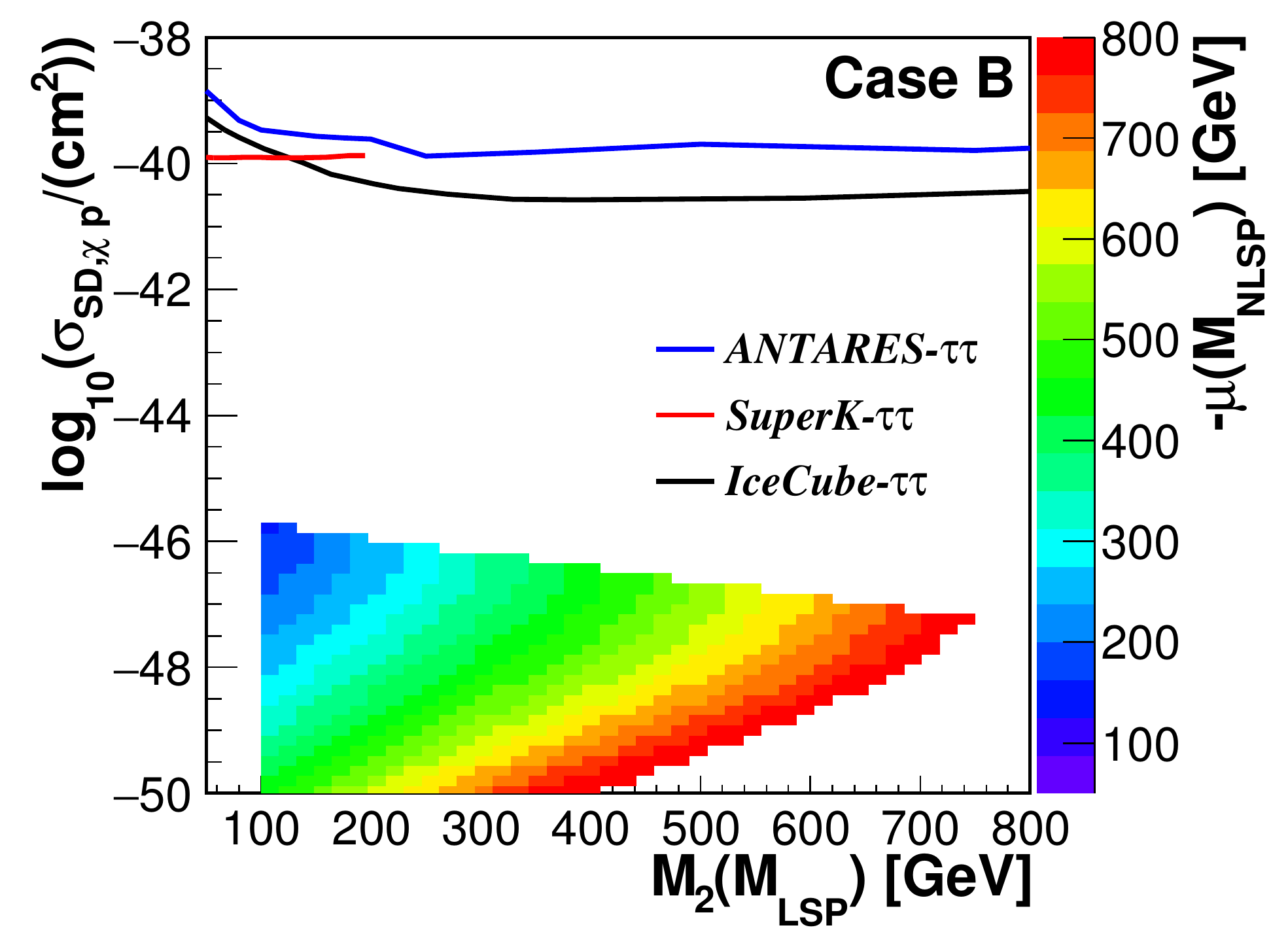}
\caption{Case B: Scaled spin-dependent DM-proton scattering cross section for DM annihilation channels: left panel $WW$ mode and right panel $\tau\tau$ mode. Experimental sensitivities from IceCube, SuperK and ANTARES are shown.
The color code indicates the $\mu$ values as labeled on the right-handed vertical axis.  }
\label{fig:SigmaSDpCaseB}
\end{figure}

The spin-dependent DM-proton scattering cross section $\sigma_{SD, \chi p}$ can also be constrained by indirect searches for neutrino signals coming from the Sun, assuming equilibrium of DM capture and annihilation. 
%
We present the theoretical prediction for the scaled\footnote{While the SD $\chi$-$p$ scattering cross section does not depend on the dark matter annihilation modes, the experimental limits of the indirect detection via neutrinos do. Since it is difficult to show the experimental limits for each parameter point, we choose to present the experimental limits assuming the observed DM relic density and 100\% annihilation fraction into a certain channel, while scaling the $\chi$-$p$ scattering cross section  for each parameter point with the corresponding DM relic density and the annihilation fraction into a particular final state.} scattering cross section $\sigma_{SD, \chi p}$ versus the LSP mass together with bounds from several neutrino telescope experiments in Fig.~\ref{fig:SigmaSDpCaseA} for Case A and Fig.~\ref{fig:SigmaSDpCaseB} for Case B, respectively. 
%
Two DM annihilation channels to $WW$ and $\tau\tau$ are shown on the left and right panels, respectively.  We see from Fig.~\ref{fig:SigmaSDpCaseA} that the bounds from SuperK and IceCube are reaching the low mass region for the Case A blind-spot scenario. In Case B, the IceCube bounds on the $WW$ mode impose the most stringent limits. However, these bounds are still about an order of magnitude away from 
the relevant blind-spot parameter space as shown in the left panel of Fig.~\ref{fig:SigmaSDpCaseB}.\footnote{In Ref.~\cite{1501-06357}, 
the authors pointed out 
that models where the dark matter is dominantly a wino-like neutralino are strongly excluded by IceCube, and that the wino relic density close to the thermally produced value satisfies the IceCube bound. This conclusion is consistent with what we found here for Case B, where the DM is typically under-abundant.
}
%
For both Case A and Case B, we have combined all contributing channels to compare with the IceCube-79 String results \cite{IC79-2016}, and find that our parameter space is still beyond the current reach, except for a small $M_1$ and $|\mu|$ region as already seen in Fig.~\ref{fig:SigmaSDpCaseA}.
 
The observational aspects for Case C are more difficult since it is both a SD and a SI blind spots. In addition to the absence of the DM direct detection signals, the indirect detection via neutrinos is also difficult given the negligible SD interactions.

%

\subsection{Gamma-ray detection}

\begin{figure}[tb]
\centering
\includegraphics[width=0.49\textwidth]{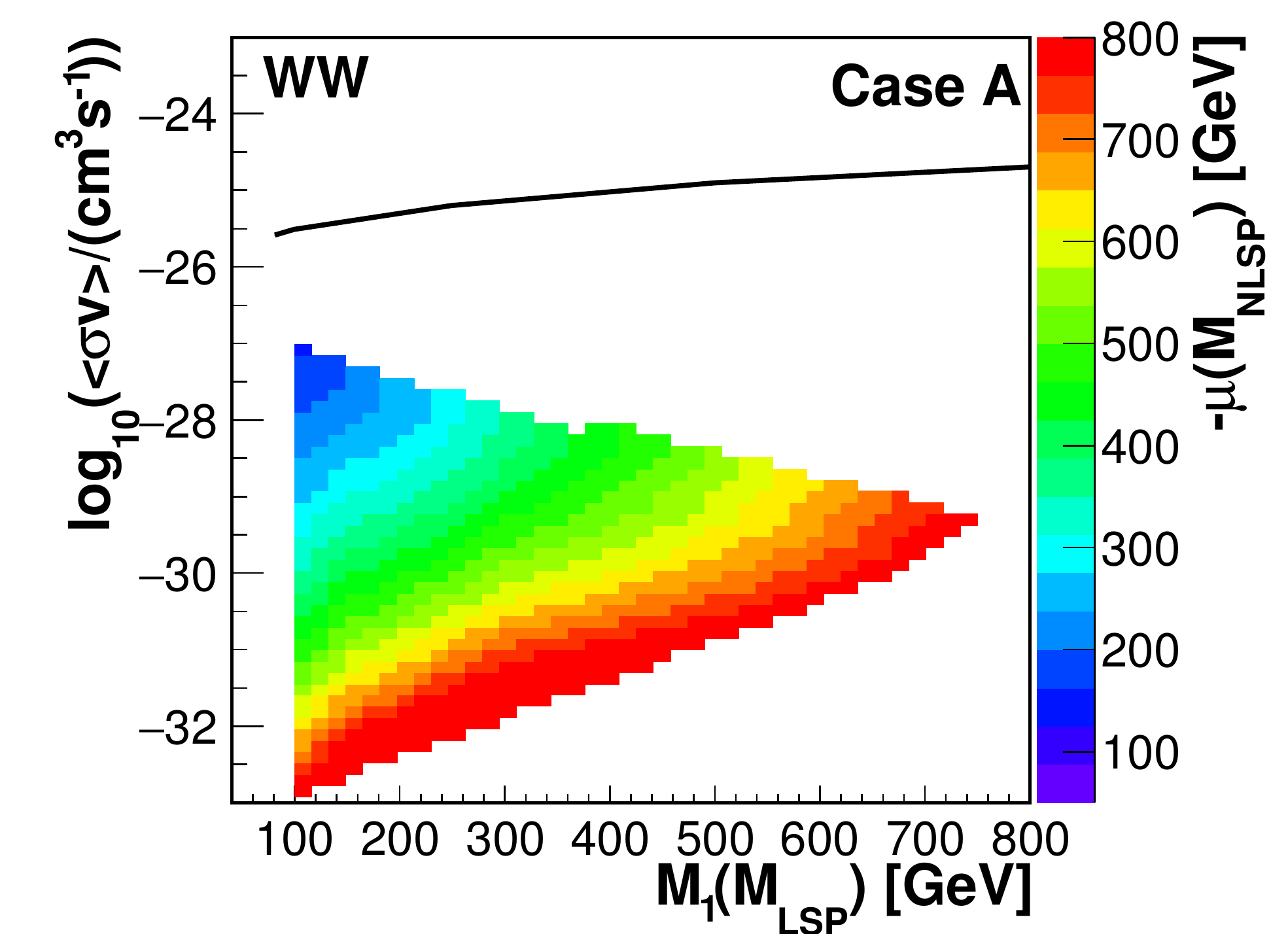}
\includegraphics[width=0.49\textwidth]{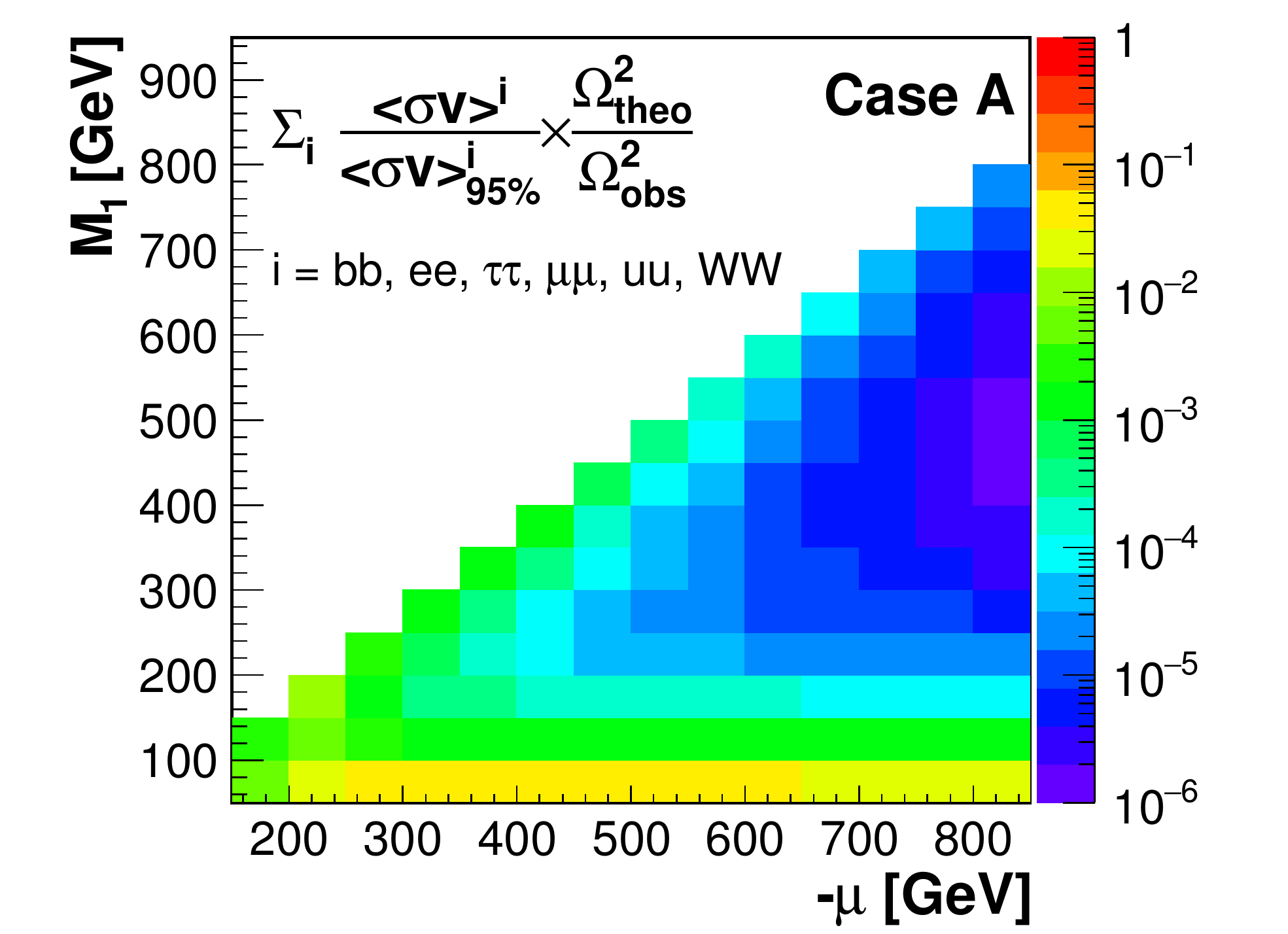}
\caption{Case A: 
Left panel shows the predicted cross section $\langle\sigma v\rangle$ versus the DM mass (colored regions) and the limits from Fermi-LAT gamma-ray observation for the $W^+W^-$ channel (the black curve).
The color code indicates the $\mu$ values.  Right panel shows the normalized multiple channel cross section by the color code in the $M_1-\mu$ plane.}
\label{fig:FermiGamma_CaseA}
\end{figure}

\begin{figure}[tb]
\centering
\includegraphics[width=0.49\textwidth]{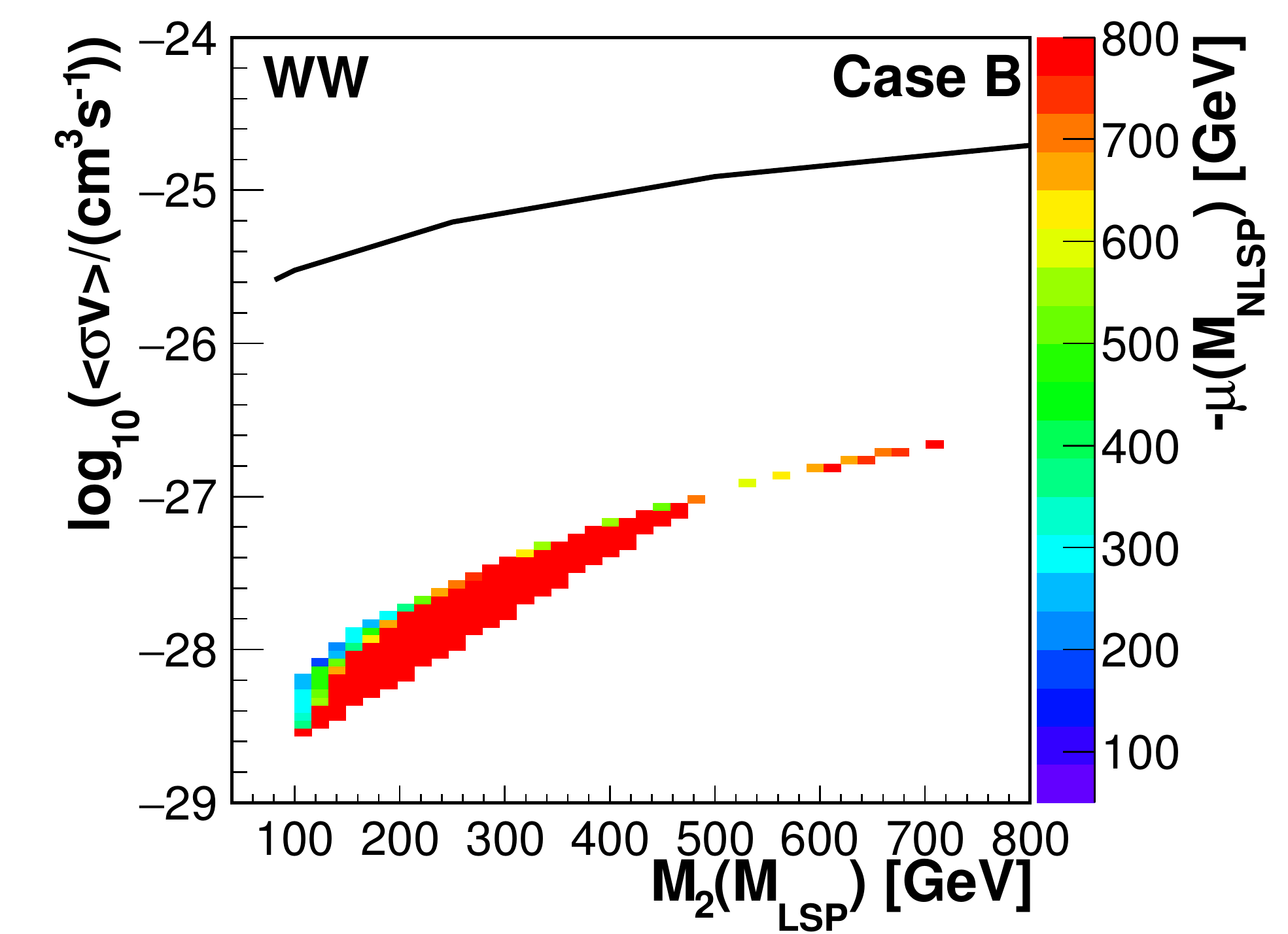}
\includegraphics[width=0.49\textwidth]{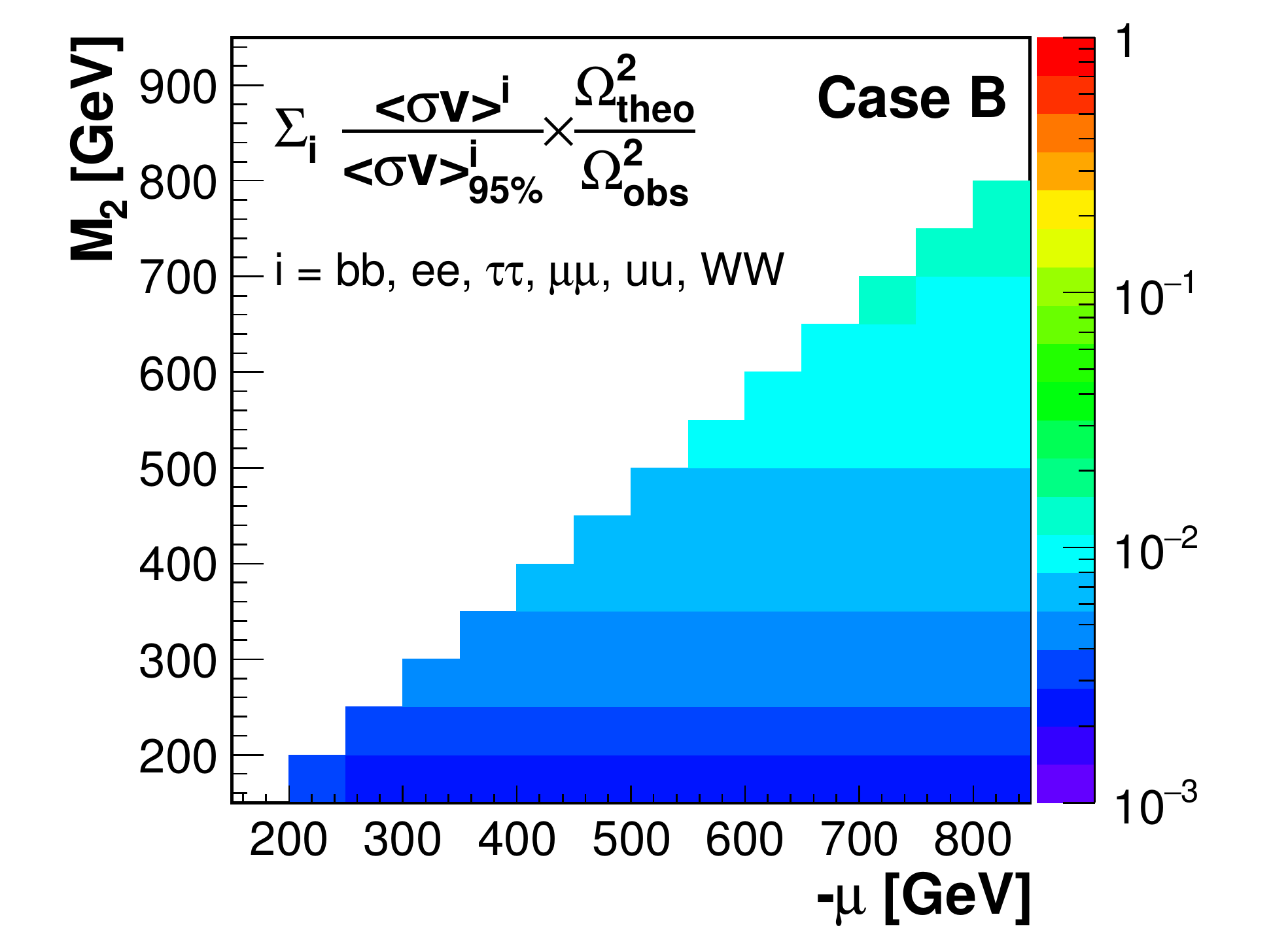}
\caption{Case B: 
Left panel shows the predicted cross section $\langle\sigma v\rangle$ versus the DM mass (colored regions) and the limits from Fermi-LAT gamma-ray observation for the $W^+W^-$ channel (the black curve).
The color code indicates the $\mu$ values.  Right panel shows the normalized multiple channel cross section by the color code in the $M_2-\mu$ plane.}
\label{fig:FermiGamma_CaseB}
\end{figure}

\begin{figure}[tb]
\centering
\includegraphics[width=0.49\textwidth]{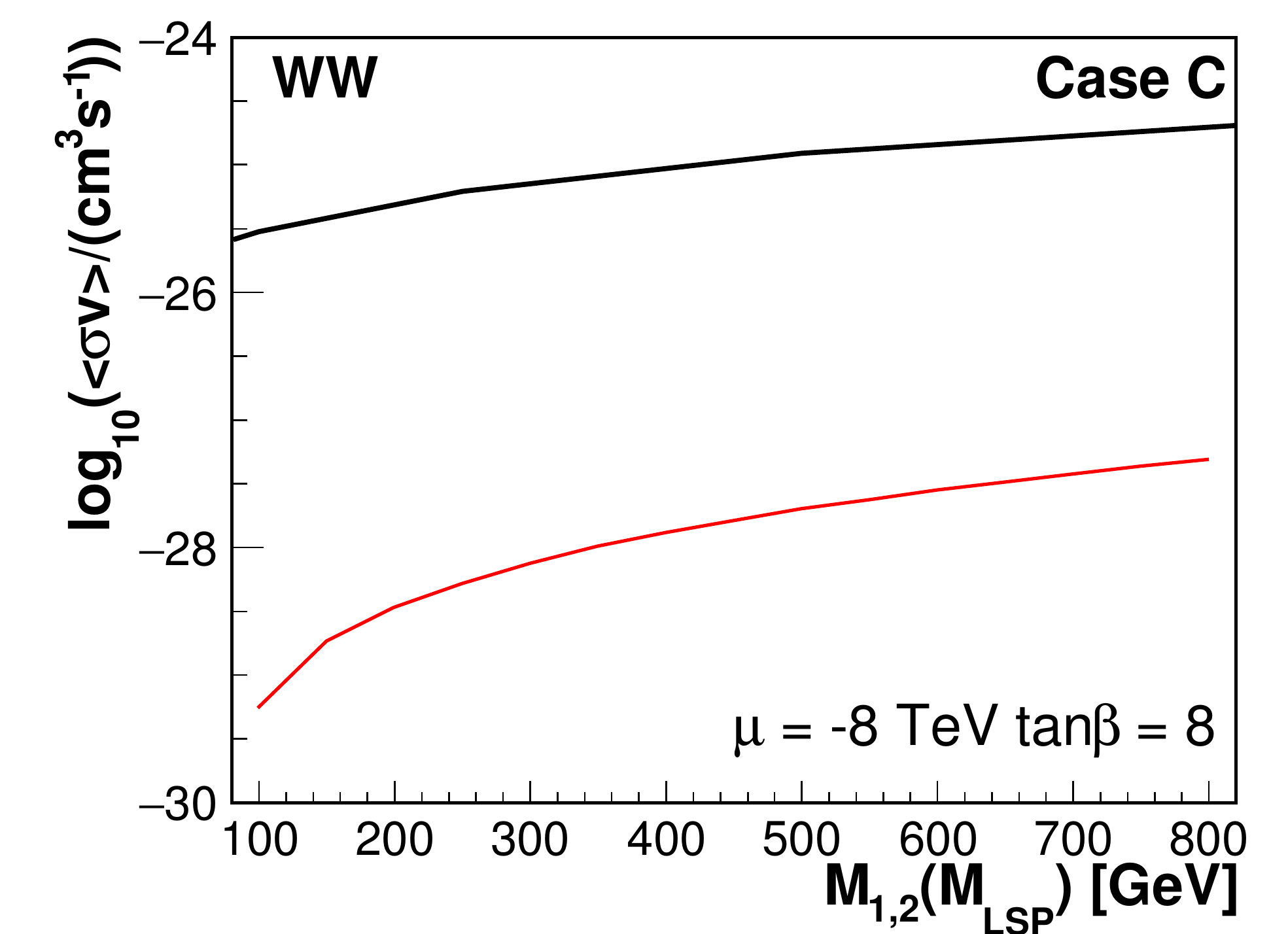}
\includegraphics[width=0.49\textwidth]{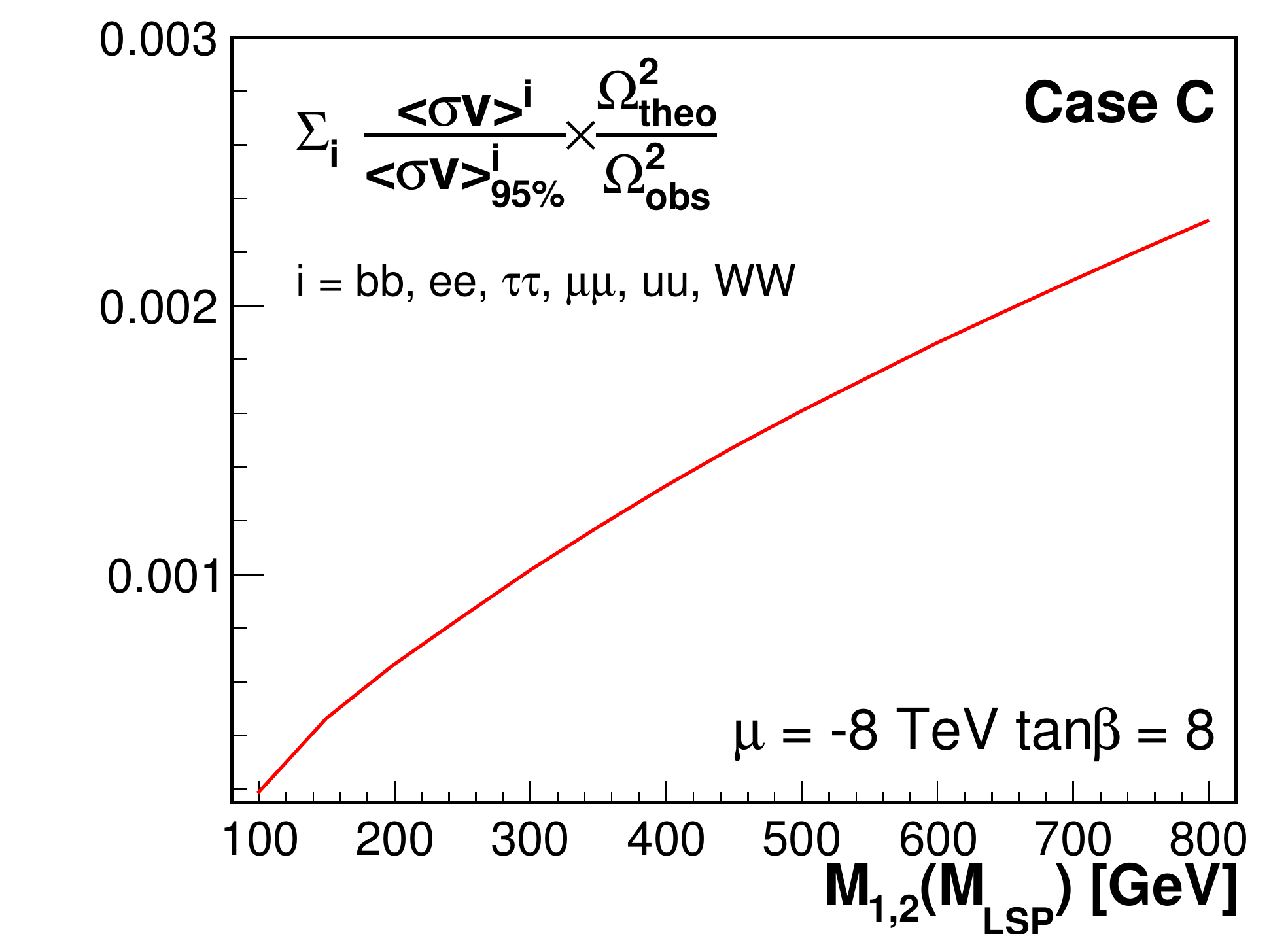}
\caption{
Case C: 
Left panel shows the limits from Fermi-LAT gamma-ray observation shown by the black (upper) curve and the predicted $\langle\sigma v\rangle$   for the $W^+W^-$ channel by the red (lower) curve.
Right panel shows the normalized multiple channel cross section versus  $M_{1,2}$. }
\label{fig:FermiGamma_CaseC}
\end{figure}

Gamma rays can be produced in dark matter annihilation through radiation, hadronization, or direct pair production, with different spectrum for different annihilation channels. 
The theoretical prediction (colored regions) for the velocity-averaged cross section $\langle \sigma v \rangle$ are presented on the left panels in Figs.~\ref{fig:FermiGamma_CaseA}, \ref{fig:FermiGamma_CaseB} and \ref{fig:FermiGamma_CaseC} for Cases A, B and C, respectively, for the most sensitive channel $W^+W^-$. 
%
The limits based on the gamma-ray observation from the Fermi-LAT results~\cite{FermiLAT-6yr-Gamma} are  shown by the solid black curves 
in all the three figures.
We see that the sensitivity from the $W^+W^-$ channel is still not reaching the theoretical parameter regions for the blind-spot scenarios.\footnote{Ref.~\cite{Agrawal:2014oha} mentioned that the Galactic center gamma-ray excess can be explained by a thermal-relic neutralino of the MSSM annihilating into $WW, ZZ, hh, t \bar t$ and the DM blind spots could be a viable region of MSSM parameter space. However, the blind-spot scenarios considered in this paper do not provide sufficient annihilation to explain this excess since the $Z$-coupling from bino-Higgsino mixing is small for bino-like DM, and  the wino-like DM is typically under-abundant.}
 
The above results are based on the assumption that the gamma-ray events in the Fermi-LAT experiment is 100\% from the $W^+W^-$. The similar assumption was also applied to the analyses for the other channels such as  $b\bar b,\ u\bar u$, as well as $\tau\tau$. A more accurate treatment would be to add those individual channels with appropriate efficiency factors and the corresponding relic abundance to account for the observed inclusive gamma-ray events.
%
For a channel labelled by $i$, denote the total efficiency by $A^{i}$ including the detector efficiency, the cosmic related factors and so on. Then the 95\% upper limit on the average cross section for this channel $\langle\sigma v\rangle_{95\%}^{i}$ satisfies:
\begin{eqnarray}
\langle\sigma v\rangle_{95\%}^{i}\times A^{i}\times \Omega_{obs}^2 = N_{obs}^{95\%},
\label{equ:FermiLAT-channel}
\end{eqnarray}
where $N_{obs}^{95\%}$ is the 95\% upper limit on the events from DM annihilation based on the observed events. The experiments will determine the efficiencies for each channel $A^i$. If we want to combine all the contributing channels, then the cross sections should satisfy
\begin{eqnarray}
N_{obs}^{95\%} \geq \sum_{i} \langle\sigma v\rangle^i\times A^{i}\times \Omega_{theo}^2 
= N_{obs}^{95\%}\times\sum_{i} \frac{\langle\sigma v\rangle^i}{\langle\sigma v\rangle_{95\%}^i}\times\frac{\Omega_{theo}^2}{\Omega_{obs}^2} \ .
\label{equ:FermiLAT-combine1}
\end{eqnarray}
This leads to an important relaltion we will use: 
\begin{eqnarray}
\sum_{i}\frac{\langle\sigma v\rangle^i}{\langle\sigma v\rangle_{95\%}^i}\times\frac{\Omega_{theo}^2}{\Omega_{obs}^2} \leq 1 ,
\label{equ:FermiLAT-combine2}
\end{eqnarray}
which we call the ``normalized multiple channel cross section''.   Regions with the normalized multiple channel cross section well below 1 are much more difficult to be probed by Fermi-LAT gamma-ray observation, while regions with the normalized multiple channel cross section   close to 1 might be covered by the next generation experiments.  

Improving the above $W^+W^-$ channel results, we present on the right panels in Figs.~\ref{fig:FermiGamma_CaseA}, \ref{fig:FermiGamma_CaseB} and \ref{fig:FermiGamma_CaseC} for Cases A, B and C, respectively, the normalized multiple channel cross section. 
For Case A, the channel combination improves the observability somewhat, but the current results from Fermi-LAT are still about one to two orders of magnitude weak to provide any relevant bound even for the low mass region of $M_1$, as seen in Fig.~\ref{fig:FermiGamma_CaseA}. 
For Case B, the relic density is far below the observation when the wino is light, so the relic density scaling governs the outcome. With the $WW$ channel dominant, the blind-spot region is at least two orders of magnitude below the sensitivity of the gamma-ray experiments, as seen in Fig.~\ref{fig:FermiGamma_CaseB}.\footnote{Ref.~\cite{Fan:2013faa} showed that wino dark matter has been strongly constrained by the Fermi-LAT and HESS data.   However, these bounds do not apply to the blind-spot scenario in Case B since the wino is typically under-abundant, which weakens the constraints.}
For Case C, the $WW$ channel is also dominant but still at least two orders of magnitude below the current sensitivity, as seen in
Fig.~\ref{fig:FermiGamma_CaseC}, for $\mu = -8$ TeV and $\tan\beta = 8$. The dependence on $\mu$ and $\tan\beta$ is very weak. 

\section{Current Bounds from LEP}
\label{sec:LEP}

\subsection{$Z$ invisible width}
\label{sec_EW}

Electroweak precision measurements at the $Z$-pole provide significant bounds on the SUSY mass parameters. 
If the neutralino is light enough, $m_Z> 2m_{\chi_1^0}$, it can be produced via the decay of a $Z$-boson. Such decays are strongly constrained by the measurements at LEP I.   The invisible decay width of $Z$ is constrained  to be $\Gamma_{inv}= 497.4 \pm 2.5$ MeV \cite{LEP-Zpeak}, 
which can be translated to 
an upper limit of non-SM contributions to the invisible decay width of the $Z$-boson
\begin{equation}
\Gamma_{inv}^{BSM} < 3.1\ {\rm MeV\ at}\ 95\% \ {\rm confidence\ level\ (CL)}.
\label{eq_51}
\end{equation}
 This is a conservative bound before the Higgs discovery. 
A stronger bound can be obtained using ZFitter \cite{Bardin:aa} with the measured Higgs mass $m_h=125$ GeV. We find the $Z$-boson invisible decay width to be $\Gamma_{inv}^{SM}= 501.7 \pm 0.2$ MeV, which sets a stronger upper bound on the BSM contributions to be 
\begin{equation}
\Gamma_{inv}^{BSM} < 1.1\ {\rm MeV\ at}\ 95\% \ {\rm CL}.  
\label{eq_52}
\end{equation}

\begin{figure}[tb]
\centering
\includegraphics[width=0.32\textwidth]{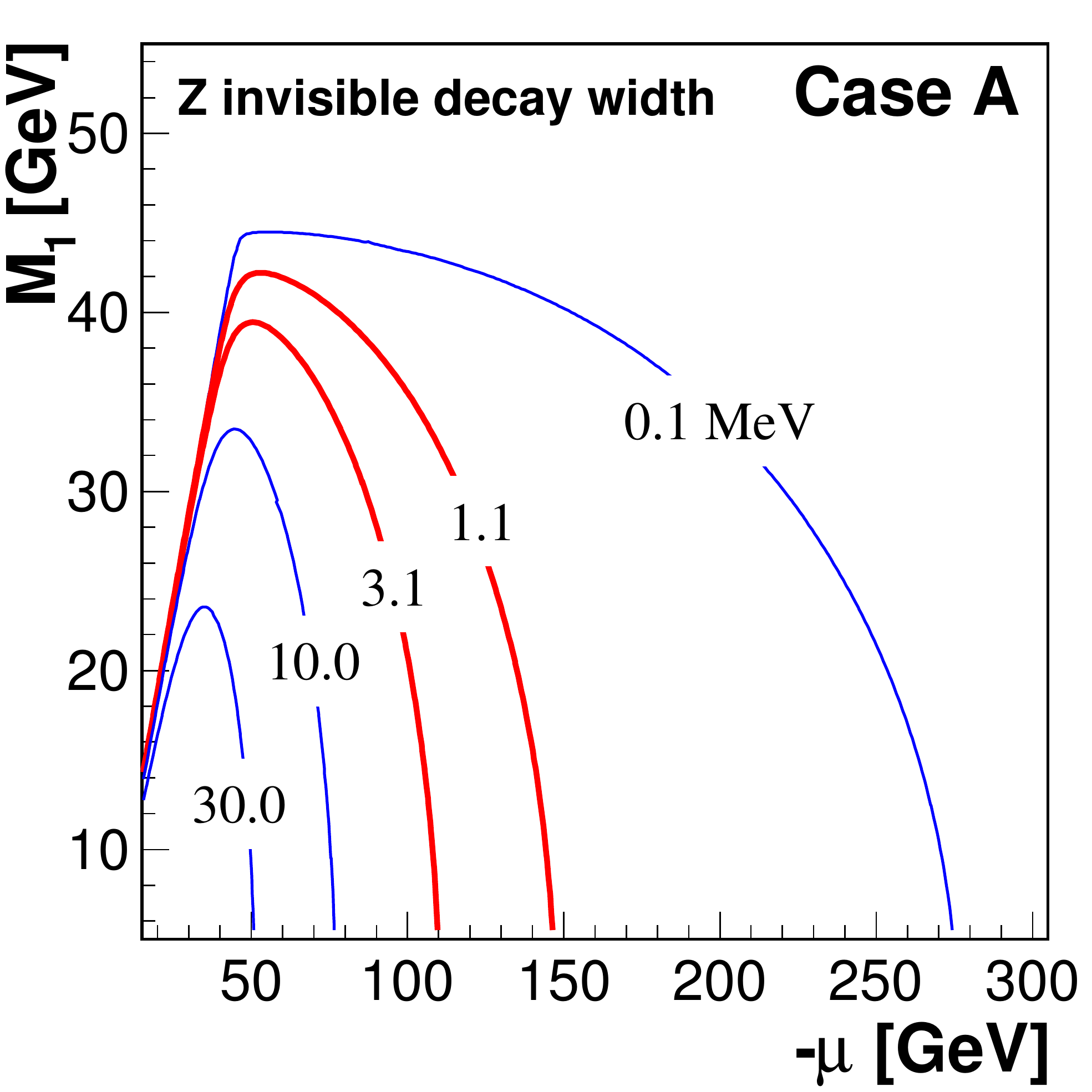}
\includegraphics[width=0.32\textwidth]{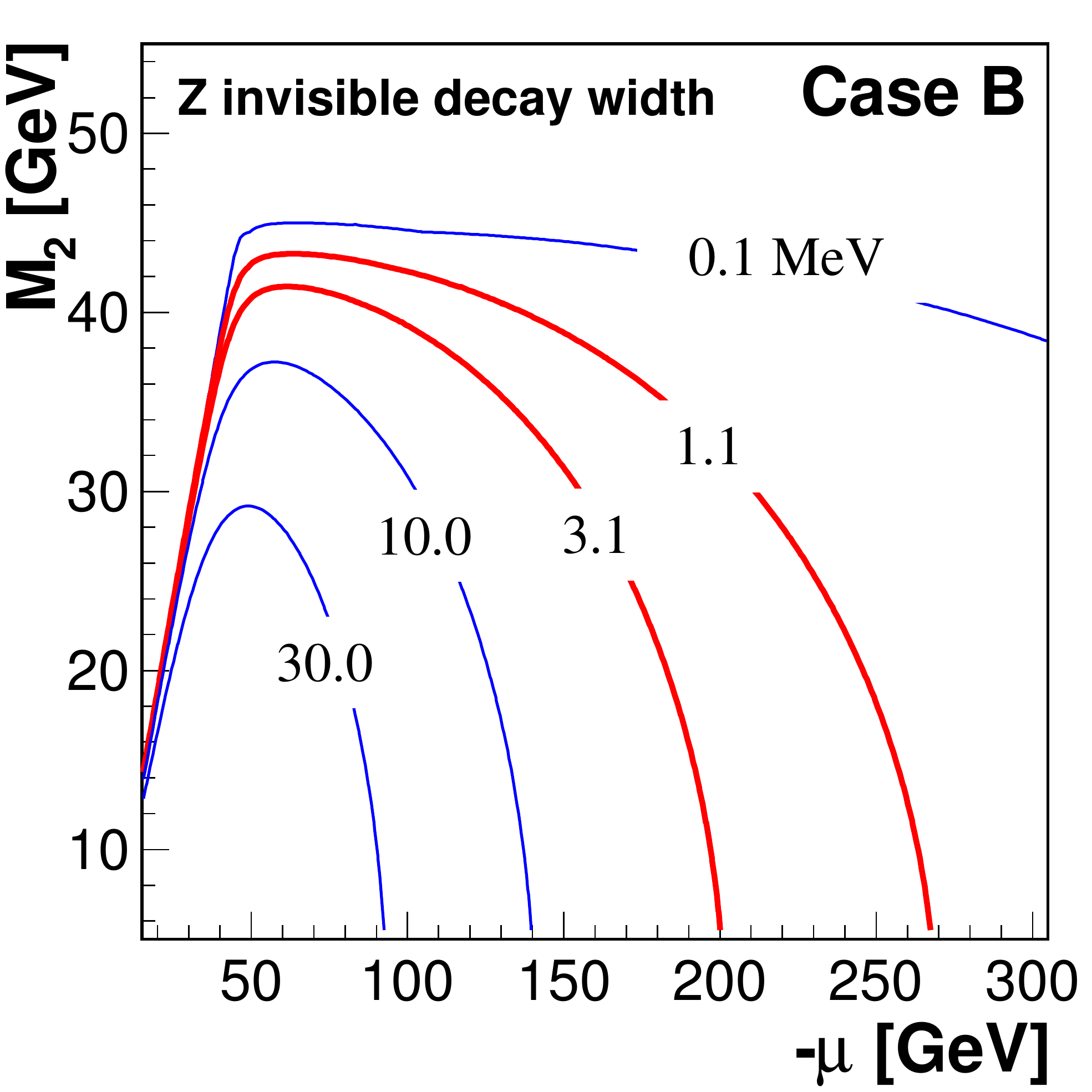}
\includegraphics[width=0.32\textwidth]{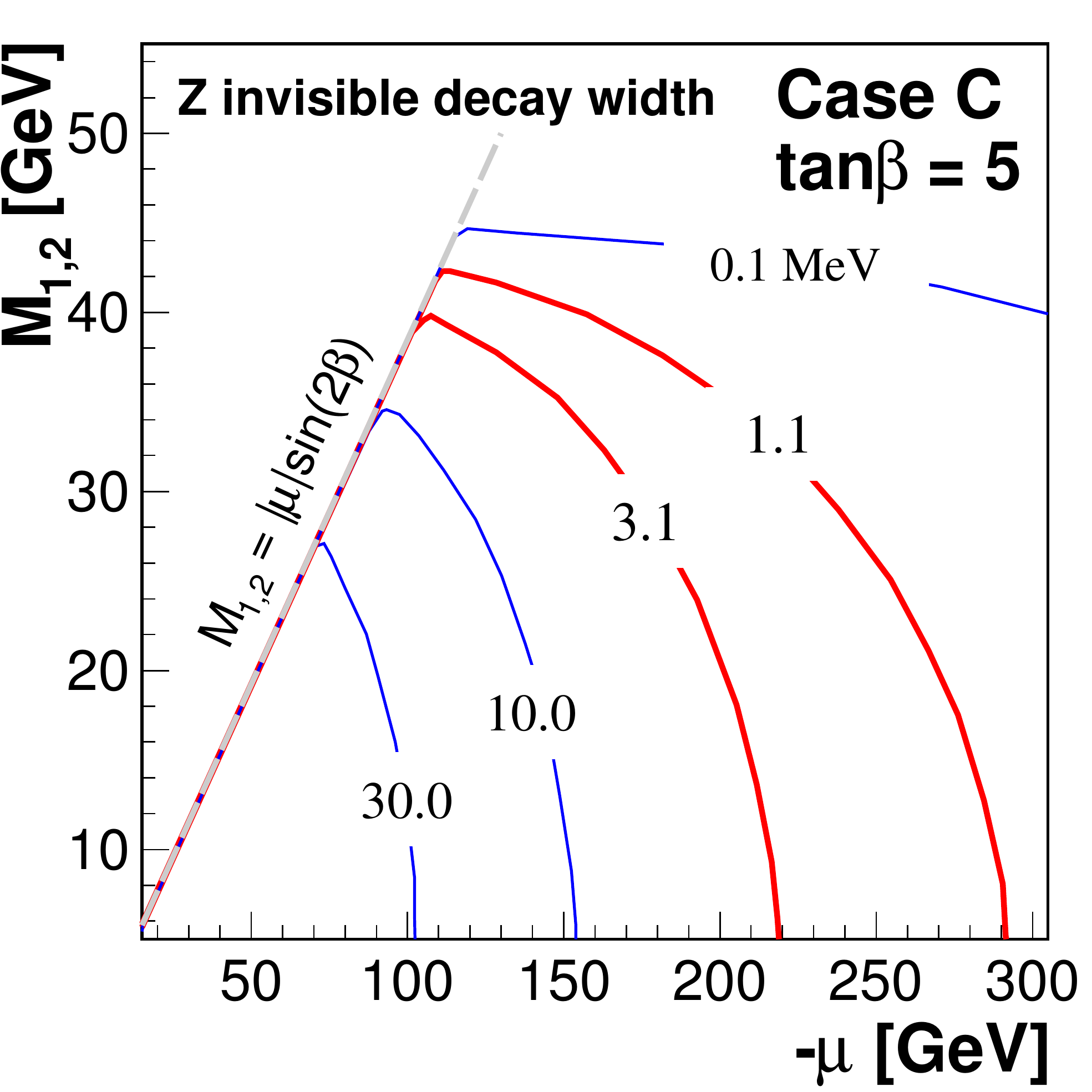}
\caption{$Z$-boson invisible decay width to neutralinos for Cases A (left), B (middle), and C (right). The two middle thick (red) contours indicate the exclusion bounds imposed by Eqs.~(\ref{eq_51}) and (\ref{eq_52}). 
}
\label{fig:Zinv}
\end{figure}

In Fig.~\ref{fig:Zinv}, we show the calculated partial decay width of the $Z$-boson into a pair of neutralinos $\Gamma(Z\to \chi_1^0\chi_1^0)$, as well as 
$\Gamma(Z\to \chi_1^0\chi_2^0,\  \chi_2^0\chi_2^0)$ when appropriate, in the SUSY mass parameter space for Cases A (left), B (middle), and C (right). The two middle thick (red) contours indicate the exclusion bounds imposed by Eqs.~(\ref{eq_51}) and (\ref{eq_52}). 
For a bino-like LSP as in Case A, the coupling of the $Z$ is proportional to the bino-Higgsino mixing $N_{13,14}$: $g_{Z \chi_1^0\chi_1^0} \sim N_{13}^2-N_{14}^2$ \cite{HABER198575}. At large Higgsino masses $|\mu|$, the mixing becomes small and therefore the coupling becomes highly
 suppressed \cite{Djouadi:aa}:\footnote{In the last step we used the blind-spot relation from Table \ref{tab:SIBS}. }
\begin{equation}
g^2_{Z \chi_1^0\chi_1^0}  = \frac{s_W^2 m_Z^4 \cos(2\beta)^2}{(\mu^2 - M_1^2)^2} = \frac{s_W^2 m_Z^4}{\mu^2 (\mu^2 - M_1^2)}.
\label{equ:gzxx}
\end{equation}
The decay width $\Gamma(Z\to \chi_1^0\chi_1^0)$ is sufficiently suppressed at large Higgsino masses and thus the LEP bound is weaker as seen in Fig.~\ref{fig:Zinv} (left). The constraints are stronger for the wino-like LSP (Case B) since the wino-Higgsino mixing is enhanced by a factor ${c_W^2}/{s_W^2}$ compared to the bino-Higgsino mixing as seen in Fig.~\ref{fig:Zinv} (middle). The constraints in Case C are similar to those in Case B but with a cut-off induced by the blind-spot relation as shown by the gray dash line in Fig.~\ref{fig:Zinv} (right).  We can see that the electroweak precision measurements exclude parameter regions 
\begin{equation}
\begin{aligned}
M_1 \lesssim\ &43 {\rm\ GeV \ \  and \ \ } |\mu|\lesssim 110-150 \text{ GeV \ \ for a bino-like LSP (Case A), and}\\
M_2 \lesssim\ &43 {\rm\ GeV \ \ and \ \   } |\mu|\lesssim 200-270 \text{ GeV \ \  for a wino-like  LSP (Case B), and}\\
M_{1,2} \lesssim\ &43 {\rm \ GeV \ \ and \ \ } |\mu|\lesssim 220-290 \text{ GeV \ \ for a bino-wino mixing LSP (Case C, }\tan\beta = 5\text{).}
\nonumber
\end{aligned}
\end{equation}

\subsection{LEP2 chargino searches}
\label{sec:LEP2}

Some of the strongest bounds on the blind-spot parameter space come from chargino searches at LEP \cite{LEP-Chargino1}. The DELPHI collaboration performed a search for charginos with a sufficient mass splitting to the lightest neutralino produced in pair production by looking for events with missing transverse momentum  in association with jets or leptons \cite{Abdallah:2003xe}. Using the data set up to 209 GeV, this search led to a mass bound
\begin{equation}
{\rm DELPHI:}\quad m_{\chi_1^\pm} > 102\ {\rm  GeV\  for}\ \Delta m = m_{\chi_1^\pm}-m_{\chi_1^0}>5\ {\rm GeV}.
\end{equation}
Complementary to the above analysis, the ALEPH collaboration performed a dedicated study to analyze the case of small mass splitting, taking into account both standard chargino searches with energetic leptons and jet as well as signatures with an ISR photon balanced by missing energy \cite{LEP-ALEPH}. This analysis found
\begin{equation}
{\rm ALEPH:}\quad m_{\chi_1^\pm} > 93\ {\rm  GeV\  for}\ \Delta m = m_{\chi_1^\pm}-m_{\chi_1^0} < 5\ {\rm GeV}.
\end{equation}
We will take those results into account in the LHC searches in the next section.

\section{Dark Matter Searches at the LHC}
\label{sec:collider}

We now consider the searches for DM at the LHC in the blind-spot scenarios in the hope to cover the parameter space that would be difficult for the direct and indirect  dark matter searches. 
 The most common signature for the DM search at colliders would be the missing transverse momentum (customarily called the ``missing energy'') carried away by the DM particles escaping from detection.  

The SUSY signatures are essentially governed by the mass different between the produced particles (mostly the chargino NLSP) and the decay final state (the neutralino LSP) $\Delta m = m_{\chi^\pm} - m_{\chi^0}$. 
In general, four different search strategies are considered:

\begin{enumerate}
	\item \textbf{Charged Track:} If the mass difference $\Delta m < m_\pi$ is small, the chargino is long-lived which leads to a charged track like a muon. The sensitivity for identifying charged tracks in collider experiments is very high. LEP searches \cite{LEP-ALEPH,collaboration:aa} for chargino pair production $e^+e^- \to \chi^+\chi^-$  excludes such particles up to $m_{\chi^\pm_1}<102$ GeV.   Searches at the LHC  for long-lived charged particles exclude the production cross section of relatively stable charginos  to be above 4 fb at 8 TeV using 20 fb$^{-1}$  data  (ATLAS) \cite{ATLAS:2014fka} or about 1 fb at 13 TeV using 13 fb$^{-1}$ data (CMS)  \cite{CMS:2016ybj}.     However, in MSSM, loop corrections to the chargino and neutralino mass typically induce a mass splitting of a few hundred MeV.    The charginos usually decay before traveling too far.  Therefore charged track searches for long-lived particles do not apply in our cases.
	\item \textbf{Disappearing Track:} For intermediate mass differences $m_\pi < \Delta m< {\rm a\ few\ GeV}$, the chargino decays inside detector with soft decay products after travelling some distance from the interaction point. This results in a disappearing track signature which we will discuss in Sec.~\ref{sec:disap}. 
	\item \textbf{Electroweakino Searches:} If the mass difference is large $\Delta m >$ few GeV, the chargino promptly decays inside the detector with energetic decay products, leading to observable leptons or jets plus missing energy. These searches will be discussed in Sec.~\ref{sec:EWsearch}. 
	\item \textbf{Monojet/monophoton Searches:}  If we only consider the pair production of the neutralino LSP (or with its degenerate charged partners), mono-jet or mono-photon searches have been a standard channel for dark matter searches at colliders.  These searches will be discussed in Sec.~\ref{sec:monojet}. 
\end{enumerate}
 
\subsection{Disappearing track searches}
\label{sec:disap}

\begin{figure}[tb]
\centering
\includegraphics[width=0.49\textwidth]{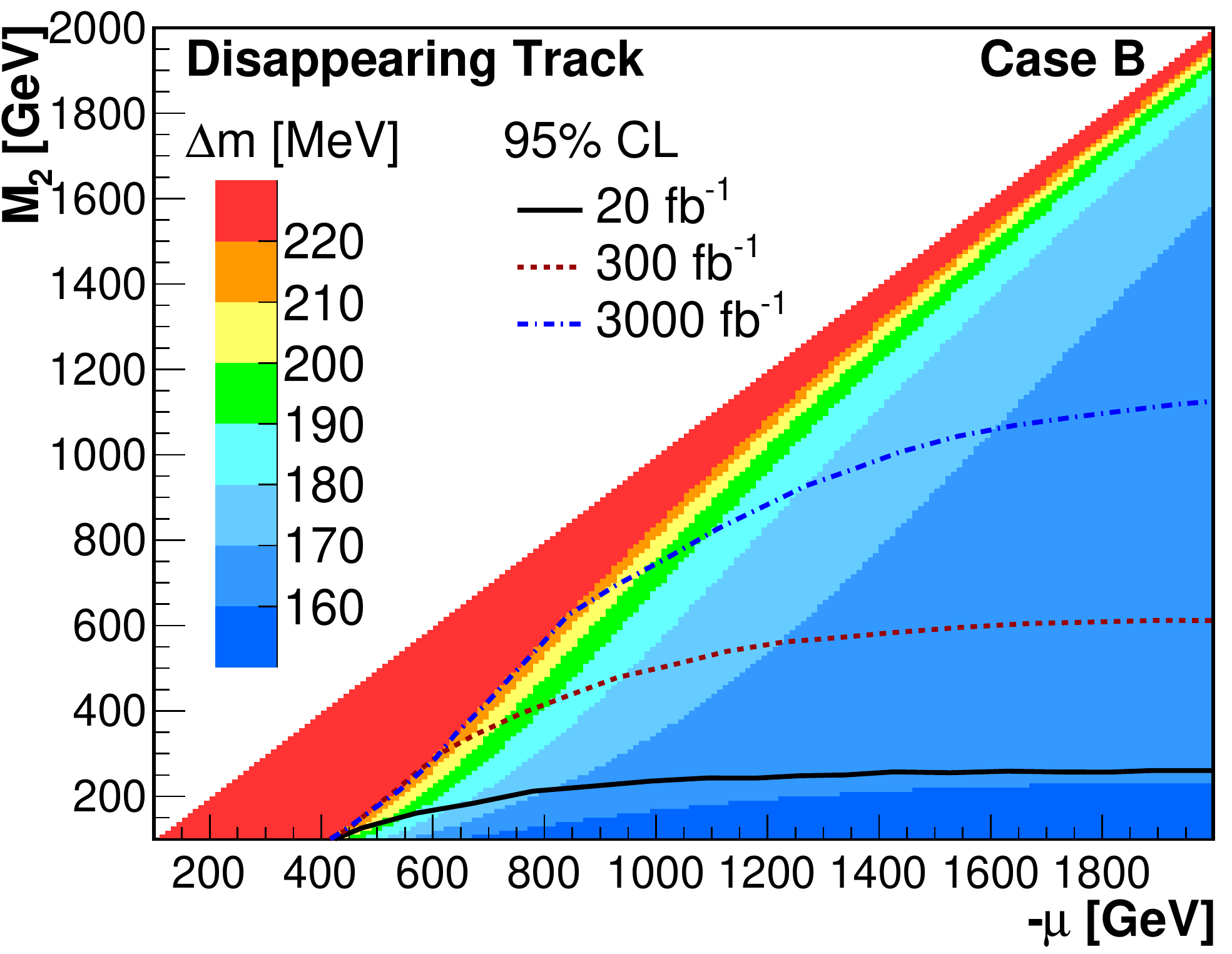}
\includegraphics[width=0.49\textwidth]{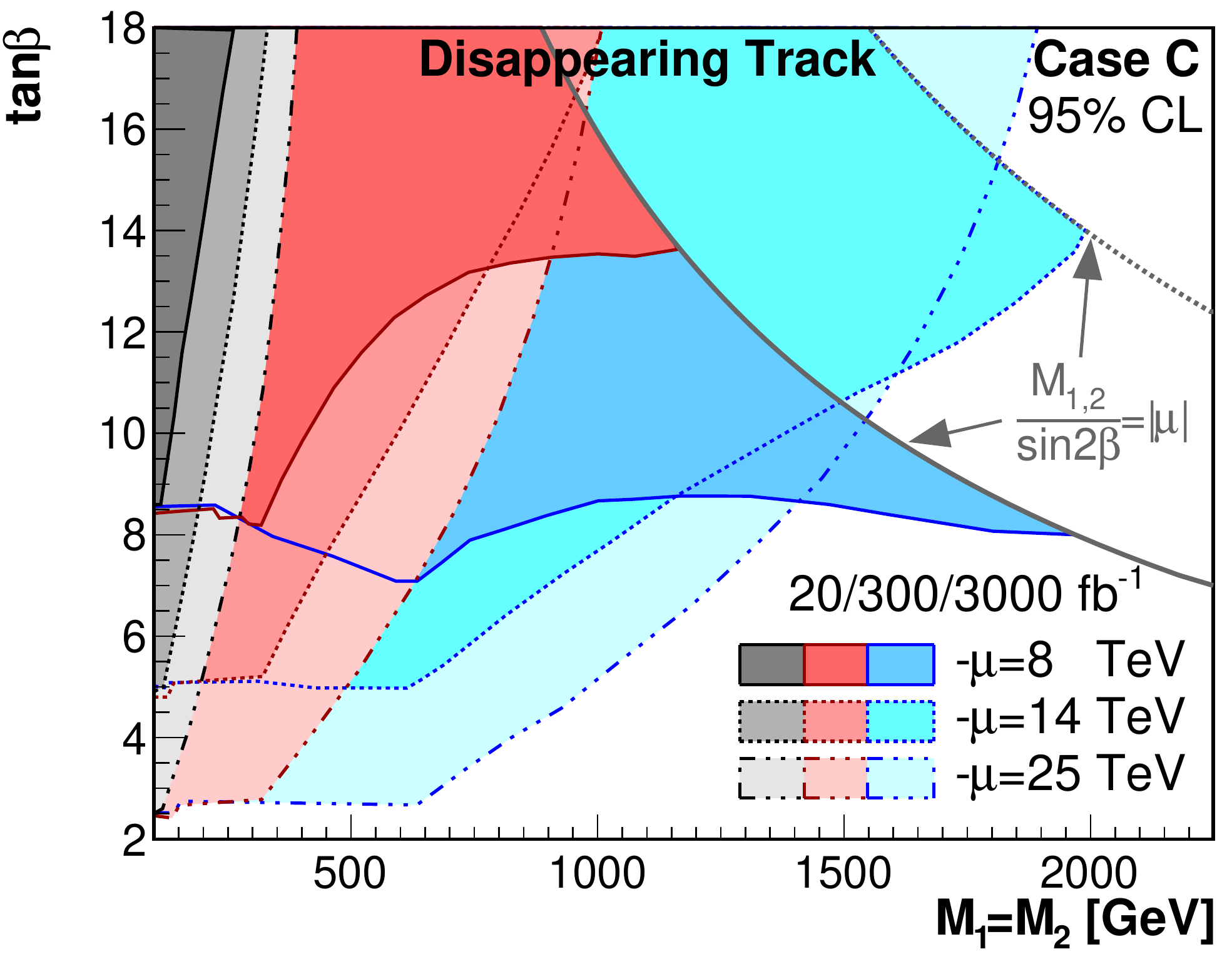}
\caption{
95\% CL exclusion from disappearing track searches at the LHC.
Left panel for Case B in the $M_2-\mu$ plane:  
ATLAS at 8 TeV with 20 fb$^{-1}$ (solid curve) and the projection for 14 TeV at 300 (dashed curve) and 3000 (dash-dotted curve) fb$^{-1}$, respectively. Mass difference $\Delta m$ is shown by  the corresponding color code. 
Right panel for Case C in the $\tan\beta-M_{1,2}$ plane:  
ATLAS at 8 TeV with 20 fb$^{-1}$ (far left black region) and the projection for 14 TeV at 300 (middle red region) and 3000 (right blue region) fb$^{-1}$, with three different values of $\mu$, respectively.   The solid and dotted gray lines in the upper-right corner indicate the upper limit on $M_{1,2}$: $M_{1,2}/\sin(2\beta)= |\mu|$ for a given value of $|\mu|$.
\label{fig:DT14TeV}
}
\end{figure}

Both Case B and Case C contain regions of parameter space with small enough wino-Higgsino mixing that permits a disappearing track (DT) signal. ATLAS and CMS performed a search for disappearing tracks using the 8 TeV data with 20 ${\rm fb}^{-1}$ integrated luminosity \cite{ATLAS-Disappearing_Track,CMS-Disappearing}.   We further projected the reach for 14 TeV with 300 or 3000 fb$^{-1}$  integrated luminosity by scaling the 8 TeV result with parton
 luminosity \cite{collider-reach}. We show the results altogether in Fig.~\ref{fig:DT14TeV} for the 95\% CL exclusion from disappearing track searches. 
Results for Case B are shown in the left panel in the $M_2-\mu$ plane, for ATLAS at 8 TeV with 20 fb$^{-1}$ (lower solid curve) and for 14 TeV at 300 (middle dotted curve) and 3000 (upper dash-dotted curve) fb$^{-1}$, respectively. 
The vertical column with the corresponding color code labels the mass difference $\Delta m = m_{\chi^\pm} - m_{\chi^0}$, including two-loop corrections \cite{twoloopdeltam}.   $M_2$ less than about 250 GeV, 600 GeV, and 1100 GeV can be excluded at 95\% CL at  large $|\mu|$ case under three different luminosities, respectively.

Results for Case C are shown in the right panel of Fig.~\ref{fig:DT14TeV} in the $\tan\beta-M_{1,2}$ plane, again for ATLAS at 8 TeV with 20 fb$^{-1}$ (far left black region) and for 14 TeV at 300 (middle red region) and 3000 (right light-blue region) fb$^{-1}$, with three different values of $-\mu=8,\ 14$ and 25 TeV, respectively, for illustration. The coverage region is on the left-side of the curves.   Note that the blind-spot requirement $|\mu|>{M_2}/{\sin2\beta}$ ensures the neutralino to be lighter than the chargino. This relation imposes an upper bound for the blind-spot parameter space on $\tan\beta$ for given $M_{1,2}$ and $\mu$ as indicated by the gray lines in the right panel of Fig. \ref{fig:DT14TeV}.
We see the wide coverage for the mass parameters: the larger $|\mu|$ is, the better coverage because of the weaker mixing and thus a more efficient disappearing track search.  The coverage also gets better for large values of $\tan\beta$. While the LSP mass is fixed by the blind-spot relation at tree level, $m_{\chi^0_1}=M_{1,2}$, the chargino mass at tree level is given by 
\begin{equation}
m_{\chi_1^+}^2 \approx m_{\chi_1^0}^2 +\frac{2 m_W^2 M_2}{\mu^2} \left(|\mu|\sin2\beta -M_2\right).
\end{equation} 
For larger $\tan\beta$, the mass splitting between the chargino and neutralino state becomes smaller 

\begin{equation}
\Delta m^2 \approx \frac{2m_W^2 M_2}{|\mu |} \left( {2 \over \tan\beta} - {M_2\over |\mu|} \right),
\end{equation} 
and therefore the sensitivity for disappearing track searches increases at large $\tan\beta$. 

We would like to emphasize the significance of the coverage of the blind-spot parameter space for  Case C by the LHC searches, since it is the ``most blind'' scenario with vanishingly weak SI and SD scattering signal, and thus hopeless for DM direct and indirect detections, as pointed out earlier. 

\subsection{Electroweakino searches}
\label{sec:EWsearch}

In the previous section, we have discussed the case in which the NLSP is almost mass-degenerate with the LSP.
Once the mass splitting is sufficiently large, typically more than a couple of GeV, the heavy electroweakino (EW) will promptly decay into the LSP and a  (virtual) $W,\ Z$ or Higgs-boson.

\begin{figure}[tb]
\centering
\includegraphics[width=1\textwidth]{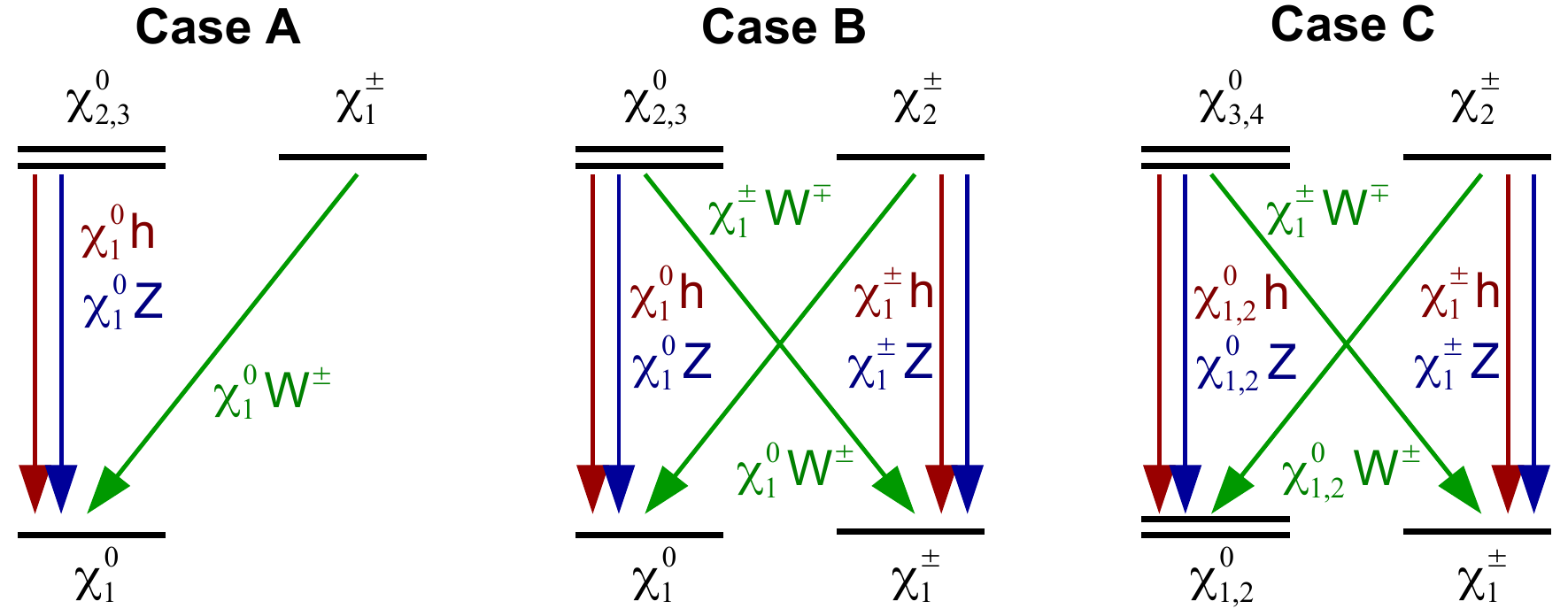}
\caption{Electroweakino mass spectrum for Case A (left), Case B (center) and Case C (right). Possible decay channels into a $Z$-boson (blue), Higgs-boson (red) and $W$-boson (green) are indicated.
}
\label{fig:spectrum}
\end{figure}

The mass spectrum and the decay patterns of Case A is illustrated in Fig.~\ref{fig:spectrum} (left panel). The LSP is a bino-like neutralino while the NLSPs are Higgsino-like neutralinos and chargino. The wino-like states are decoupled and are assumed not contribute to the collider phenomenology. The main NLSP production channels are 
\begin{equation}
q\bar q\to \chi_{2}^0 \chi_{3}^0,\ \ \chi_{2,3}^0 \chi_1^\pm\ \ {\rm and}\ \  \chi_1^+ \chi_1^-.
\end{equation}
The heavy gaugino decays in Case A include
\begin{equation}
\chi_{2,3}^0 \to Z\chi_1^0,\ \  h\chi_1^0 \ \ {\rm and}\ \ \chi_1^\pm \to  W^\pm \chi_1^0. 
\end{equation}  
We therefore expect to observe the signatures 
\begin{equation}
W^+ W^- + \met,\ WZ+\met,\ Wh+\met,\ ZZ+\met,\ \  Zh+\met\ \ {\rm  and}\ \  hh+\met. 
\label{eq:casea}
\end{equation}
Note that introducing a right-handed stau to achieve correct relic density will not affect the results of the collider search, as the decay rate of Higgsino-like NLSP into stau is negligible.

The center panel of Fig.~\ref{fig:spectrum} shows the mass spectrum and the decay patterns for Case B with a wino LSP. The mass difference among the wino triplet is very small and therefore the decay products of the wino-like chargino state will be too soft to be appreciable. 
Hence, the three wino-triplet states can be treated as LSPs. The Higgsino-like neutralinos and chargino form the NLSP states while the bino-like states is decoupled. The main NLSP production channels are
\begin{equation}
 q\bar q \to \chi_{2}^0 \chi_{3}^0,\ \ \chi_{2,3}^0 \chi_2^\pm\ \ {\rm and}\ \ \chi_2^+ \chi_2^-. 
 \end{equation}
 The heavy gaugino decays in Case B include
\begin{equation}
\chi_{2,3}^0 \rightarrow Z\chi_1^0,\  h\chi_1^0, \  W^\pm\chi_1^\mp\  {\rm and}\  \chi_2^\pm \rightarrow W^\pm\chi_1^0,\ Z\chi_1^\pm,\ h\chi_1^\pm. 
\end{equation}  
Comparing with Case A in Eq.~(\ref{eq:casea}), we therefore have the additional final state
\begin{equation}
W^\pm W^\pm+\met ,
\end{equation}  
coming from the process $\chi_{2}^0 \chi_{3}^0 \to W^\pm W^\pm \chi_1^\mp \chi_1^\mp$ or  $\chi_{2,3}^0 \chi_{2}^\pm \to W^\pm W^\pm \chi_1^\mp \chi_1^0$.  This provides a very clean same-sign dilepton final state with very low SM backgrounds. 

In Case C as seen in Fig.~\ref{fig:spectrum} (right panel), the mass spectrum of the bino-wino-like states is compressed ($M_1=M_2$) and therefore these states can be treated as (nearly degenerate) LSPs. The Higgsino-like neutralinos and chargino form the NLSP states. If the Higgsino mass is decoupled, no NLSP would be produced and therefore direct electroweakino searches would not provide any
 constraint \cite{EWK}. For smaller Higgsino masses within the reach of LHC,  all the final states will be similar to those in Case B and they could be observed as well.  In the discussion of the collider reaches below, we focus on Case A and Case B only.  We will comment on Case C at the end of this section.

\begin{figure}
\centering
\includegraphics[width=0.49\textwidth]{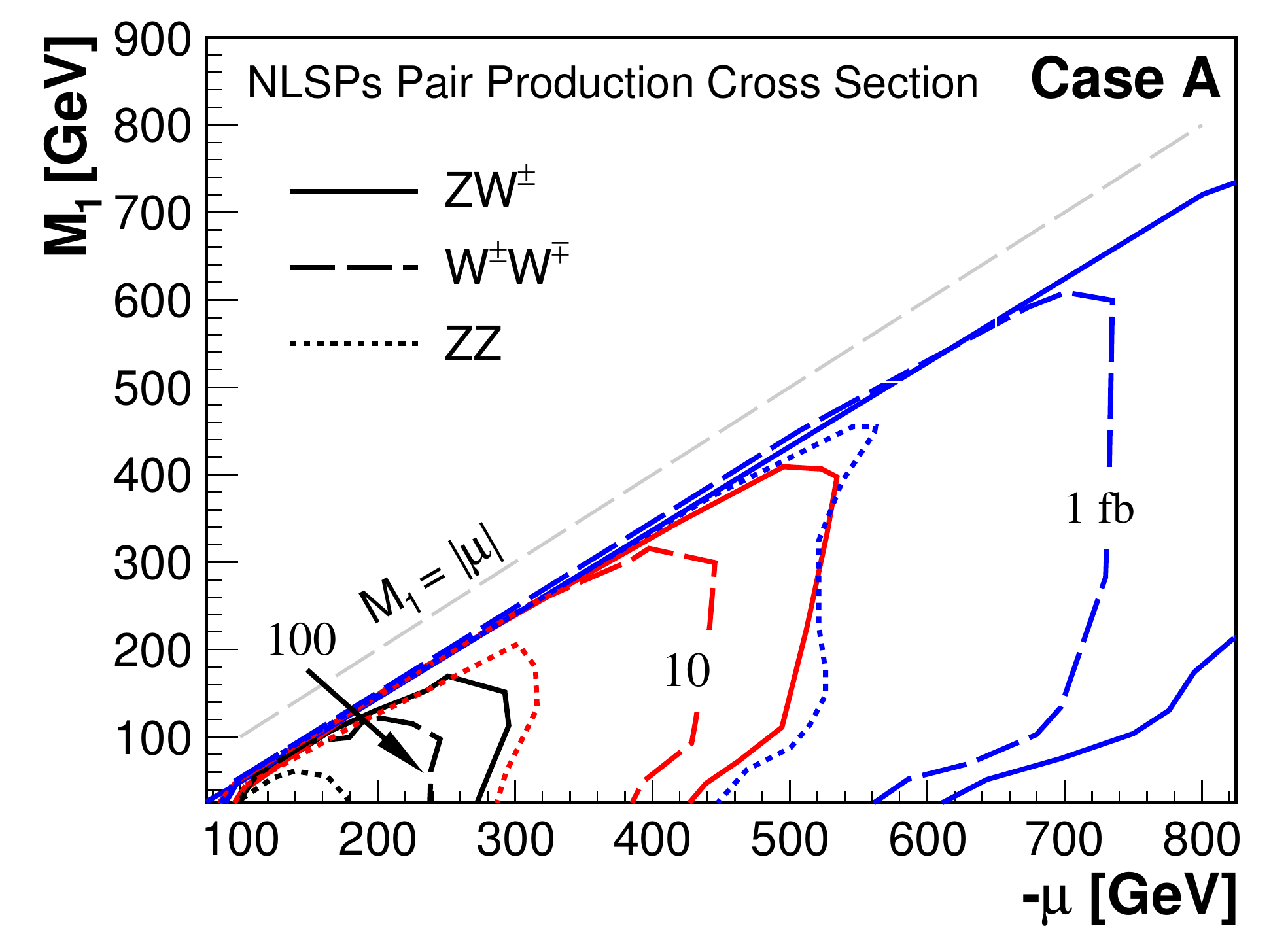}
\includegraphics[width=0.49\textwidth]{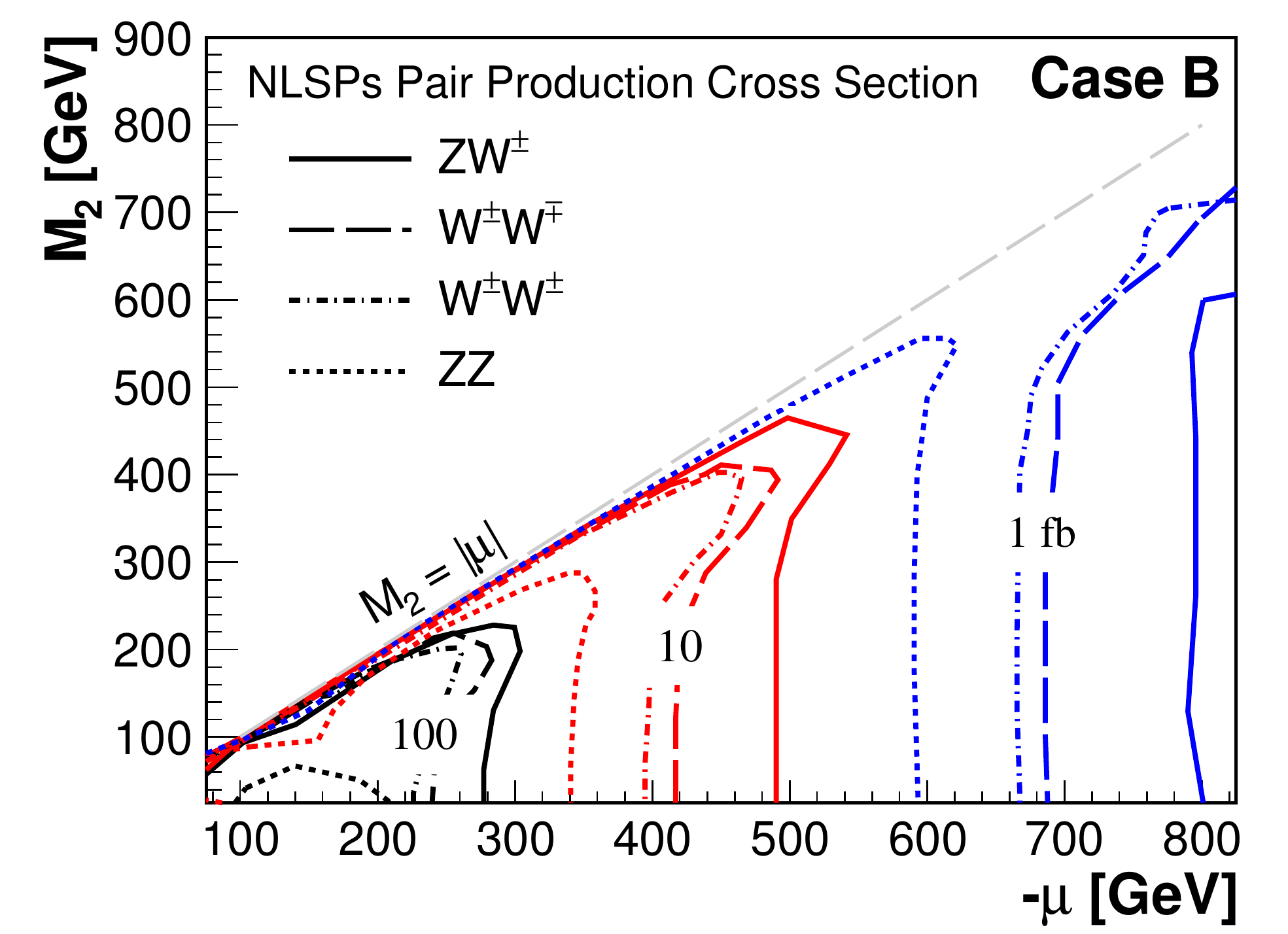}
\caption{Cross section contours for Higgsino-like NLSPs  pair production with subsequent decays to the corresponding gauge-boson final states, 
$ZW^\pm$ (solid curves),
$W^\pm W^\mp$ (dashed),
$W^\pm W^\pm$ (dash-dotted), and 
$ZZ$ (dotted), at 14 TeV LHC for Case A in the $M_1-\mu$ plane (left), and Case B in the $M_2-\mu$ plane (right). 
}
\label{fig:proCS_AB}
\end{figure}

In Fig.~\ref{fig:proCS_AB}, we present the production cross section contours of the dominant channels with gauge boson final states ($ZW^\pm ,\ W^\pm W^\mp ,\ W^\pm W^\pm$ and $ZZ$), for Case A in the $M_1-\mu$ plane and Case B in the $M_2-\mu$ plane, respectively. These cross sections take into account all possible NLSP pair productions $\chi\chi$ with their decays to gauge bosons including the decay branching fractions. We see that cross sections may reach the level of 1 fb for  the NLSP mass parameter $|\mu| \sim 800$ GeV. 

\begin{figure}[tb]
\centering
\includegraphics[width=0.49\textwidth]{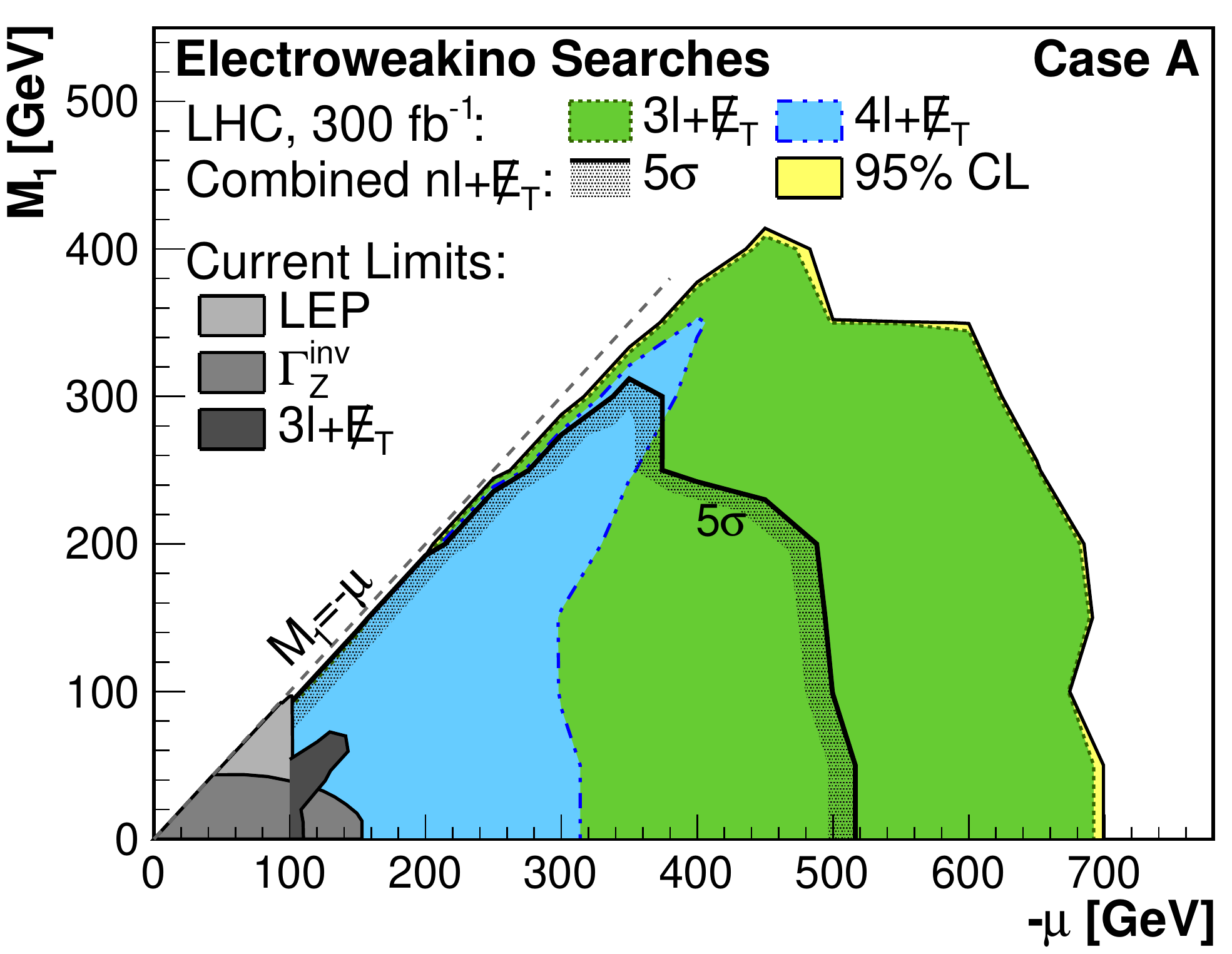}
\includegraphics[width=0.49\textwidth]{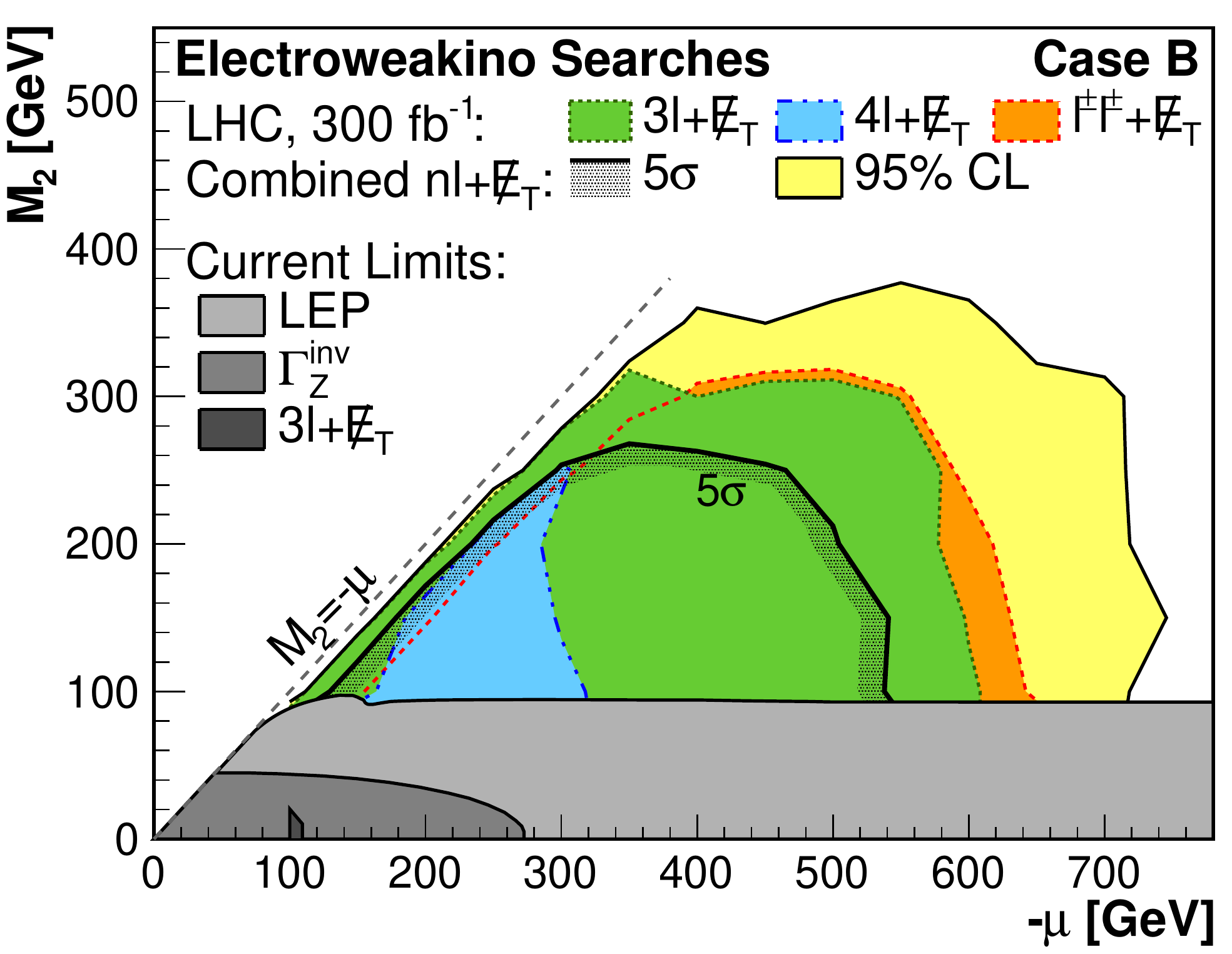}
\caption{Estimated 95\% CL. exclusion reach for eletroweakino searches for Case A (left) and Case B (right). The colored regions show the same-sign di-lepton (orange), tri-lepton (green) and four-lepton (blue) channel as well as their combination (yellow). The combined 5$\sigma$ discovery reach is indicated by the thick solid line next to the hatched region. 
Existing constraints from Electroweak Precision Measurements, LEP~\cite{Abdallah:2003xe,LEP-ALEPH} and LHC~\cite{CMS-SUS-13-006} are shown in gray shaded regions. }
\label{fig:ElectroweakinoSummary}
\end{figure}

Both ATLAS and CMS performed direct electroweakino searches using NLSP pair production. Current experiment limits of several combined channels are shown in Table \ref{tab:ExpResCombined}, for $m_{\n}=0$ with strong assumptions that $m_{\cha}=m_{\nn}$, and the NLSP decays to $\n$ with $100\%$ branching fraction. The results in Table \ref{tab:ExpResCombined} are meant to be a general guidance for the current situation.  However,   they cannot be easily translated to the blind-spot scenarios because of the hidden information such as the values of $\tan\beta$. 
The best sensitivity is obtained in the $WZ$-channel, considering both the $3\ell$ and $2\ell 2j$ final state. The CMS limits for the combined channel ($2\ell,\ 3\ell$) are slightly weaker than the  ATLAS limits due to a $2\sigma$ deviation in the CMS observed data from the expected results.   We apply the CMS bounds from \cite{CMS-SUS-13-006} to the blind-spot scenarios Case A and Case B. The corresponding limits can be seen in Fig. \ref{fig:ElectroweakinoSummary} as dark gray shaded region.

\begin{table}[tb]
\centering
\resizebox{\textwidth}{!}{
\begin{tabular}{|c|c|r l |}
\toprule[1pt]
Production and Decay &  Search Channel & \multicolumn{2}{c|}{95\% CL Exclusion Region}  \\
\midrule[1pt]
$pp\to\nn\cha \to ZW^\pm \n \n$ 
& $2\ell+2j+\met$
& ATLAS \cite{1403-5294}:&$100<m_{\nn,\cha}<420\text{ GeV}$ \\ 
& $3\ell+\met$
& CMS \cite{CMS-SUS-13-006}:&$100<m_{\nn,\cha}<270\text{ GeV}$ \\
\midrule[1pt]
$pp\to\nn\cha \to hW^\pm \n \n$ 
& $\ell+b\bar{b}+\met$, $\ell+\gamma\gamma +\met$ 
& ATLAS \cite{1501-07110}:&$125<m_{\nn,\cha}<272\text{ GeV}$ \\ 
& $\ell^\pm\ell^\pm+\met$, $3\ell+\met$  
& CMS \cite{CMS-SUS-14-002,CMS-SUS-13-006}:&$130<m_{\nn,\cha}<215\text{ GeV}$ \\
\bottomrule[1pt]
\end{tabular}
}
\caption{Current combined exclusion regions on the electroweakinos at the 8 TeV LHC, assuming $m_{\n}=0,\ m_{\cha}=m_{\nn}$, and the NLSP decays to $\n$ with $100\%$ branching fraction.}
\label{tab:ExpResCombined}
\end{table}

There have already been many studies analyzing the reach of electroweakino searches in the context of MSSM at the LHC \cite{Bino-Higgsino1,Bino-Higgsino2,Bino-Higgsino3,Wino-Higgsino1,Wino-Higgsino2,Wino-Higgsino3,1310-4274,1511-05386}.   Those results cannot be directly translated into bounds on the blind-spot parameter space described by the parameter relations given in Table \ref{tab:SIBS}.
In the following we estimate the reach for electroweakino searches at 14 TeV LHC for the blind-spot regions by performing a detailed collider study. 
We use Madgraph 5/MadEvent v1.5.11 \cite{Alwall:2011aa} to generate signal events in the relevant mass parameter plane. Each signal sample consists 250,000 events and contains up to 1 additional jet. These events are passed to Pythia 6.4 \cite{Sjostrand:2006aa} to simulate initial and final state radiation, showering and hadronization. The events are further passed through Delphes 3.1 \cite{Ovyn:2009aa,Favereau:2013aa} with the Snowmass combined LHC detector card \cite{Anderson:2013aa} to simulate detector effects. 
We use the backgrounds generated for the Snowmass Energy Frontier Simulations \cite{Anderson:2013aa} without pile-up. We include the main backgrounds with two or more leptons in the final state: $VV$ (boson pair), $VVV$ (triple boson), $t \bar t$ (top pair) and $t\bar tV$ (top pair plus boson) production. 
To simulate the trigger, we require the events to pass one of the three selections:
\begin{equation}
{\rm either}\ p_{T,\ell}>30\ {\rm GeV};\quad {\rm or\ two\ leptons}\ p_{T,\ell_1}, p_{T,\ell_2} >20,\ 10\ {\rm  GeV};\quad {\rm or}\  \met > 100\ {\rm GeV}. 
\end{equation}
In this section, we do not consider a situation containing a compressed spectrum. The trigger efficiencies are high for the lepton selections. 

We separate these events into different signal regions which are discussed below. Each signal region has a set of observables. These will be passed to a Boosted Decision Tree (BDT) which is implemented in the Root package TMVA \cite{Hoecker:aa}.\footnote{For each benchmark point we train a set of 1000 randomized BDT with a maximal depth of 3.} Using HistFactory \cite{Cranmer:2012sba}, RooFit \cite{Verkerke:2003ir} and RooStats \cite{Moneta:2010pm}, the resulting BDT distribution is used to perform a hypothesis test yielding the significance. We assume a 10\% systematic error on the background cross sections. 
Although this is a rather conservative choice, our conclusions would not change much if we vary this value since the observability is dominated by the statistics.

For all signal regions, we include the following variables as input for the BDT: lepton transverse momentum $p_{T,\ell_i}$ for all leptons $\ell_i$,  missing energy  $\met$ and the scalar sum of hadronic transverse momentum $H_{T}$. We distinguish different signal regions based on lepton content of the final states: 
\begin{itemize}
\item \textbf{Same-Sign Dilepton ($W^\pm W^\pm + \met$):} This signal region consists of events with two same-sign leptons. This mainly targets the processes $\chi_{2}^0\ \chi_{3}^0 \to W^\pm W^\pm \chi_1^\mp \chi_1^\mp$ and $\chi_{2,3}^0 \chi_{2}^\pm \to W^\pm W^\pm \chi_1^\mp \chi_1^0$  (Case B) which have a same-sign dilepton plus missing energy final state. The dominating backgrounds are vector boson-pair, top-pair and top-pair vector boson associate production with leptonic decays. We use the following variables as additional input for the BDT: 
\begin{itemize}
\item $R_{\ell\ell}$, $p_{T,\ell\ell}, m_{\ell\ell}$: separation, transverse momentum and invariant mass of the two leptons.
\end{itemize}

\item \textbf{Opposite-Sign Dilepton ($W^\pm W^\mp + \met$):} This signal region consists of events with two opposite-sign leptons. However, this channel suffers from large $tt$, $WW$ and mono-$Z$ backgrounds and therefore is insensitive in most parts of the parameter space.

\item \textbf{Trilepton ($W^\pm Z + \met$):} This signal region consists of events with three leptons. This mainly targets the processes like $\chi_{1,2}^\pm\ \chi^0_{2,3} \to W^\pm Z \chi_1^0 \chi_1^0$ which has a tri-lepton plus missing energy final state.   We require at least one opposite-sign, same flavor lepton pair which will be identified as $Z$ originated. The dominating backgrounds are $WZ$ production with sub-leading contributions from triple vector boson and top-pair vector boson associate production.  We use the following variables as additional input for the BDT: 
\begin{itemize}
\item $\Delta R_Z$, $p_{T,Z}, m_{Z}$: separation, transverse momentum and invariant mass of the two leptons which are considered to be $Z$ originated.
\item $M_T$, $\Delta R_{Zl}$, $m_{Z,\ell}$: transverse mass of the $W$-originated lepton, separation and invariant mass of the $Z$-candidate and the third lepton.
\end{itemize}

\item \textbf{Four Leptons ($ZZ + \met$):} This signal region consists of events with four leptons. This mainly targets the processes like $\chi^0_2\ \chi^0_{3} \to Z Z \chi_1^0 \chi_1^0$ which has a four-lepton plus missing energy final state. We require   two opposite-sign, same flavor lepton pairs which will be assumed to be $Z$ originated. The dominating backgrounds are $ZZ$, triple vector boson and top-pair vector boson associate production. We use the following variables as additional input for the BDT: 
\begin{itemize}
\item $\Delta R_{Z_i}$, $p_{T,Z_i}, m_{Z_i} \ (i=1,2)$: separation, transverse momentum and invariant mass of the two leptons which are considered to be $Z$ originated .
\item $\Delta \phi_{ZZ}$, $\Delta R_{ZZ}$, $m_{ZZ}$: angular separation, separation and invariant mass of the two $Z$-candidates.
\end{itemize}

\end{itemize}
For each signal region we independently perform an analysis for final states with one and two lepton flavors, and combine the results afterwards. While the identification of the $Z$-candidate is straightforward in the case of a two lepton-flavor final state, we choose the opposite-sign lepton pair closest to the $Z$ mass in the one lepton flavor final state case. We do not include a search strategy targeting the decay channels with $h$ in the final states due to large backgrounds. 

The results of our analysis are shown in Fig.~\ref{fig:ElectroweakinoSummary}. The LEP limits from electroweak precision measurement and chargino searches, as discussed in Sec.~\ref{sec:LEP2}, are shown as light gray shaded area in Fig.~\ref{fig:ElectroweakinoSummary}. While chargino searches exclude $M_2 \approx m_{\chi^+_1} \lesssim 100$ GeV for Case B, only   a small region with $-\mu\lesssim 100$ GeV is excluded in Case A. 
Bounds from current LHC searches at 8 TeV \cite{CMS-SUS-13-006} are shown as dark gray shaded area and exceed the LEP  bounds only in Case A for $-\mu=140$ GeV and $M_1 = 70$ GeV. 
We consider 14 TeV LHC at an integrated luminosity of 300 fb$^{-1}$ and take into account a 10\% systematic error on the background cross sections. 
The colored region shows the estimated 95\% CL exclusion reach for the same-sign dilepton channel (orange region with dashed line in the right panel), tri-lepton channel (green region with dotted line), four lepton channel (blue region with dash-dotted line) and a combination of all channels (yellow region with solid line). We can see that the three lepton channel (green) provide the best reach in Case A. A similar reach is obtained by the same-sign dilepton channel (orange) in Case B. 
The combined reach in Case B is therefore about 100 GeV better than either the trilepton channel or the same-sign dilepton channel.   The reach in the four lepton final channel (blue) is limited due to the smaller $\chi\chi\to ZZ+\met$ cross section (see Fig.~\ref{fig:proCS_AB}) and the small branching fraction $Z \to \ell \bar\ell$. 
%
The thick solid line with next to the hatched area encloses the estimated 5$\sigma$ discovery reach for 14 TeV LHC with integrated luminosity of 300 fb$^{-1}$.  At the 14 TeV LHC with 300 ${\rm fb}^{-1}$ luminosity, the 95\% CL reach with the combination of all channels in $|\mu|$ is about 700 GeV for both  Case A  and Case B.    The 5$\sigma$ discovery reach is about 500 GeV for both cases. 

We do not perform a detailed collider study for electroweakino searches for Case C. The NLSP production rates vanish for a decoupled Higgsino masses when  the NLSPs are out of the LHC reach. If the Higgsino is   within the reach of LHC, then the collider phenomenology will be similar to that of Case B. In the most optimistic case, in which the Higgsino has the lightest possible mass allowed by the blind-spot relation, $\mu + M_2\sin2\beta=0$, the reach is expected to be very similar to Case B.   Thus we do not further study the collider phenomenology of Case C in detail.


\subsection{Mono-jet searches}
\label{sec:monojet}

While the production of only an LSP pair is unobservable at colliders, the LSP pair production in association with a hard ISR jet (or photon) provides a clean and distinctive signature in dark matter searches. The main SM background to this mono-jet final state comes from $Z+$jet production in which the $Z$-boson decays into neutrinos. The rate for this process is small and hence making searches for such events with large missing transverse momentum balanced by an energetic jet (or photon) a promising tool for neutralino LSP at colliders. 

If the LSP is bino-like, its coupling to the $Z$-boson is proportional to the bino-Higgsino mixing,  which quickly decreases at large Higgsino masses (see discussion in Sec.~\ref{sec_EW}). Therefore the LSP pair production rate is strongly suppressed and no limit based on mono-jet searches can be obtained in Case A.


Both ATLAS \cite{ATLASmonojet} and CMS \cite{CMSmonojetPublished} have performed a monojet search at 8 TeV LHC using 20.3 fb$^{-1}$ and 19.7 fb$^{-1}$ integrated luminosity respectively. For Case B, using the same cuts used by ATLAS and the upper limit set by ATLAS \cite{ATLASmonojet},   the current exclusion bounds are rather weak.  Only the region where LSP mass is less than about 100 GeV have been excluded by current 8 TeV results.   These limits have a similar reach as the mono-photon search at LEP.   

To estimate the collider reach and expected bounds at 14 TeV LHC, we perform a MC analysis.\footnote{More details about the methodology can be found in Sec.~\ref{sec:EWsearch}.}  As signal we consider the pair production of both wino-like states generated with up to two additional jets. We use the backgrounds generated by the Snowmass Energy Frontier Simulations \cite{Anderson:2013aa} without pile-up. We consider the main backgrounds including large missing energy: $W/Z$ + jets and vector boson pair production.

For jet identification, we use the anti-$k_t$ jet algorithm with $R=0.4$ to find jets with $p_T>30$ GeV and $|\eta_j|<4.5$. Similar to the CMS 8 TeV analyses \cite{CMSmonojetPublished}, we require events to either pass the missing energy trigger $\met>120$ GeV, or the MET+Jet trigger with $\met>105$ GeV and a leading jet with $p_T>80$ GeV and $|\eta|<2.6$. Events with more than two jets are rejected. Events with leptons ($e,\mu,\tau$), photons and tagged jets ($b,\tau$) are vetoed. Following the analysis in \cite{Wino-Higgsino2} we assume a systematic error of $1-2\%$ on the background cross section.
Two signal regions are defined based on the number of jets: 1 jet + $\met$ and 2 jets + $\met$. As input for the BDT we use $\met$, $H_T$, the jets transverse momentum and rapidity $p_{T,j}$ and $\eta_{j}$ as well as the angular separation between jet and missing energy or the two jets,  $\Delta \phi (\met,j) $ and $\Delta \phi (j,j)$.

\begin{figure}[tb]
\centering
\includegraphics[width=0.49\textwidth]{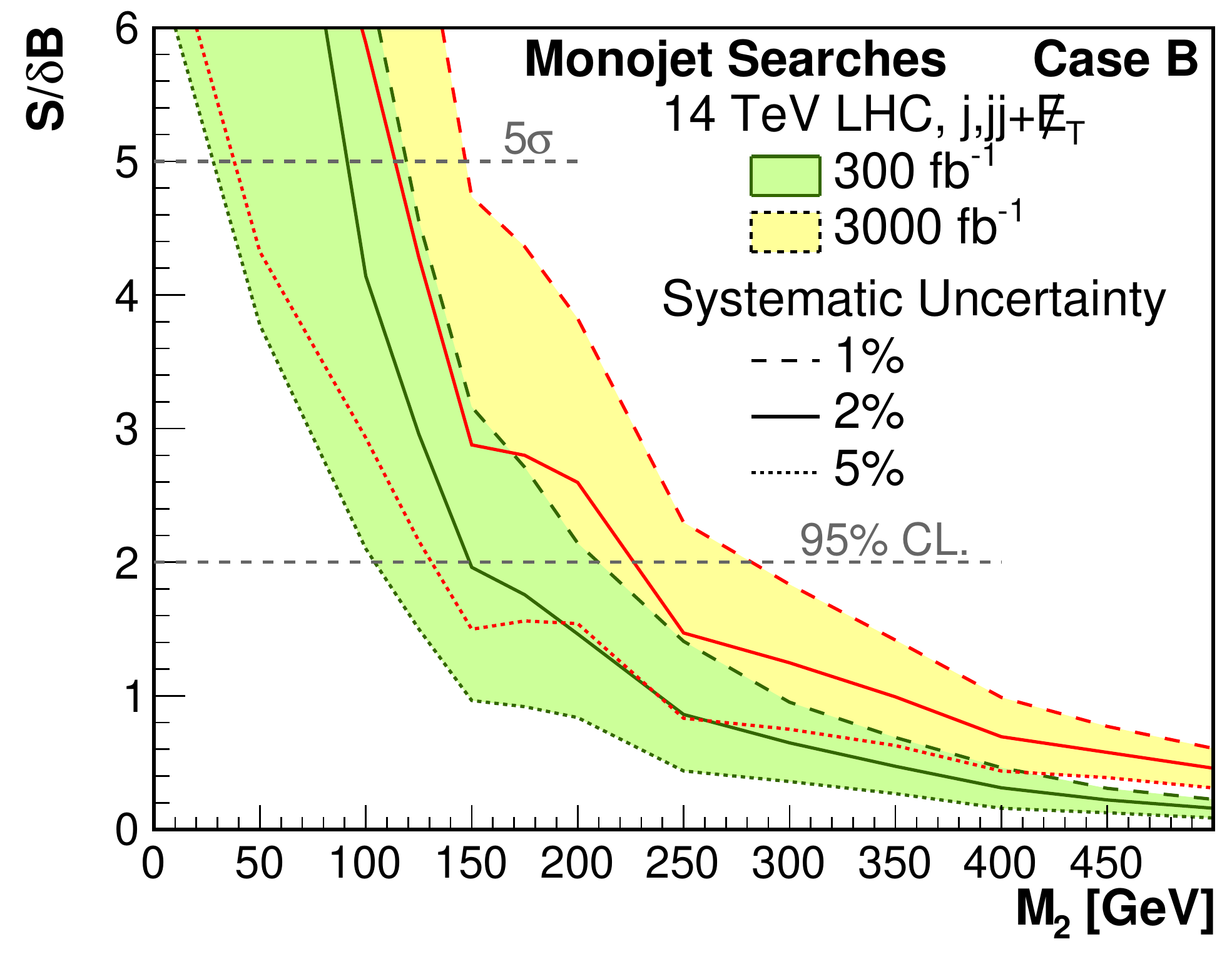}
\includegraphics[width=0.49\textwidth]{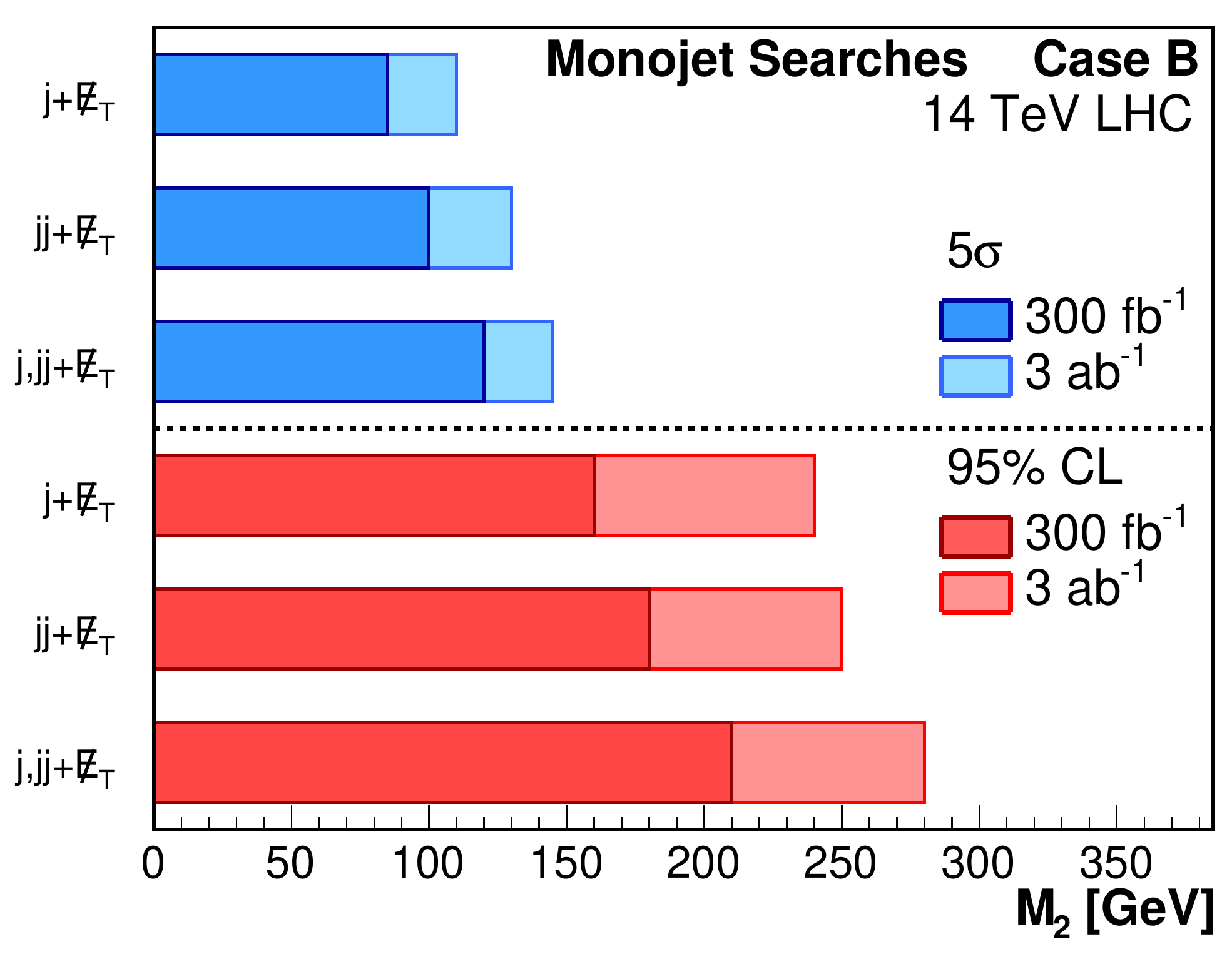}
\caption{Left: 
Mass reach in the monojet channel with 1-jet and 2-jets events at 14 TeV LHC with  300 fb$^{-1}$  (green) and 3000 fb$^{-1}$  (yellow) integrated luminosity.  The bands are generated by varying the background systematics  from $1\%$ (dashed far-right curves), $2\%$ (solid middle  curves), to $5\%$ (dotted far-left curves).  
Right: Exclusion and discover reach at 14 TeV LHC assuming 1\% systematic uncertainty on the background cross section.}
\label{fig:MonojetSummary}
\end{figure}

We find that typically the $W/Z$+jets backgrounds dominate while the sub-leading vector boson pair background has a similar cross section as the signal. The mass reach is shown in Fig.~\ref{fig:MonojetSummary}.  Both the one and two jet channels contribute about equally. Combining both channels, we obtain a 95\% CL exclusion reach of 
$100-210$  GeV ($130-280$ GeV) at 14 TeV LHC with 300 (3000) fb$^{-1}$ assuming a systematic error on the background cross section of $1\%-5\%$, as indicated by the wide bands in Fig.~\ref{fig:MonojetSummary} (the left panel).   Increasing the systematic error from 1\% to $5\%$ would lower the mass coverage by almost 100 GeV or more.
These limits are in agreement with the results found for a pure wino LSP in \cite{Wino-Higgsino2}. The mono-jet search limits for Case C is very similar to that of Case B. 

 A scenario similar to our Case C, with $M_1 \approx M_2$ and decoupled $\mu$ resulting in a compressed spectrum, has been studied in the literature \cite{Schwaller:2013baa} using signatures of  large missing energy and soft leptons.  At 95\% CL, the reach is $M_1\approx M_2 > 150-200$ GeV at LHC with 300 fb$^{-1}$ luminosity assuming a systematic error of 1\% on the background cross section. Ref.~\cite{vbf-tao} proposed a search for VBF production of bino-wino LSP pairs with two forward jets.  At 14 TeV LHC with 300 fb$^{-1}$ luminosity,  a 95\% CL reach of $M_1=M_2> 200$ GeV for $\tan\beta=5$ can be obtained. However, given that $S/B\sim 2-3$\%, the results heavily depends on effective control of the systematic errors.

\section{Summary and Conclusions}
\label{sec:conclusion}

In this paper, we carried out comprehensive analyses for the dark matter (DM) searches under the
 blind-sport scenarios in MSSM. We first summarized all the theoretical conditions for blind spots, and listed  three distinctive situations (Cases A, B, C) at the end of Sec.~\ref{sec:BS}. We identified a new blind-spot condition for both spin-independent (SI) and spin-dependent (SD)  scattering (Case C) that was not studied in the literature \cite{blindspots}. We then quantified the requirements of acceptable DM relic abundance for the blind-spot scenarios as shown in Sec.~\ref{sec:RD}. We found that Case A was disfavored by the relic density considerations because of the bino nature of the DM candidate with a low annihilation rate, as seen in Fig.~\ref{fig:Relic_CaseA}. However, including co-annihilation effects with a light slepton could help to enhance the annihilation cross section and render the relic density at an acceptable level, as shown in Fig.~\ref{fig:Relic_CaseA_stau}.
Cases B and C can readily provide an acceptable relic density since winos are typically under-abundant as long as they are not too heavy, as seen in Fig.~\ref{fig:Relic_CaseBC}.
Co-annihilation could also help to extend to heavier winos by enhancing the annihilation cross section and thus to yield a desirable relic density.
Note that including light slepton, e.g. $\tilde{\tau}_R$, will not have notable effects in the DM direct and indirect detections, as well as collider searches for the gauginos. 

We set out to explore the complementary coverage for the blind spots in the MSSM theory parameters, for the projection of the future underground DM direct searches, the indirect searches from the relic DM annihilation into photons and neutrinos, and for collider DM searches including the current bounds from the existing LEP and LHC results, as well as the future LHC upgrade to higher luminosities (HL-LHC). We found that
\begin{itemize}
\item
The SI blind spots for Cases A and B may be rescued by the SD direct detections, as seen in Figs.~\ref{fig:SigmaSDn} and \ref{fig:SigmaSDp}, with $\chi$-$n$ scattering more promising than that of $\chi$-$p$ scattering.
%
\item
The neutrino detections from IceCube and SuperK are approaching the sensitivity on the SD scattering cross section for the blind-spot region for Case A in the best $WW$ channel and the next $\tau\tau$ channel, as seen in Fig.~\ref{fig:SigmaSDpCaseA}, but still about an order of magnitude away for Case B, as in seen Fig.~\ref{fig:SigmaSDpCaseB}.
\item
The detection of gamma-rays from Fermi-LAT may not reach the desirable sensitivity for searching for the DM blind-spot regions, as shown in Figs.~\ref{fig:FermiGamma_CaseA}, \ref{fig:FermiGamma_CaseB} and \ref{fig:FermiGamma_CaseC}.
\item
The $Z$-invisible decay searches at LEP1 already excluded the small-mass region for $m^{}_{\text{DM}}<45$ GeV as discussed in Sec.~\ref{sec_EW} and in Fig.~\ref{fig:Zinv}.
The chargino searches at LEP2 also imposed some bounds as seen in Sec.~\ref{sec:LEP2}.
\item
The Disappearing Track (DT) search of winos at the LHC experiments are particularly sensitive to the large values of $|\mu|$ and $\tan\beta$ when the mixing with Higgsinos are small. The projected $95\%$ CL  sensitivity at the 14 TeV LHC could reach $M_{2}\sim 600\ (1100)$ GeV with 300 fb$^{-1}$ (3000 fb$^{-1}$) for Case B, and $M_{1,2}\sim 1\ (2)$ TeV with 300 fb$^{-1}$ (3000 fb$^{-1}$) for Case C, 
as shown in the two panels of Fig.~\ref{fig:DT14TeV}, respectively.  
\item
Cross sections for the electroweakino (EW) pair production and decay to $WW/WZ/ZZ$ channels at the 14 TeV LHC are plotted in Fig.~\ref{fig:proCS_AB}. The SUSY search sensitivity with 300 fb$^{-1}$ may  cover the blind-spot regions of Case A 
up to $M_{1}\sim 300$ GeV (400 GeV) for $5\sigma$ ($95\%$ CL), 
and Case B 
up to $M_{2}\sim 260$ GeV (380 GeV) for $5\sigma$ ($95\%$ CL), 
and $|\mu| \sim 500$ GeV (700 GeV) for $5\sigma$ ($95\%$ CL), 
as shown in Fig.~\ref{fig:ElectroweakinoSummary}.   With 3000 ${\rm fb}^{-1}$ luminosity, the reach in $|\mu|$ is about 150 GeV better.
\item
The searches of mono-jet signal at the 14 LHC with 300 fb$^{-1}$ luminosity  may reach a sensitivity to cover the blind-spot regions of Case B up to $M_{2}\sim 130$ GeV (210 GeV) for $5\sigma$ ($95\%$ CL) as shown in Fig.~\ref{fig:MonojetSummary}. The coverage can be improved by about $30\%$ in the mass reach with 3000 fb$^{-1}$.  
\end{itemize}

\begin{figure}[tb]
\centering
\includegraphics[width=0.49\textwidth]{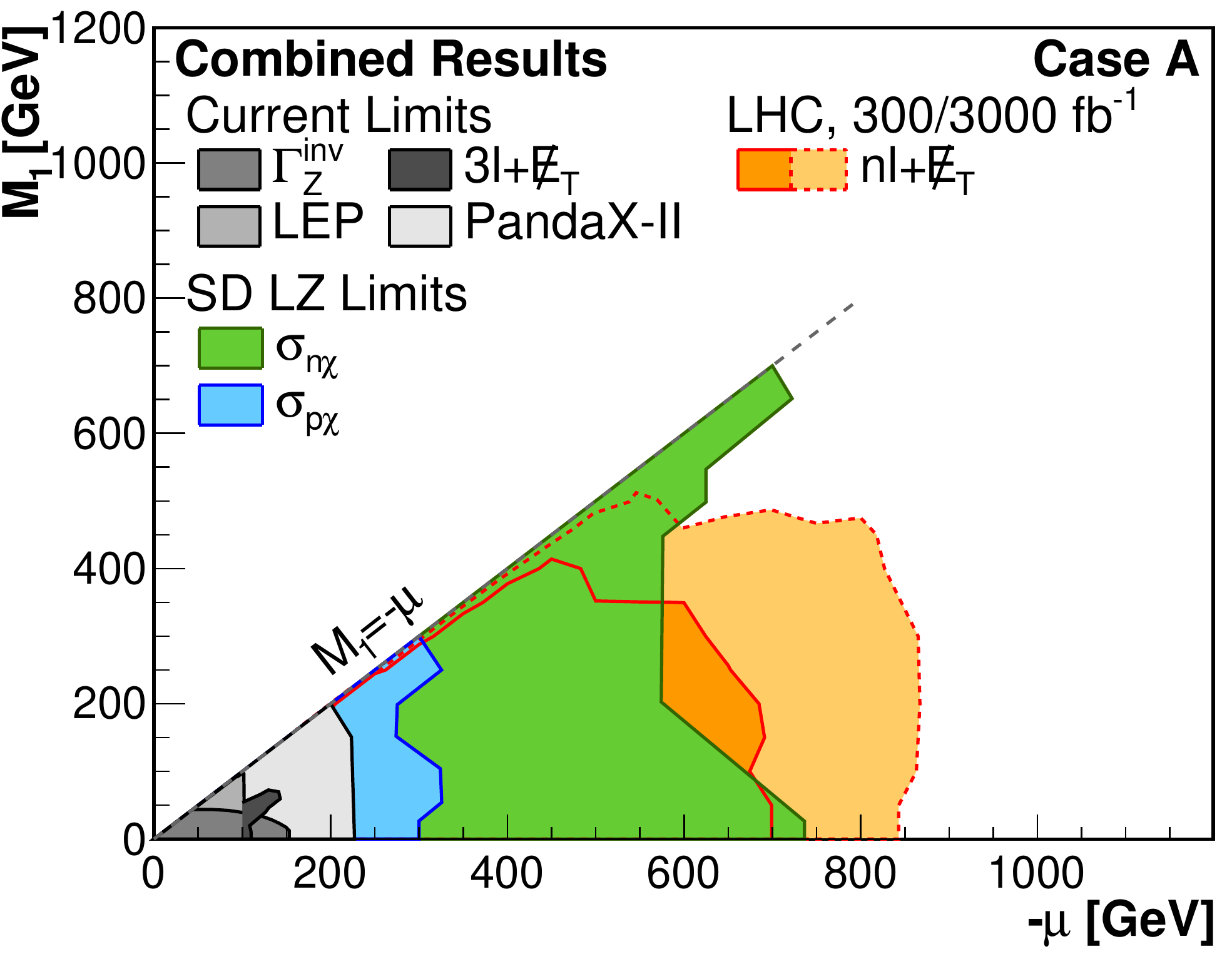}
\includegraphics[width=0.49\textwidth]{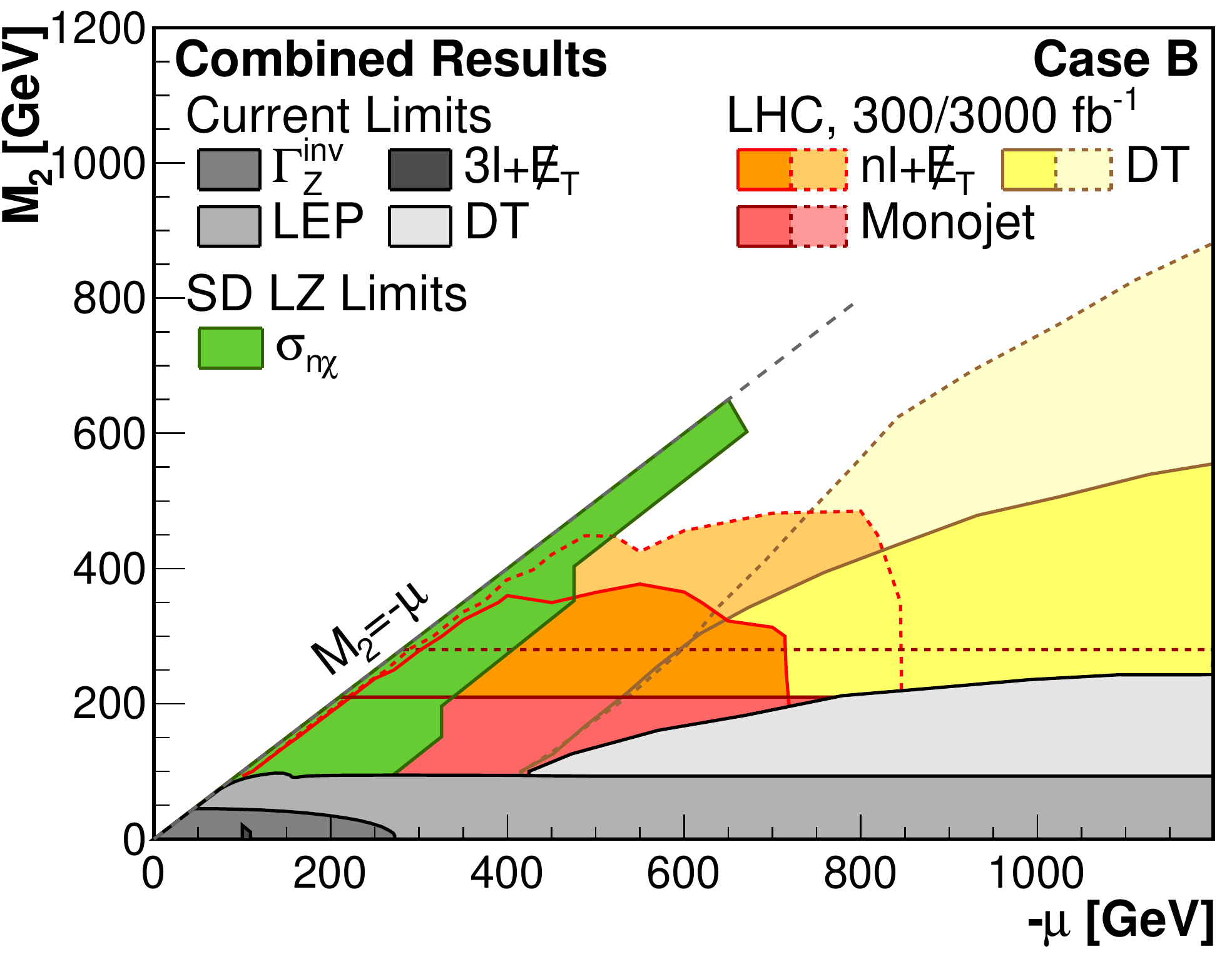}
\caption{Summary plot for the blind-spot reach including the current bounds from LEP and LHC, future DM search projection, and 95\% CL exclusion reach at the 300 ${\rm fb}^{-1}$ 14 TeV LHC and the HL-LHC. 
Left panel for Case A in the $M_1-\mu$ plane and right panel for Case B in the $M_2-\mu$ plane. ``DT'' denotes the disappearing track search. }
\label{fig:Summary}
\end{figure}

\begin{table}[tbh]
\centering
\begin{tabular}{|c|c|c|c|}
\hline
Blind spots in Sec.~\ref{sec:bs}/Searches: & Direct detection & Indirect detection & LHC searches \\ \hline
Case A (SI) & yes (SD) & maybe (SD) & yes (EW) \\ \hline
Case B (SI) & yes (SD) & maybe not (SD) & yes (DT, EW, jets) \\ \hline
Case C (SI  $\&$ SD) & no & no & yes (DT, EW, jets) \\ \hline
\end{tabular}
\caption{
Blind spots as defined in Sec.~\ref{sec:bs} and their search sensitivities in various means as discussed in the text. Notations and details can be also found in this summary section. }
\label{tab:sum}
\end{table}

\noindent
Our results are summarized in Fig.~\ref{fig:Summary} for Cases A and B and tabulated in Table  \ref{tab:sum}, where we see the complementarity for the blind-spot coverage, including the current bounds from LEP and LHC, future DM search via the SD scattering, and collider reach at the 14 TeV LHC and the HL-LHC extension for the SUSY blind-spot parameters. 
Proper treatments of the DM relic densities have been taken into account, as stated at the end of Sec.~\ref{sec:RD}.

In conclusion, 
 the SUSY WIMP dark matter would be difficult to discover in the blind-spot regions via the leading direct detection means via the SI scattering, as well as via the indirect detections with gamma rays. 
We point out that the SI scattering blind spots \cite{blindspots} may be rescued by SD scattering searches in the future direct detection experiments. The neutrino detections from IceCube and SuperK are approaching the sensitivity on the SD scattering cross section for the blind-spot Case A.
Furthermore, the SUSY searches at the LHC will substantially extend the coverage for all the three blind-spot scenarios to large parameter regions, in particular including the ``most blind'' scenario Case C. 
After all, the dark matter blind spots may be unblinded with the collective efforts in future DM searches. 
In the optimistic situation with a discovery for the SUSY signals at the LHC, it is of ultimate important to determine the parameters both for the masses and couplings to check against the DM properties as well as the blind-spot relations. Only with those achievements of a fully consistency check, can one reach the conclusion for the identification of the SUSY dark matter.

\acknowledgments
We would like to thank Xerxes Tata for discussions. 
The work of TH is supported in part by the Department of Energy under Grant No.~DE-FG02-95ER40896, and in part by PITT PACC.  The work of FK is supported by NSF under Grant PHY-1620638. FK also acknowledges support from the Fermilab Graduate Student Research Program in Theoretical Physics operated by Fermi Research Alliance, LLC under Contract No.~DE-AC02-07CH11359 with the United States Department of Energy. The work of SS is supported by the Department of Energy under Grant DE-FG02-13ER41976/de-sc0009913, and partly by the National Science Foundation of China (NSFC) under Grant No.~11428511.   The work of YW is supported by Chinese Scholarship Council.  We would also like to thank the Aspen Center for Physics for hospitality, where part of the work was completed. The Aspen Center for Physics is supported by the NSF under Grant No.~PHYS-1066293. 
 
\bibliographystyle{JHEP}
\bibliography{references}

\providecommand{\href}[2]{#2}\begingroup\raggedright\begin{thebibliography}{10}

\bibitem{Lee:1977ua}
B.~W. Lee and S.~Weinberg, {\it {Cosmological Lower Bound on Heavy Neutrino
  Masses}},  {\em Phys. Rev. Lett.} {\bf 39} (1977) 165--168.

\bibitem{Goldberg:1983nd}
H.~Goldberg, {\it {Constraint on the Photino Mass from Cosmology}},  {\em Phys.
  Rev. Lett.} {\bf 50} (1983) 1419. [Erratum: Phys. Rev.
  Lett.103,099905(2009)].

\bibitem{CDEX}
{\bf CDEX} Collaboration, Q.~Yue et~al., {\it {Limits on light WIMPs from the
  CDEX-1 experiment with a p-type point-contact germanium detector at the China
  Jingping Underground Laboratory}},  {\em Phys. Rev.} {\bf D90} (2014) 091701,
  [\href{http://arxiv.org/abs/1404.4946}{{\tt arXiv:1404.4946}}].

\bibitem{XENON1T}
{\bf XENON1T} Collaboration, E.~Aprile, {\it {The XENON1T Dark Matter Search
  Experiment}},  {\em Springer Proc. Phys.} {\bf 148} (2013) 93--96,
  [\href{http://arxiv.org/abs/1206.6288}{{\tt arXiv:1206.6288}}].

\bibitem{XENON1T1512}
{\bf XENON} Collaboration, E.~Aprile et~al., {\it {Physics reach of the XENON1T
  dark matter experiment}},  {\em JCAP} {\bf 1604} (2016), no.~04 027,
  [\href{http://arxiv.org/abs/1512.07501}{{\tt arXiv:1512.07501}}].

\bibitem{PICO-2L}
{\bf PICO} Collaboration, C.~Amole et~al., {\it {Dark Matter Search Results
  from the PICO-2L C$_3$F$_8$ Bubble Chamber}},  {\em Phys. Rev. Lett.} {\bf
  114} (2015), no.~23 231302, [\href{http://arxiv.org/abs/1503.00008}{{\tt
  arXiv:1503.00008}}].

\bibitem{PICO-60}
{\bf PICO} Collaboration, C.~Amole et~al., {\it {Dark Matter Search Results
  from the PICO-60 CF$_3$I Bubble Chamber}},  {\em Submitted to: Phys. Rev. D}
  (2015) [\href{http://arxiv.org/abs/1510.07754}{{\tt arXiv:1510.07754}}].

\bibitem{LUX-SD}
{\bf LUX} Collaboration, D.~S. Akerib et~al., {\it {Results on the
  Spin-Dependent Scattering of Weakly Interacting Massive Particles on Nucleons
  from the Run 3 Data of the LUX Experiment}},  {\em Phys. Rev. Lett.} {\bf
  116} (2016), no.~16 161302, [\href{http://arxiv.org/abs/1602.03489}{{\tt
  arXiv:1602.03489}}].

\bibitem{LUX-SI}
{\bf LUX} Collaboration, D.~S. Akerib et~al., {\it {Results from a search for
  dark matter in the complete LUX exposure}},
  \href{http://arxiv.org/abs/1608.07648}{{\tt arXiv:1608.07648}}.

\bibitem{PandaX-II-SI}
{\bf PandaX-II} Collaboration, A.~Tan et~al., {\it {Dark Matter Results from
  First 98.7 Days of Data from the PandaX-II Experiment}},  {\em Phys. Rev.
  Lett.} {\bf 117} (2016), no.~12 121303,
  [\href{http://arxiv.org/abs/1607.07400}{{\tt arXiv:1607.07400}}].

\bibitem{PandaX-II-SD}
{\bf PandaX-II} Collaboration, C.~Fu et~al., {\it {Spin-dependent WIMP-nucleon
  cross section limits from first data of PandaX-II experiment}},
  \href{http://arxiv.org/abs/1611.06553}{{\tt arXiv:1611.06553}}.

\bibitem{LZ-CDR}
{\bf LZ} Collaboration, D.~S. Akerib et~al., {\it {LUX-ZEPLIN (LZ) Conceptual
  Design Report}},  \href{http://arxiv.org/abs/1509.02910}{{\tt
  arXiv:1509.02910}}.

\bibitem{1509-08767}
{For a recent review, see, i.e., T. Marrod{\'a}n Undagoitia} and L.~Rauch, {\it
  {Dark matter direct-detection experiments}},  {\em J. Phys.} {\bf G43}
  (2016), no.~1 013001, [\href{http://arxiv.org/abs/1509.08767}{{\tt
  arXiv:1509.08767}}].

\bibitem{CMSmonojetPublished}
{\bf CMS} Collaboration, V.~Khachatryan et~al., {\it {Search for dark matter,
  extra dimensions, and unparticles in monojet events in proton--proton
  collisions at $\sqrt{s} = 8$ TeV}},  {\em Eur. Phys. J.} {\bf C75} (2015),
  no.~5 235, [\href{http://arxiv.org/abs/1408.3583}{{\tt arXiv:1408.3583}}].

\bibitem{ATLASmonojet}
{\bf ATLAS} Collaboration, G.~Aad et~al., {\it {Search for new phenomena in
  final states with an energetic jet and large missing transverse momentum in
  pp collisions at $\sqrt{s}=$8 TeV with the ATLAS detector}},  {\em Eur. Phys.
  J.} {\bf C75} (2015), no.~7 299, [\href{http://arxiv.org/abs/1502.01518}{{\tt
  arXiv:1502.01518}}]. [Erratum: Eur. Phys. J.C75,no.9,408(2015)].

\bibitem{AMS02-Ele1}
{\bf AMS} Collaboration, L.~Accardo et~al., {\it {High Statistics Measurement
  of the Positron Fraction in Primary Cosmic Rays of 0.5--500 GeV with the
  Alpha Magnetic Spectrometer on the International Space Station}},  {\em Phys.
  Rev. Lett.} {\bf 113} (2014) 121101.

\bibitem{AMS02-Ele2}
{\bf AMS} Collaboration, M.~Aguilar et~al., {\it {Electron and Positron Fluxes
  in Primary Cosmic Rays Measured with the Alpha Magnetic Spectrometer on the
  International Space Station}},  {\em Phys. Rev. Lett.} {\bf 113} (2014)
  121102.

\bibitem{AMS02-Ele3}
{\bf AMS} Collaboration, M.~Aguilar et~al., {\it {Precision Measurement of the
  ($e^+ + e^-$) Flux in Primary Cosmic Rays from 0.5 GeV to 1 TeV with the
  Alpha Magnetic Spectrometer on the International Space Station}},  {\em Phys.
  Rev. Lett.} {\bf 113} (2014) 221102.

\bibitem{IC79-2016}
{\bf IceCube} Collaboration, M.~G. Aartsen et~al., {\it {Improved limits on
  dark matter annihilation in the Sun with the 79-string IceCube detector and
  implications for supersymmetry}},  {\em JCAP} {\bf 1604} (2016), no.~04 022,
  [\href{http://arxiv.org/abs/1601.00653}{{\tt arXiv:1601.00653}}].

\bibitem{SuperK}
{\bf Super-Kamiokande} Collaboration, K.~Choi et~al., {\it {Search for
  neutrinos from annihilation of captured low-mass dark matter particles in the
  Sun by Super-Kamiokande}},  {\em Phys. Rev. Lett.} {\bf 114} (2015), no.~14
  141301, [\href{http://arxiv.org/abs/1503.04858}{{\tt arXiv:1503.04858}}].

\bibitem{ANTARES-SUN-DM}
{\bf ANTARES} Collaboration, S.~Adrian-Martinez et~al., {\it {First results on
  dark matter annihilation in the Sun using the ANTARES neutrino telescope}},
  {\em JCAP} {\bf 1311} (2013) 032, [\href{http://arxiv.org/abs/1302.6516}{{\tt
  arXiv:1302.6516}}].

\bibitem{FermiLAT-6yr-Gamma}
{\bf Fermi-LAT} Collaboration, M.~Ackermann et~al., {\it {Searching for Dark
  Matter Annihilation from Milky Way Dwarf Spheroidal Galaxies with Six Years
  of Fermi Large Area Telescope Data}},  {\em Phys. Rev. Lett.} {\bf 115}
  (2015), no.~23 231301, [\href{http://arxiv.org/abs/1503.02641}{{\tt
  arXiv:1503.02641}}].

\bibitem{Conrad:2014tla}
{For a recent review, see, i.e., J. Conrad}, {\it {Indirect Detection of WIMP
  Dark Matter: a compact review}},  in {\em {Interplay between Particle and
  Astroparticle physics (IPA2014) London, United Kingdom, August 18-22, 2014}},
  2014.
\newblock \href{http://arxiv.org/abs/1411.1925}{{\tt arXiv:1411.1925}}.

\bibitem{blindspots}
C.~Cheung, L.~Hall, D.~Pinner, and J.~Ruderman, {\it {Prospects and blind spots
  for neutralino dark matter}},  {\em Journal of High Energy Physics} {\bf
  2013} (2013), no.~5 [\href{http://arxiv.org/abs/1211.4873}{{\tt
  arXiv:1211.4873}}].

\bibitem{1405-6716}
M.~Cahill-Rowley, R.~Cotta, A.~Drlica-Wagner, S.~Funk, J.~Hewett, A.~Ismail,
  T.~Rizzo, and M.~Wood, {\it {Complementarity of dark matter searches in the
  phenomenological MSSM}},  {\em Phys. Rev.} {\bf D91} (2015), no.~5 055011,
  [\href{http://arxiv.org/abs/1405.6716}{{\tt arXiv:1405.6716}}].

\bibitem{1409-6322}
T.~A.~W. Martin and D.~Morrissey, {\it {Electroweakino constraints from LHC
  data}},  {\em JHEP} {\bf 12} (2014) 168,
  [\href{http://arxiv.org/abs/1409.6322}{{\tt arXiv:1409.6322}}].

\bibitem{1412-5952}
G.~Grilli~di Cortona, {\it {Hunting electroweakinos at future hadron colliders
  and direct detection experiments}},  {\em JHEP} {\bf 05} (2015) 035,
  [\href{http://arxiv.org/abs/1412.5952}{{\tt arXiv:1412.5952}}].

\bibitem{1501-06357}
H.~Baer, V.~Barger, P.~Huang, D.~Mickelson, M.~Padeffke-Kirkland, and X.~Tata,
  {\it {Natural SUSY with a bino- or wino-like LSP}},  {\em Phys. Rev.} {\bf
  D91} (2015), no.~7 075005, [\href{http://arxiv.org/abs/1501.06357}{{\tt
  arXiv:1501.06357}}].

\bibitem{1508-01173}
E.~A. Bagnaschi et~al., {\it {Supersymmetric Dark Matter after LHC Run 1}},
  {\em Eur. Phys. J.} {\bf C75} (2015) 500,
  [\href{http://arxiv.org/abs/1508.01173}{{\tt arXiv:1508.01173}}].

\bibitem{1509-05076}
K.~Freese, A.~Lopez, N.~R. Shah, and B.~Shakya, {\it {MSSM A-funnel and the
  Galactic Center Excess: Prospects for the LHC and Direct Detection
  Experiments}},  {\em JHEP} {\bf 04} (2016) 059,
  [\href{http://arxiv.org/abs/1509.05076}{{\tt arXiv:1509.05076}}].

\bibitem{1509-05771}
A.~Choudhury, K.~Kowalska, L.~Roszkowski, E.~M. Sessolo, and A.~J. Williams,
  {\it {Less-simplified models of dark matter for direct detection and the
  LHC}},  {\em JHEP} {\bf 04} (2016) 182,
  [\href{http://arxiv.org/abs/1509.05771}{{\tt arXiv:1509.05771}}].

\bibitem{blindspotsHeavyH}
P.~Huang and C.~E.~M. Wagner, {\it {Blind Spots for neutralino Dark Matter in
  the MSSM with an intermediate $m_A$}},  {\em Phys. Rev.} {\bf D90} (2014),
  no.~1 015018, [\href{http://arxiv.org/abs/1404.0392}{{\tt arXiv:1404.0392}}].

\bibitem{blindspotsStops}
A.~Crivellin, M.~Hoferichter, M.~Procura, and L.~C. Tunstall, {\it {Light
  stops, blind spots, and isospin violation in the MSSM}},  {\em JHEP} {\bf 07}
  (2015) 129, [\href{http://arxiv.org/abs/1503.03478}{{\tt arXiv:1503.03478}}].

\bibitem{blindspotsNMSSM}
M.~Badziak, M.~Olechowski, and P.~Szczerbiak, {\it {Blind spots for neutralino
  dark matter in the NMSSM}},  {\em JHEP} {\bf 03} (2016) 179,
  [\href{http://arxiv.org/abs/1512.02472}{{\tt arXiv:1512.02472}}].

\bibitem{suspect}
A.~Djouadi, J.-L. Kneur, and G.~Moultaka, {\it {SuSpect: A Fortran code for the
  supersymmetric and Higgs particle spectrum in the MSSM}},  {\em Comput. Phys.
  Commun.} {\bf 176} (2007) 426--455,
  [\href{http://arxiv.org/abs/hep-ph/0211331}{{\tt hep-ph/0211331}}].

\bibitem{PlanckXIII}
{\bf Planck} Collaboration, P.~A.~R. Ade et~al., {\it {Planck 2015 results.
  XIII. Cosmological parameters}},  {\em Astron. Astrophys.} {\bf 594} (2016)
  A13, [\href{http://arxiv.org/abs/1502.01589}{{\tt arXiv:1502.01589}}].

\bibitem{micromegas1305}
G.~Belanger, F.~Boudjema, A.~Pukhov, and A.~Semenov, {\it {micrOMEGAs 3: A
  program for calculating dark matter observables}},  {\em Comput. Phys.
  Commun.} {\bf 185} (2014) 960--985,
  [\href{http://arxiv.org/abs/1305.0237}{{\tt arXiv:1305.0237}}].

\bibitem{micromegas1005}
G.~Belanger, F.~Boudjema, A.~Pukhov, and A.~Semenov, {\it {micrOMEGAs: A Tool
  for dark matter studies}},  {\em Nuovo Cim.} {\bf C033N2} (2010) 111--116,
  [\href{http://arxiv.org/abs/1005.4133}{{\tt arXiv:1005.4133}}].

\bibitem{micromegas0803}
G.~Belanger, F.~Boudjema, A.~Pukhov, and A.~Semenov, {\it {Dark matter direct
  detection rate in a generic model with micrOMEGAs 2.2}},  {\em Comput. Phys.
  Commun.} {\bf 180} (2009) 747--767,
  [\href{http://arxiv.org/abs/0803.2360}{{\tt arXiv:0803.2360}}].

\bibitem{micromegas0607}
G.~Belanger, F.~Boudjema, A.~Pukhov, and A.~Semenov, {\it {MicrOMEGAs 2.0: A
  Program to calculate the relic density of dark matter in a generic model}},
  {\em Comput. Phys. Commun.} {\bf 176} (2007) 367--382,
  [\href{http://arxiv.org/abs/hep-ph/0607059}{{\tt hep-ph/0607059}}].

\bibitem{Ellis:1999mm}
J.~R. Ellis, T.~Falk, K.~A. Olive, and M.~Srednicki, {\it {Calculations of
  neutralino-stau coannihilation channels and the cosmologically relevant
  region of MSSM parameter space}},  {\em Astropart. Phys.} {\bf 13} (2000)
  181--213, [\href{http://arxiv.org/abs/hep-ph/9905481}{{\tt hep-ph/9905481}}].
  [Erratum: Astropart. Phys.15,413(2001)].

\bibitem{Han:2013gba}
T.~Han, Z.~Liu, and A.~Natarajan, {\it {Dark matter and Higgs bosons in the
  MSSM}},  {\em JHEP} {\bf 11} (2013) 008,
  [\href{http://arxiv.org/abs/1303.3040}{{\tt arXiv:1303.3040}}].

\bibitem{Agrawal:2014oha}
P.~Agrawal, B.~Batell, P.~J. Fox, and R.~Harnik, {\it {WIMPs at the Galactic
  Center}},  {\em JCAP} {\bf 1505} (2015) 011,
  [\href{http://arxiv.org/abs/1411.2592}{{\tt arXiv:1411.2592}}].

\bibitem{Fan:2013faa}
J.~Fan and M.~Reece, {\it {In Wino Veritas? Indirect Searches Shed Light on
  Neutralino Dark Matter}},  {\em JHEP} {\bf 10} (2013) 124,
  [\href{http://arxiv.org/abs/1307.4400}{{\tt arXiv:1307.4400}}].

\bibitem{LEP-Zpeak}
{\bf SLD Electroweak Group, DELPHI, ALEPH, SLD, SLD Heavy Flavour Group, OPAL,
  LEP Electroweak Working Group, L3} Collaboration, S.~Schael et~al., {\it
  {Precision electroweak measurements on the $Z$ resonance}},  {\em Phys.
  Rept.} {\bf 427} (2006) 257--454,
  [\href{http://arxiv.org/abs/hep-ex/0509008}{{\tt hep-ex/0509008}}].

\bibitem{Bardin:aa}
D.~Bardin, M.~Bilenky, P.~Christova, M.~Jack, L.~Kalinovskaya, A.~Olchevski,
  S.~Riemann, and T.~Riemann, {\it Zfitter v.6.21 - a semi-analytical program
  for fermion pair production in e+e- annihilation},
  \href{http://arxiv.org/abs/hep-ph/9908433}{{\tt hep-ph/9908433}}.

\bibitem{HABER198575}
H.~Haber and G.~Kane, {\it The search for supersymmetry: Probing physics beyond
  the standard model},  {\em Physics Reports} {\bf 117} (1985), no.~2 75 --
  263.

\bibitem{Djouadi:aa}
A.~Djouadi, M.~Drees, P.~F. Perez, and M.~M{\"u}hlleitner, {\it Loop induced
  higgs and z boson couplings to neutralinos and implications for collider and
  dark matter searches},  \href{http://arxiv.org/abs/hep-ph/0109283}{{\tt
  hep-ph/0109283}}.

\bibitem{LEP-Chargino1}
{\bf LEP2 SUSY Working Group} Collaboration, N.~Filippis, K.~Desch, G.~Grenier,
  C.~Hensel, A.~Perrotta, and S.~Rosier-Lees, {\it {Combined LEP Chargino
  Results, up to 208 GeV for low DM}},  {\em LEPSUSYWG/02-04.1} (2002).
  {Available at \url{http://lepsusy.web.cern.ch/lepsusy}}.

\bibitem{Abdallah:2003xe}
{\bf DELPHI} Collaboration, J.~Abdallah et~al., {\it {Searches for
  supersymmetric particles in e+ e- collisions up to 208-GeV and interpretation
  of the results within the MSSM}},  {\em Eur. Phys. J.} {\bf C31} (2003)
  421--479, [\href{http://arxiv.org/abs/hep-ex/0311019}{{\tt hep-ex/0311019}}].

\bibitem{LEP-ALEPH}
{\bf ALEPH} Collaboration, A.~Heister et~al., {\it {Search for charginos nearly
  mass degenerate with the lightest neutralino in e+ e- collisions at
  center-of-mass energies up to 209-GeV}},  {\em Phys. Lett.} {\bf B533} (2002)
  223--236, [\href{http://arxiv.org/abs/hep-ex/0203020}{{\tt hep-ex/0203020}}].

\bibitem{collaboration:aa}
{\bf OPAL} Collaboration, G.~Abbiendi et~al., {\it {Search for stable and
  longlived massive charged particles in e+ e- collisions at sqrt(s) = 130 to
  209 GeV}},  {\em Phys. Lett.} {\bf B572} (2003) 8--20,
  [\href{http://arxiv.org/abs/hep-ex/0305031}{{\tt hep-ex/0305031}}].

\bibitem{ATLAS:2014fka}
{\bf ATLAS} Collaboration, G.~Aad et~al., {\it {Searches for heavy long-lived
  charged particles with the ATLAS detector in proton-proton collisions at $
  \sqrt{s}=8 $ TeV}},  {\em JHEP} {\bf 01} (2015) 068,
  [\href{http://arxiv.org/abs/1411.6795}{{\tt arXiv:1411.6795}}].

\bibitem{CMS:2016ybj}
{\bf CMS} Collaboration, {\it {Search for heavy stable charged particles with
  $12.9~\mathrm{fb}^{-1}$ of 2016 data}},  Tech. Rep. CMS-PAS-EXO-16-036, CERN,
  Geneva, 2016.

\bibitem{ATLAS-Disappearing_Track}
{\bf ATLAS} Collaboration, G.~Aad et~al., {\it {Search for charginos nearly
  mass degenerate with the lightest neutralino based on a disappearing-track
  signature in pp collisions at $\sqrt{s}$=8 TeV with the ATLAS detector}},
  {\em Phys. Rev.} {\bf D88} (2013), no.~11 112006,
  [\href{http://arxiv.org/abs/1310.3675}{{\tt arXiv:1310.3675}}].

\bibitem{CMS-Disappearing}
{\bf CMS} Collaboration, V.~Khachatryan et~al., {\it {Search for disappearing
  tracks in proton-proton collisions at $ \sqrt{s}=8 $ TeV}},  {\em JHEP} {\bf
  01} (2015) 096, [\href{http://arxiv.org/abs/1411.6006}{{\tt
  arXiv:1411.6006}}].

\bibitem{collider-reach}
G.~Salam and A.~Weiler, {\em Collider reach $\beta$}.
\newblock {Available at \url{http://collider-reach.web.cern.ch}}.

\bibitem{twoloopdeltam}
M.~Ibe, S.~Matsumoto, and R.~Sato, {\it {Mass Splitting between Charged and
  Neutral Winos at Two-Loop Level}},  {\em Phys. Lett.} {\bf B721} (2013)
  252--260, [\href{http://arxiv.org/abs/1212.5989}{{\tt arXiv:1212.5989}}].

\bibitem{EWK}
T.~Han, S.~Padhi, and S.~Su, {\it {Electroweakinos in the Light of the Higgs
  Boson}},  {\em Phys.Rev.} {\bf D88} (2013), no.~11 115010,
  [\href{http://arxiv.org/abs/1309.5966}{{\tt arXiv:1309.5966}}].

\bibitem{CMS-SUS-13-006}
{\bf CMS Collaboration} Collaboration, CMS, {\it {Searches for electroweak
  production of charginos, neutralinos, and sleptons decaying to leptons and W,
  Z, and Higgs bosons in pp collisions at 8 TeV}},  {\em Eur. Phys. J. C} {\bf
  74} (May, 2014) 3036. 61 p.

\bibitem{1403-5294}
{\bf ATLAS} Collaboration, G.~Aad et~al., {\it {Search for direct production of
  charginos, neutralinos and sleptons in final states with two leptons and
  missing transverse momentum in $pp$ collisions at $\sqrt{s} =$ 8 TeV with the
  ATLAS detector}},  {\em JHEP} {\bf 05} (2014) 071,
  [\href{http://arxiv.org/abs/1403.5294}{{\tt arXiv:1403.5294}}].

\bibitem{1501-07110}
{\bf ATLAS} Collaboration, G.~Aad et~al., {\it {Search for direct pair
  production of a chargino and a neutralino decaying to the 125 GeV Higgs boson
  in $\sqrt{s} = 8$ TeV ${pp}$ collisions with the ATLAS detector}},  {\em Eur.
  Phys. J.} {\bf C75} (2015), no.~5 208,
  [\href{http://arxiv.org/abs/1501.07110}{{\tt arXiv:1501.07110}}].

\bibitem{CMS-SUS-14-002}
{\bf CMS Collaboration} Collaboration, CMS, {\it {Searches for electroweak
  neutralino and chargino production in channels with Higgs, Z, and W bosons in
  pp collisions at 8 TeV}},  {\em Phys. Rev. D} {\bf 90} (Sep, 2014) 092007. 61
  p, [\href{http://arxiv.org/abs/1409.3168}{{\tt 1409.3168}}].

\bibitem{Bino-Higgsino1}
D.~Barducci, A.~Belyaev, A.~K.~M. Bharucha, W.~Porod, and V.~Sanz, {\it
  {Uncovering Natural Supersymmetry via the interplay between the LHC and
  Direct Dark Matter Detection}},  {\em JHEP} {\bf 07} (2015) 066,
  [\href{http://arxiv.org/abs/1504.02472}{{\tt arXiv:1504.02472}}].

\bibitem{Bino-Higgsino2}
C.~Han, D.~Kim, S.~Munir, and M.~Park, {\it {Accessing the core of naturalness,
  nearly degenerate higgsinos, at the LHC}},  {\em JHEP} {\bf 04} (2015) 132,
  [\href{http://arxiv.org/abs/1502.03734}{{\tt arXiv:1502.03734}}].

\bibitem{Bino-Higgsino3}
N.~Nagata and S.~Shirai, {\it {Higgsino Dark Matter in High-Scale
  Supersymmetry}},  {\em JHEP} {\bf 01} (2015) 029,
  [\href{http://arxiv.org/abs/1410.4549}{{\tt arXiv:1410.4549}}].

\bibitem{Wino-Higgsino1}
A.~Anandakrishnan, L.~M. Carpenter, and S.~Raby, {\it {Degenerate gaugino mass
  region and mono-boson collider signatures}},  {\em Phys. Rev.} {\bf D90}
  (2014), no.~5 055004, [\href{http://arxiv.org/abs/1407.1833}{{\tt
  arXiv:1407.1833}}].

\bibitem{Wino-Higgsino2}
M.~Low and L.-T. Wang, {\it {Neutralino dark matter at 14 TeV and 100 TeV}},
  {\em JHEP} {\bf 08} (2014) 161, [\href{http://arxiv.org/abs/1404.0682}{{\tt
  arXiv:1404.0682}}].

\bibitem{Wino-Higgsino3}
M.~Badziak, A.~Delgado, M.~Olechowski, S.~Pokorski, and K.~Sakurai, {\it
  {Detecting underabundant neutralinos}},  {\em JHEP} {\bf 11} (2015) 053,
  [\href{http://arxiv.org/abs/1506.07177}{{\tt arXiv:1506.07177}}].

\bibitem{1310-4274}
C.~Han, A.~Kobakhidze, N.~Liu, A.~Saavedra, L.~Wu, and J.~M. Yang, {\it
  {Probing Light Higgsinos in Natural SUSY from Monojet Signals at the LHC}},
  {\em JHEP} {\bf 02} (2014) 049, [\href{http://arxiv.org/abs/1310.4274}{{\tt
  arXiv:1310.4274}}].

\bibitem{1511-05386}
J.~Cao, Y.~He, L.~Shang, W.~Su, and Y.~Zhang, {\it {Testing the light dark
  matter scenario of the MSSM at the LHC}},  {\em JHEP} {\bf 03} (2016) 207,
  [\href{http://arxiv.org/abs/1511.05386}{{\tt arXiv:1511.05386}}].

\bibitem{Alwall:2011aa}
J.~Alwall, M.~Herquet, F.~Maltoni, O.~Mattelaer, and T.~Stelzer, {\it Madgraph
  5 : Going beyond},  \href{http://arxiv.org/abs/1106.0522}{{\tt
  arXiv:1106.0522}}.

\bibitem{Sjostrand:2006aa}
T.~Sjostrand, S.~Mrenna, and P.~Skands, {\it Pythia 6.4 physics and manual},
  {\em JHEP} {\bf 0605} (2006) 026,
  [\href{http://arxiv.org/abs/hep-ph/0603175}{{\tt hep-ph/0603175}}].

\bibitem{Ovyn:2009aa}
S.~Ovyn, X.~Rouby, and V.~Lemaitre, {\it Delphes, a framework for fast
  simulation of a generic collider experiment},
  \href{http://arxiv.org/abs/0903.2225}{{\tt arXiv:0903.2225}}.

\bibitem{Favereau:2013aa}
J.~de~Favereau, C.~Delaere, P.~Demin, A.~Giammanco, V.~Lema{\^\i}tre,
  A.~Mertens, and M.~Selvaggi, {\it Delphes 3, a modular framework for fast
  simulation of a generic collider experiment},
  \href{http://arxiv.org/abs/1307.6346}{{\tt arXiv:1307.6346}}.

\bibitem{Anderson:2013aa}
J.~Anderson, A.~Avetisyan, R.~Brock, S.~Chekanov, T.~Cohen, N.~Dhingra,
  J.~Dolen, J.~Hirschauer, K.~Howe, A.~Kotwal, T.~LeCompte, S.~Malik,
  P.~Mcbride, K.~Mishra, M.~Narain, J.~Olsen, S.~Padhi, M.~E. Peskin, J.~S.
  III, and J.~G. Wacker, {\it Snowmass energy frontier simulations},
  \href{http://arxiv.org/abs/1309.1057}{{\tt arXiv:1309.1057}}.

\bibitem{Hoecker:aa}
A.~Hoecker, P.~Speckmayer, J.~Stelzer, J.~Therhaag, E.~von Toerne, H.~Voss,
  M.~Backes, T.~Carli, O.~Cohen, A.~Christov, D.~Dannheim, K.~Danielowski,
  S.~Henrot-Versille, M.~Jachowski, K.~Kraszewski, A.~K. Jr., M.~Kruk,
  Y.~Mahalalel, R.~Ospanov, X.~Prudent, A.~Robert, D.~Schouten, F.~Tegenfeldt,
  A.~Voigt, K.~Voss, M.~Wolter, and A.~Zemla, {\it Tmva - toolkit for
  multivariate data analysis},
  \href{http://arxiv.org/abs/physics/0703039}{{\tt physics/0703039}}.

\bibitem{Cranmer:2012sba}
{\bf ROOT} Collaboration, K.~Cranmer, G.~Lewis, L.~Moneta, A.~Shibata, and
  W.~Verkerke, {\it {HistFactory: A tool for creating statistical models for
  use with RooFit and RooStats}}, .

\bibitem{Verkerke:2003ir}
W.~Verkerke and D.~P. Kirkby, {\it {The RooFit toolkit for data modeling}},
  {\em eConf} {\bf C0303241} (2003) MOLT007,
  [\href{http://arxiv.org/abs/physics/0306116}{{\tt physics/0306116}}].
  [,186(2003)].

\bibitem{Moneta:2010pm}
L.~Moneta, K.~Belasco, K.~S. Cranmer, S.~Kreiss, A.~Lazzaro, D.~Piparo,
  G.~Schott, W.~Verkerke, and M.~Wolf, {\it {The RooStats Project}},  {\em PoS}
  {\bf ACAT2010} (2010) 057, [\href{http://arxiv.org/abs/1009.1003}{{\tt
  arXiv:1009.1003}}].

\bibitem{Schwaller:2013baa}
P.~Schwaller and J.~Zurita, {\it {Compressed electroweakino spectra at the
  LHC}},  {\em JHEP} {\bf 03} (2014) 060,
  [\href{http://arxiv.org/abs/1312.7350}{{\tt arXiv:1312.7350}}].

\bibitem{vbf-tao}
G.~F. Giudice, T.~Han, K.~Wang, and L.-T. Wang, {\it {Nearly Degenerate
  Gauginos and Dark Matter at the LHC}},  {\em Phys. Rev.} {\bf D81} (2010)
  115011, [\href{http://arxiv.org/abs/1004.4902}{{\tt arXiv:1004.4902}}].

\end{thebibliography}\endgroup

\end{document}